\documentclass[nofootinbib]{revtex4-1}
\usepackage[utf8]{inputenc}
\usepackage{amsmath,latexsym}
\usepackage{mathtools}
\usepackage{hyperref}
\usepackage{subcaption}
\usepackage{amssymb}
\usepackage{graphicx}
\RequirePackage[a4paper,top=3.0cm,left=3cm,right=2cm,bottom=2.5cm]{geometry}
\usepackage{multirow}

\let\oldref\ref
\renewcommand{\ref}[1]{(\oldref{#1})}
\begin{document}
\title{The physics of gravitational waves}
\author{Enrico Barausse}
\affiliation{SISSA, Via Bonomea 265, 34136 Trieste, Italy and INFN Sezione di Trieste}
\affiliation{IFPU - Institute for Fundamental Physics of the Universe, Via Beirut 2, 34014 Trieste, Italy}

\begin{abstract}
These lecture notes collect the material that I have been using over the years for various short courses on the physics of gravitational waves, first at the Institut d'Astrophysique de Paris (France), and then at SISSA (Italy) and various summer/winter schools. 
The level should be appropriate for PhD students in physics or for MSc students that have taken a first course in general relativity. I try as much as possible to derive results from first principles and focus on the physics, rather than on astrophysical applications. The reason is not only that the latter require a solid understanding of the physics, but it also lies in the trove of data that are being uncovered by the LIGO-Virgo-KAGRA collaboration after the first direct detection of gravitational waves~\cite{GW150914}. Any attempt to summarize such a rich and fast changing landscape and its evolving astrophysical interpretation is bound to become obsolete before the ink hits the page. 
\end{abstract}

\maketitle
\tableofcontents


\section{Prerequisites}
\label{sec:prerequisites}
These notes assume familiarity with Einstein's equations, which in units with $G=c=1$
can be written as 
\begin{equation}
\label{eq:Einsteinequations}
    G_{\mu \nu} = 8 \pi  T_{\mu \nu}\,,
\end{equation}
with $G^{\mu\nu}$ the Einstein tensor and $T^{\mu\nu}$ the matter stress energy tensor.
It may be useful to recall that because of the Bianchi identity $\nabla_\mu G^{\mu \nu}=0$, the stress energy tensor  satisfies the ``conservation'' equation $\nabla_\mu T^{\mu \nu}=0$ on shell.

For a perfect fluid,
in the $(-+++)$ signature that we will use throughout these notes, 
the stress energy tensor takes the form
\begin{equation}
    \label{eq:perfectfluidstressenergy}
    T^{\mu \nu} = (\rho+p)u^\mu u^\nu + p g^{\mu \nu},
\end{equation}
where $u^\mu$ is the 4-velocity of the fluid element, $\rho$ the energy density and $p$ the pressure. 
With this ansatz, the conservation of the stress energy tensor implies
\begin{equation}
\label{eq:perfectfluidconservation}
    \frac{\mathrm{d \rho}}{\mathrm{d}\tau} = -(p+\rho) \nabla_\mu u^\mu\,,
\end{equation}
with $\tau$ the fluid element's proper time,
and \begin{equation}
\label{eq:fluidacceleration}
    a^\mu = -\frac{\gamma^{\mu \nu} \partial_\nu p}{\rho+p}\,,
\end{equation}
with $a^\mu=u^\nu \nabla_\nu u^\mu$ the 4-acceleration and 
\begin{equation}
    \gamma^{\mu \nu} = g^{\mu \nu}+u^\mu u^\nu
\end{equation}
the projector on the hypersurface orthogonal to $u^\mu$.
Let us recall that Eq.~\ref{eq:perfectfluidconservation} simply encodes the conservation of energy, while
Eq.~\ref{eq:fluidacceleration} generalizes the Newtonian Euler equation. In particular, for $p=0$
the relativistic Euler equation reduces to the geodesic equation $a^\mu=0$.

It is worth recalling that the stress energy tensor can be defined 
in terms of the functional derivative of the matter action, i.e.
\begin{equation}
    T^{\mu \nu} = \frac{2}{\sqrt{-g}}\frac{\delta S_{\mathrm{pp}}}{\delta g_{\mu \nu}}.
\end{equation}
For a point particle of mass $m$, the action is simply given by
\begin{equation}
\label{eq:actionpointparticle}
    S_{\mathrm{pp}} =  -m \int \mathrm{d}\tau\,,
\end{equation}
where the integral is along the trajectory. By varying this action with respect to the trajectory, one
obtains the geodesic equation $a^\mu=0$, while the functional derivative with respect to the metric yields
\begin{equation}
\label{eq:pointparticlestressenergy}
     T^{\mu \nu}_{\mathrm{pp}} = 
     \frac{m}{\sqrt{-g}} \delta^{(3)} \left( \vec{x}-\vec{X}(t) \right)\frac{u^\mu u^\nu}{u^t}\,,
\end{equation}
with $\vec{X}(t)$ the trajectory.
This stress energy tensor can be mapped into that of a perfect fluid with $p=0$ (dust).
The same clearly applies to a collection of point particles. We can therefore conclude that for 
point particles, the geodesic equation is implied by the conservation of the stress energy tensor.

Finally, let us recall that 
the ``divergence'' or ``focusing'' of neighboring geodesics is described by the geodesic deviation equation. More precisely, given a family of geodesics $x^\mu(v,\tau)$ labeled by a parameter $v$ and the proper time $\tau$, let us introduce the separation vector $v^\mu=(\partial x^\mu/\partial v) \delta v$ joining two neighboring geodesics with parameters $v$ and $v+\delta v$. This vector then satisfies the geodesic deviation equation
\begin{equation}
\frac{\mathrm{D}^2 v^\mu}{\mathrm{d}\tau^2}=R^{\mu}_{\alpha\beta\gamma}u^\alpha u^\beta v^\gamma\,,    
\end{equation}
with $R^{\mu}_{\alpha\beta\gamma}$ the Riemann tensor and $\mathrm{D}/\mathrm{d}\tau$ the covariant derivative along the four velocity.
\\
\\
{\bf Exercise 1:} {\it From the action of a point particle, $S=-m\int d\tau$, derive the geodesics equations by varying with respect to the trajectory, and the stress energy tensor by varying with respect to the metric.}

\section{The propagation and generation of gravitational waves}
    \subsection{Linear  perturbations on flat space}\label{flat_prop}
Let us start by considering generic vacuum perturbations $h_{\mu \nu}$ of a flat background spacetime. The perturbed spacetime's metric  at linear order is therefore given by
\begin{equation}
    g_{\mu \nu} = \eta_{\mu \nu} + h_{\mu \nu}\,, \qquad g^{\mu \nu} = \eta^{\mu \nu} - h^{\mu \nu}\,, 
\end{equation}
where indices are meant to be  raised and lowered with the Minkowski metric $\eta_{\mu \nu}$. We will now derive 
the (linearized) Einstein equations for the metric perturbation $h_{\mu\nu}$.

Let us first recall that the gauge group of GR is given by diffeomorphisms, i.e. coordinate transformations. Under
a transformation  $\Tilde{x}^\mu = \Tilde{x}^\mu(x)$, the  metric transforms as 
\begin{equation}\label{diffeo}
    \Tilde{g}_{\mu \nu} \left(\Tilde{x}(x) \right) = \frac{\partial x^\alpha}{\partial \Tilde{x}^\mu} \frac{\partial x^\beta}{\partial \Tilde{x}^\nu} g_{\alpha \beta}(x)\,.
\end{equation}
If the coordinate transformation is ``infinitesimal'', $\Tilde{x} ^\mu = x^\mu + \xi^\mu$ with $\partial_\nu\xi^\mu \ll 1$, 
a Taylor expansion of Eq.~\ref{diffeo} implies that the metric perturbation transforms as
\begin{equation}
\label{eq:metricdiffeomorphism}
    \Tilde{h}_{\mu \nu} = h_{\mu \nu} - \mathcal{L}_\xi \eta_{\mu \nu} = h_{\mu \nu} - \partial_\mu \xi_\nu- \partial_\nu \xi_\mu.
\end{equation}
where $\mathcal{L}_\xi$ is the Lie derivative along  the vector field $\xi$. 

Let us use this gauge freedom to impose the Lorenz gauge condition
\begin{equation}
    \label{eq:lorenz}
    \partial_\mu \bar{h}^{\mu \nu} =0\,,
\end{equation}
where we have introduced the trace-reversed metric perturbation
$\bar{h}_{\mu \nu} = h_{\mu \nu}-\frac{1}{2}h \eta_{\mu \nu}$, with $h=\eta_{\mu\nu} h^{\mu\nu}$ the trace.
Note that this condition is also known as de Donder gauge condition, or also as harmonic gauge condition (as it can be easily proven that it is equivalent to $\Box x^\nu=0$, where $\Box = \eta^{\mu \nu} \partial_\mu \partial_\nu$ is the flat space d'Alembertian and the $x^\nu$ are scalar functions defining the coordinates).
In this gauge, it is straightforward to show that
the linearized Ricci tensor is
\begin{equation}
    R_{\mu \nu} = -\frac{1}{2} \Box h_{\mu \nu}\,.
\end{equation}

The Einstein equations can be written in the two equivalent forms
\begin{equation}
   R_{\mu \nu} -\frac{1}{2}Rg_{\mu \nu} = 8 \pi T_{\mu \nu} \quad \Leftrightarrow \quad R_{\mu \nu} = 8 \pi \left( T_{\mu \nu} -\frac{1}{2}Tg_{\mu \nu} \right).
\end{equation}
From the second expression, the linearized equations can therefore be written as
\begin{equation}
    \Box h_{\mu \nu} = -16 \pi  \left( T_{\mu \nu} -\frac{1}{2}Tg_{\mu \nu} \right)  \quad \Leftrightarrow \quad \Box \bar{h}_{\mu \nu} = -16 \pi   T_{\mu \nu}\,,
\end{equation}
respectively in terms of $ h_{\mu \nu} $ and $ \bar{h}_{\mu \nu} $. In vacuum ($T_{\mu\nu}=0$), 
both quantities satisfy the homogeneous wave equation
\begin{equation}
    \Box \bar{h}_{\mu \nu}=\Box {h}_{\mu \nu}=0.
\end{equation}
From this, it can already be seen that the metric perturbations (i.e. gravitational waves) on flat space travel at the speed of light.

Let us now further investigate how many independent components/propagating degrees of freedom the metric perturbation has. In principle, a symmetric rank-2 tensor has 10 independent components. The
Lorenz gauge condition is a vector equation and removes 4 of them, and hence one would expect 6 degrees of freedom.
However, let us note that the Lorenz gauge condition does not fix completely the gauge. In fact, let us consider a metric perturbation $h_{\mu \nu}$ that respects the Lorenz condition, and let us perform an infinitesimal change of coordinates. According to Eq.~\ref{eq:metricdiffeomorphism}, one has
\begin{equation}
\begin{aligned}
    & \tilde{h}_{\mu \nu}= h_{\mu \nu} - \partial_\mu \xi_\nu- \partial_\nu \xi_\mu, \\
    & \tilde{h} = h-2\partial_\mu \xi^\mu,
\end{aligned}    
\end{equation}
and thus
\begin{equation}
\label{eq:bartransformation}
    \bar{\tilde{h}}_{\mu \nu}= \bar{h}_{\mu \nu} - \partial_\mu \xi_\nu- \partial_\nu \xi_\mu + \eta_{\mu \nu} \partial_\alpha \xi^\alpha.
\end{equation}
We therefore see that
\begin{equation}
    \partial_\mu \bar{\tilde{h}}^{\mu \nu} = \partial_{\mu} \bar{h}^{\mu \nu}- \Box \xi^\nu\,.
\end{equation}
Clearly, if one starts with a perturbation in the Lorenz gauge, any gauge related to the original one by harmonic generators (i.e. ones such that $\Box \xi^\mu=0$\footnote{This terminology derives from the fact that the d'Alembertian is the Minkowskian generalization of the (Euclidean) Laplacian $\nabla^2 = \delta^{ij}\partial_i \partial_j$. In vector analysis, a function that satisfies the Laplace equation $\nabla^2 f=0$ is called harmonic. This is because the eigenfunctions of the Laplacian on the sphere are called ``spherical harmonics'', since they are the higher dimensional analog of the Fourier basis, consisting of sines and cosines, which describes  harmonic motion.})
still satisfies the Lorenz gauge condition. In other words, the Lorenz
gauge is defined up to a harmonic gauge generator.\footnote{This is also obvious because as mentioned above, the harmonic gauge condition can also be written as $\Box x^\mu=0$.}

Let us now exploit this residual gauge freedom to simplify the (trace reversed) metric perturbation
$\bar{h}_{\mu \nu}$ in vacuum. Since the latter satisfies $\Box \bar{h}_{\mu \nu}=0$, 
it can be decomposed, without loss of generality, in planes waves:
\begin{equation}
    \bar{h}_{\mu \nu}(x) = A_{\mu \nu} e^{ik_\alpha x^\alpha}+ \ \mathrm{c.c.}\text{    ,}
\end{equation}
where $\mathrm{c.c.}$ denotes the complex conjugate, $A_{\mu \nu}$ is a constant polarization tensor, and
$k^\mu$ is a null-vector ($k_\alpha k^\alpha=0$). Similarly, the (residual) gauge generator must satisfy
$\Box \xi^\mu=0$, and thus
\begin{equation}
    \xi^\mu (x) = B^\mu e^{ik_\alpha x^\alpha}+ \ \mathrm{c.c.}  \text{    ,}
\end{equation}
where again $k_\alpha k^\alpha=0$ and $B^\mu$ is a constant vector.
Imposing now the condition \ref{eq:lorenz}, one finds that
\begin{equation}
\label{eq:polarizationorthogonal}
    k^\mu A_{\mu \nu}=0,
\end{equation}
i.e. the wavevector must belong to the null space of the polarization tensor.

Considering now for simplicity a wave propagating along the $z$ axis, i.e.
\begin{equation}
    k = \frac{\partial}{\partial t} + \frac{\partial}{\partial z}\,,
\end{equation}
Eq.~\ref{eq:polarizationorthogonal} yields
\begin{equation}
\label{eq:trasversality}
    A^{t \nu} = A^{z \nu}.
\end{equation}
Evaluating the transformation given by Eq.~\ref{eq:bartransformation} for $\mu=t$ and $\nu =i$, one also finds
\begin{equation}
    \bar{\tilde{h}}_{t i} = \bar{h}_{ti} - \partial_t \xi_i- \partial_i \xi_t = \bar{h}_{ti} - i k_t B_i e^{i k_\alpha x^\alpha}- i k_i B_t e^{i k_\alpha x^\alpha}\,,
\end{equation}
from which it follows that we can choose $B_i$ such that $ \bar{\tilde{h}}_{ti}=0$.
Moreover, the condition \ref{eq:trasversality} for $\nu=i$ implies $A_{zi}=0$ (since
$ \bar{\tilde{h}}_{ti}=0$ and thus $A_{ti}=0$). Evaluating the same condition for $\nu=t$, one has instead  $A^{tt}=A^{zt}=0$.
Looking the gauge transformation for the trace,
\begin{equation}
    \tilde{h}=h-2 \partial_\mu \xi^\nu = h-2i k_\mu B^\mu e^{ik_\alpha x^\alpha},
\end{equation}
one can finally see  that by an appropriate choice of $B^t$ one can set $\tilde{h}=0$.

In  light of the above, the residual gauge freedom   allows one to write the polarization tensor as
\begin{equation}
    \left(A_{\mu \nu}\right) = \begin{pmatrix}
0 & 0 & 0 & 0\\
0 & h_+ & h_\times & 0\\
0 & h_\times & -h_+ & 0\\
0 & 0 & 0 & 0\\
\end{pmatrix}.
\end{equation}
The conclusion is that gravitational waves 
propagating on flat space
have only two independent transverse polarizations, i.e. there exist only two propagating degrees of freedom.

\subsection{Linear  perturbations on curved space}\label{curved_space_pert}
Let us now  generalize the previous calculation to a generic curved background. Again, the spacetime metric is given by the background metric $g_{\mu \nu}$ and a perturbation $h_{\mu \nu}$,
and the trace reversed perturbation can  be defined as  $\bar{h}_{\mu \nu} = h_{\mu \nu} -\frac{1}{2}h g_{\mu \nu}$, with $h=h_{\mu\nu} g^{\mu\nu}$. Indices are understood to be raised and lowered with the background metric. Choosing the Lorenz gauge condition $\nabla_\mu \bar{h}^{\mu \nu}=0$, where $\nabla$ is the covariant derivative
compatible with the background metric $g_{\mu \nu}$, 
and linearizing the Einstein equations, one finds~\cite{poisson}
\begin{equation}
\label{eq:eomcurvedspacetime}
    \Box \bar{h}^{\alpha \beta} + 2 R_{\mu \ \nu}^{\ \alpha \ \beta} \bar{h}^{\mu \nu} + S_{\mu \ \nu}^{\ \alpha \ \beta} \bar{h}^{\mu \nu} = -16 \pi T^{\alpha \beta},
\end{equation}
where $\Box = g^{\mu \nu}\nabla_\mu \nabla_\nu$, $R_{\mu \nu \rho \sigma}$ is the (background) Riemann tensor and
\begin{equation}
    S_{\mu \alpha \nu \beta} = 2G_{\mu (\alpha} g_{\beta) \nu}-R_{\mu \nu} g_{\alpha \beta}-2g_{\mu \nu} G_{\alpha \beta}\,.
\end{equation}
In vacuum $R_{\mu \nu}=T_{\mu \nu} = G_{\mu \nu} = S_{\mu \alpha \nu \beta}=0$, and Eq.~\ref{eq:eomcurvedspacetime} becomes
\begin{equation}\label{eins}
    \Box \bar{h}^{\alpha \beta} + 2 R_{\mu \ \nu}^{\ \alpha \ \beta} \bar{h}^{\mu \nu}=0.
\end{equation}
Note in particular the coupling between gravitational waves and the background's Riemann tensor, which affects  their propagation (i.e. gravitational waves can be scattered by the curvature and can therefore also travel inside an observer's lightcone).

Let us again count the physical (propagating) degrees of freedom in vacuum. After fixing the gauge,  
$h_{\mu \nu}$ still has, in principle, 6 independent components. 
However, proceeding like in the flat background case, one easily finds that in vacuum the Lorenz gauge leaves a residual gauge freedom, namely one can always perform a gauge transformation with a harmonic generator (i.e. a generator $\xi^\mu$ obeying $\Box \xi^\mu=0$) and still preserve the harmonic gauge condition. 
Under a gauge transformation, the metric perturbation transforms as
\begin{equation}
\label{eq:metricdiffeomorphism2}
    \Tilde{h}_{\mu \nu} = h_{\mu \nu} - \mathcal{L}_\xi g_{\mu \nu} = h_{\mu \nu} - \nabla_\mu \xi_\nu- \nabla_\nu \xi_\mu\,,
\end{equation}
hence the trace transforms according to $  \Tilde{h}=h-2\nabla_\mu \xi^\mu$. Let us first try to set the trace
$  \Tilde{h}=0$ using the residual gauge freedom of the Lorenz gauge. That would require choosing $\xi^\mu$ such that 
\begin{equation}\label{nulltrace}
 \nabla_\mu \xi^\mu-\frac{h}{2}=0
\end{equation}
 throughout the whole spacetime.

To see if this requirement is compatible with the Lorenz gauge condition's residual freedom, let us note that  
the generators of the latter must obey $\Box \xi^\mu=0$ and are therefore completely characterized  by the initial conditions to this wave equation, i.e.
$\xi^\mu(t=0,x^i)$ and $\partial_t\xi^\mu(t=0,x^i)$. 
It is not a priori obvious that by choosing these initial conditions properly, Eq.~\ref{nulltrace}
can be satisfied in the whole spacetime. However, taking a d'Alembertian of
Eq.~\ref{nulltrace}, an involved calculation using the trace of Eq.~\ref{eins} and $\Box \xi^\mu=0$ yields
\begin{equation}
    \Box \left(\nabla_\mu \xi^\mu - \frac{h}{2} \right) = 0.
\end{equation}
Therefore, for Eq.~\ref{nulltrace}
to be satisfied in the whole spacetime, we just need 
to impose $\nabla_\mu \xi^\mu - {h}/{2}=\partial_t(\nabla_\mu \xi^\mu - {h}/{2})=0$ at $t=0$.
This can be attained by choosing the initial conditions 
$\xi^\mu(t=0,x^i)$ and $\partial_t\xi^\mu(t=0,x^i)$ characterizing the residual gauge freedom~\cite{Flanagan:2005yc}.

In conclusion,  also in curved spacetime
the trace of the metric perturbations can be set to zero in the Lorenz gauge. However,
showing that only
two non-zero components ($h_+$ and $h_\times$) survive, like in Minkowski space, is in general not possible. We will shed light on this fact in the next section, where we will do perturbation theory in a slightly different way, 
by exploiting  a scalar-vector-tensor-decomposition of the metric perturbation. As we will see, there will still be just two {\it propagating} degrees of freedom for the gravitational field, but additional 
 {\it non-propagating} potentials will be present, including and generalizing the Newtonian potential.

\subsection{Linear perturbations on flat space: a scalar-vector-tensor decomposition}

\label{sec:2}

To gain more insight on the degrees of freedom of the metric perturbation, let us go back to the case of a flat background spacetime. 
Let us introduce a book-keeping parameter $\epsilon \ll 1$
and write $g_{\mu \nu}= \eta_{\mu \nu}+\epsilon h_{\mu \nu}$. Moreover, let us describe  matter by the perfect fluid stress energy tensor 
\begin{equation}
    T^{\mu \nu} = \epsilon\left[(\rho+p)u^\mu u^\nu + p g^{\mu \nu} \right].
\end{equation}

One can then split the components of the metric perturbation 
 according to their transformation properties under spatial rotations. For instance, $h_{tt}$ is a scalar under rotations, $h_{ti}$ a vector,
 $h_{ij}$ a tensor. Moreover, we can perform a Helmholtz decomposition of
 the vector into the gradient of a scalar plus a divergenceless vector (i.e. a curl), and similarly decompose the tensor into two scalars, a divergenceless vector and a transverse traceless tensor.
 As a result, one has~\cite{Flanagan:2005yc,Bardeen,Mukhanov}
\begin{equation}
\begin{aligned}
    & h_{tt} = 2 \phi, \\
    & h_{ti} = \partial_i \gamma + \beta_i \quad \mathrm{with} \ \partial_i \beta^i = 0,\\
    & h_{ij} = \frac{1}{3} H \delta_{ij} + \partial_{(i}\varepsilon_{j)}+ \left(\partial_i \partial_j-\frac{1}{3}\delta_{ij}\nabla^2 \right) \lambda + h_{ij}^{\mathrm{TT}},\\
    & \quad \quad \ \mathrm{with} \ \partial_i \varepsilon^i = 0 \ \mathrm{and} \  \partial_i h^{\mathrm{TT}ij}=0=h^{\mathrm{TT}i}_{\ \ \ \ i} .
\end{aligned}
\end{equation}
Here, spatial indices are understood to be raised and lowered with the Euclidean metric $\delta_{ij}$; $\phi$, $\gamma$, $H$ and $\lambda$ are scalars under spatial rotations; $\beta$ and $\varepsilon$ are divergenceless vectors and $h^{\mathrm{TT}}$ is a transverse ($\partial_i h^{\mathrm{TT}ij}=0$) and traceless tensor. One can easily verify that the number of degrees of freedom of this decomposition is correct. For instance, $h_{ti}$
has three independent components, which correspond to $\gamma$ (one degree of freedom) and $\beta_i$ (two degrees of freedom). Similarly,
$h_{ij}$ has 6 independent components, which correspond
to the scalars $\lambda$ and $H$ (one degree of freedom each), the divergenceless vector $\varepsilon_i$ (two degrees of freedom) and the 
transverse traceless tensor $ h^{\mathrm{TT}ij}$ (two degrees of freedom). Moreover, one can verify that these decompositions are uniquely defined (up to boundary conditions). For instance,  to obtain $\gamma$ we can compute
\begin{equation}
    \partial_i h_{ti} = \nabla^2 \gamma + \partial_i \beta_i =  \nabla^2 \gamma,
\end{equation}
and we can formally invert this expression to obtain
\begin{equation}
    \gamma = \nabla^{-2} \left(\partial_i h_{ti} \right).
\end{equation}
Here $ \nabla^{-2}$ is the inverse of the (Euclidean) Laplacian, which is well defined if boundary conditions are given for $\gamma$ (e.g. it is reasonable to assume that  $\gamma$ decays ``fast'' at large distances to preserve asymptotic flatness) and which can be expressed explicitly in terms of a Green function (see below). Once $\gamma$ is determined, 
$\beta_i$ can be computed as
\begin{equation}
    \beta_i = h_{ti}-\partial_i \left[ \nabla^{-2} \left(\partial_i h_{ti} \right) \right].
\end{equation}
Similarly, one can show that the decomposition of $h_{ij}$ is well defined and unique by computing $h^i_i$, $\partial_i h^{ij}$ and 
$\partial_i\partial_j h^{ij}$.

A similar decomposition can be performed on the
stress energy tensor~\cite{Flanagan:2005yc},
\begin{equation}
\begin{aligned}
    & T_{tt} = \rho ,\\
    & T_{ti} = \partial_i S + S_i \quad \mathrm{with} \ \partial_i S^i = 0 ,\\
    & T_{ij} = p \delta_{ij} + \partial_{(i}\sigma_{j)}+ \left(\partial_i \partial_j-\frac{1}{3}\delta_{ij}\nabla^2 \right) \sigma + \sigma_{ij} ,\\
    \label{eq:stressenergypoisson}
    & \quad \quad \ \mathrm{with} \ \partial_i \sigma^i = 0 \ \mathrm{and} \  \partial_i \sigma^{ij}=0=\sigma^{i}_{ \ i},
\end{aligned}
\end{equation}
where $\rho$, $S$, $p$, $\sigma$ are scalars, $S_i$ and $\sigma_i$ divergenceless vectors and $\sigma_{ij}$ a transverse traceless tensor. Similarly, the generator $\xi^\mu$ of infinitesimal coordinate transformations can be expressed as
\begin{equation}
\begin{aligned}
    & \xi^t = A ,\\
    & \xi^i = \partial_i C + B_i, \quad \mathrm{with} \ \partial_i B^i = 0\,
\end{aligned}    
\end{equation}
with $A$ and $C$ scalars and $B_i$ a divergenceless vector.

By using this decomposition for the generator in Eq.~\ref{eq:metricdiffeomorphism}, we obtain~\cite{Flanagan:2005yc}
\begin{eqnarray}
\tilde{\phi} &=&\phi - \partial_t A\;,
\label{eq:phi_gauge}\\
\tilde{\beta}_i&=& \beta_i - \partial_t B_i\;,
\label{eq:beta_gauge}\\
\tilde{\gamma} &=& \gamma - A - \partial_t C\;,
\label{eq:gamma_gauge}\\
\tilde{H} &=& H - 2\nabla^2C\;,
\label{eq:h_gauge}\\
\tilde{\lambda}&=& \lambda - 2 C\;,
\label{eq:lambda_gauge}\\
\tilde{\varepsilon}_i &=& \varepsilon_i - 2 B_i\;,
\label{eq:epsilon_gauge}\\
\tilde{h}_{ij}^{\rm TT} &=& h_{ij}^{\rm TT}\;.
\label{eq:hTT_gauge}
\end{eqnarray}
First, let us notice that $h_{ij}^{\rm TT}$ is gauge invariant. Moreover, one can remove two scalars and one divergenceless vector 
 by a suitable choice of $A,C$ and $B_i$. For instance, one can  choose to remove $\gamma,\lambda$ and $\varepsilon_i$, so that
\begin{equation}\label{eq:ansatz}
\begin{aligned}
    & h_{tt} = 2 \phi, \\
    & h_{ti} = \beta_i,\\
    & h_{ij} = \frac{1}{3} H \delta_{ij} + h_{ij}^{\mathrm{TT}}.
\end{aligned}    
\end{equation}
This particular choice is called Poisson gauge, and unlike
the Lorenz gauge, it fixes completely the coordinates at linear order (i.e. there is no residual gauge freedom).

Alternatively, one can construct particular combinations of the scalar and vector potentials that are gauge invariant (recall that $h_{ij}^{\rm TT}$ is already gauge invariant). These are 
 Bardeen's gauge invariant variables~\cite{Bardeen}, i.e.
\begin{equation}
\begin{aligned}
    & \psi = -\phi + \partial_t{\gamma} -\frac{1}{2} \partial^2_t{\lambda} \,,\\
    & \theta = \frac{1}{3} \left(H-\nabla^2 \lambda \right) \,,\\
    & \Sigma_i = \beta_i -\frac{1}{2} \partial_t{\varepsilon}_i\,,
\end{aligned}
\end{equation}
which reduce respectively to $-\phi$, $H/3$ and $ \beta_i$ in the Poisson gauge. Thus, using the Poisson gauge is exactly equivalent to using Bardeen's gauge invariant variables.

We can now express the linearized Einstein equations in terms of the Bardeen variables (or alternatively compute them in the Poisson gauge). For the Einstein tensor we obtain~\cite{Flanagan:2005yc}
\begin{equation}
\begin{aligned}
    & G_{tt} = - \nabla^2 \theta ,\\
    & G_{ti} = - \frac{1}{2}\nabla^2 \Sigma_i - \partial_i \partial_t{\theta},\\
    & G_{ij} = -\frac{1}{2} \Box h_{ij}^\mathrm{TT} - \partial_{(i}\partial_t{\Sigma}_{j)}-\frac{1}{2}\partial_i \partial_j \left(2 \psi+\theta \right) + \delta_{ij} \left[\frac{1}{2}\nabla^2 \left(2 \psi + \theta \right) - \partial_t^2{\theta}\right]\,,
\end{aligned}    
\end{equation}
with $\Box$ the flat d'Alembertian.
Decomposing also the right-hand side (i.e. the stress energy tensor), we find that for the $tt$-component of the equations one has
\begin{equation}\label{Ett}
    G_{tt} = 8 \pi T_{tt} \quad \Leftrightarrow \quad \nabla^2 \theta = -8 \pi \rho.
\end{equation}
From the $ti$-components one then has
\begin{equation}
   0= G_{ti}-8 \pi T_{ti} = \left(-\partial_i \partial_t\theta -8\pi \partial_i S \right)+\left(-\frac{1}{2} \nabla^2 \Sigma_i -8\pi S_i \right).
\end{equation}
where the first term in round brackets is the scalar part and the
second is the (divergenceless) vector part.
Since the Helmholtz decomposition of a vector is unique, 
in order for this equation to be satisfied both terms must vanish, i.e.
\begin{equation}\label{Etj}
\begin{cases}
      \partial_t\theta +8\pi  S=0 \,,\\
     \frac{1}{2} \nabla^2 \Sigma_i + 8\pi S_i=0 \,.
\end{cases}    
\end{equation}
The same procedure can be applied to the spatial components:
\begin{align}
   &0= G_{ij}-8 \pi T_{ij} = 
    -\frac{1}{2} (\Box h_{ij}^\mathrm{TT}+16\pi \sigma_{ij}) - [\partial_{(i}\partial_t{\Sigma}_{j)}+8\pi \partial_{(i} \sigma_{j)}]\nonumber\\&-\partial_i \partial_j \left(\psi+\frac{\theta}{2} +8\pi\sigma\right) + \delta_{ij} \left[\frac{1}{2}\nabla^2 \left(2 \psi + \theta \right) - \partial_t^2{\theta}+\frac83 \pi\nabla^2\sigma-8\pi p\right]\,,\,
\end{align}
leading to
\begin{equation}\label{Eij}
    \begin{cases}
     \Box h_{ij}^{\mathrm{TT}} = -16 \pi \sigma_{ij}\,, \\
\partial_t{\Sigma}_{j}+8\pi  \sigma_{j}=0\,,\\
\psi+\frac{\theta}{2} +8\pi\sigma=0\,,\\
\nabla^2 \left( \psi + \frac{\theta}{2} \right) - \frac32\partial_t^2{\theta}-12 \pi p=0\,.
\end{cases} 
\end{equation}
Let us consider now the energy conservation and relativistic Euler equations, which we have seen to follow from 
the conservation of the matter stress-energy tensor,  $\partial_\mu T^{\mu \nu }=0$. Decomposing this vector equations in two scalar equations and one equation for a divergenceless vector in the (by now) usual way, one gets~\cite{Flanagan:2005yc}
\begin{equation}
\begin{aligned}
\label{eq:stressenergyconservationbardeen}
    & \nabla^2 S = \partial_t{\rho}, \\
    & \nabla^2 \sigma = -\frac{3}{2}p + \frac{3}{2}\partial_t{S},\\
    & \nabla^2 \sigma_i =2\partial_t{S}_i.
\end{aligned}    
\end{equation}
These equations can be used to simplify the Einstein equations \ref{Ett}, \ref{Etj} and \ref{Eij}. In particular, as expected from the Bianchi identify, one can show explicitly that three of those equations (one involving a divergenceless vector and two involving scalars) are automatically satisfied on shell (i.e. if the matter stress energy conservation is enforced). The remaining Einstein equations can then be written as~\cite{Flanagan:2005yc}
\begin{equation}
\begin{cases}
     \nabla^2 \theta = -8 \pi \rho \quad & \text{(1 d.o.f.)} \,,\\
     \nabla^2 \psi = 4 \pi \left(\rho+3p-3\partial_t{S} \right) \quad & \text{(1 d.o.f.)}\,,\label{laplPsi}\\
     \nabla^2 \Sigma_i =- 16\pi S_i \quad & \text{(2 d.o.f.'s)}\,,\\
     \Box h_{ij}^{\mathrm{TT}} = -16 \pi \sigma_{ij} \quad & \text{(2 d.o.f.'s)}\,.
\end{cases}    
\end{equation}
These are six independent equations for the six gauge invariant degrees of freedom of the metric perturbation. 
As can be seen, only the transverse traceless tensor modes (i.e. the gravitational waves) are propagating, while the scalar and vector modes are not, as they satisfy elliptic (Poisson-like) equations.

The solution of the Poisson equation
can be easily written in terms of
the Laplacian's Green function. Let us recall indeed the
distributional identity,
\begin{equation}
    \nabla^2 \frac{1}{\left| \vec{x}\right|} = - 4 \pi \delta^{(3)} \left( \vec{x} \right),
\end{equation}
which implies that the Laplacian's Green function is proportional to $1/|\vec{x}-\vec{x}'|$. We can then write, for instance,
\begin{equation}
    \theta \left(t,\vec{x} \right) = 2 \int \frac{\rho \left(t,\vec{x}' \right)}{\left| \vec{x}-\vec{x}' \right|} \mathrm{d}^3 x'\,,
\end{equation}
which resembles the Newtonian potential (as we will further discuss later on). Far from a localized source, we can then
approximate $\left|\vec{x}-\vec{x}' \right| \approx \left|\vec{x}\right| = r$ and write
\begin{equation}
    \theta \left(t,\vec{x} \right) \approx \frac{2}{r} \int \mathrm{d}^3 x' \rho \left(t,\vec{x}' \right) .
\end{equation}
It is tempting to call ``mass'' the integral of $\rho$,
$M=\int \mathrm{d}^3 x \rho$.  We can in fact see that this quantity is conserved:
\begin{equation}
    \frac{\mathrm{d}}{\mathrm{d}t}  \int \rho \,\mathrm{d}^3 x  = \int \partial_t{\rho} \,\mathrm{d}^3 x,
\end{equation}
and using the conservation of the stress-energy tensor (Eq.~\ref{eq:stressenergyconservationbardeen}) one has
\begin{equation}
    \frac{\mathrm{d}}{\mathrm{d}t} \left( \int \rho \,\mathrm{d}^3 x \right) = \int \nabla^2 S \,\mathrm{d}^3 x = \int  \vec{\nabla} S \cdot \vec{n}\, \mathrm{d}^2S,
\end{equation}
where in the last equality we used Gauss' theorem to reduce the integral to the flux of $\vec{\nabla}S$ through  a surface at infinity with normal unit vector $\vec{n}$. If there is no matter at infinity, $\vec{\nabla} S $ vanishes, proving the conservation of the ``mass''.
Solving in a similar way the equation for $\psi$ one gets
\begin{equation}
    \psi \left(t,\vec{x} \right) \approx -\frac{1}{r} \int \mathrm{d}^3 x' \left( \rho+3p-3\partial_t{S} \right) = \frac{\bar{M}}{r}\,,
\end{equation}
with $\bar{M}=\int \mathrm{d}^3 x \left( \rho+3p-3\partial_t{S} \right)$. 
Using the conservation of the stress-energy tensor (Eq.~\ref{eq:stressenergyconservationbardeen}), we can prove that
\begin{equation}
    \bar{M}-M = \int \mathrm{d}^3 x' \left(3p-3\dot{S} \right) = -2 \int \nabla^2 \sigma\, \mathrm{d}^3 x' = 0,
\end{equation}
where in the last step we used again the absence of matter at infinity. This proves that $M=\bar{M}$.
For $ \Sigma_i$ we obtain
\begin{equation}
    \Sigma_i \approx \frac{4}{r} \int S_i \mathrm{d}^3 x',
\end{equation}
and with similar steps one finds that 
\begin{equation}
    \pi_i = \int S_i \mathrm{d}^3 x',
\end{equation}
which physically describes the linear momentum of the source, is conserved.

In summary, the scalar and vector gauge invariant Bardeen variables present 
a Newton-like behavior, i.e.
\begin{equation}\label{pnfields}
\begin{aligned}
    & \psi \sim \frac{M}{r}, \\
        & \theta \sim \frac{M}{r}, \\
    & \Sigma_i \sim \frac{\pi_i}{r}.
\end{aligned}    
\end{equation}
As will become clearer from the post-Newtonian (PN) formalism, these three non-propagating degrees of freedom  generalize 
the Newtonian potential ($\psi$), and
encode relativistic effects such as periastron precession and light bending ($\theta$) and frame dragging ($\Sigma_i$).

\subsection{Generation of gravitational waves: a first derivation of the quadrupole formula}
\label{sec:quadrupolewrong}
Let us consider now not the propagation, but the generation of gravitational waves from matter sources, by solving 
\begin{equation}\label{gweq}
    h_{ij}^\mathrm{TT} = -16 \pi \sigma_{ij}\,.
\end{equation}
This will lead us to a first
 derivation of the quadrupole formula. We will then highlight some shortcomings of this derivation, which will be amended in section~\ref{sec:postNewtonian}.

The solution of Eq.~\ref{gweq} 
can be obtained in terms of retarded potentials. The Green function of the flat space d'Alembertian $\Box = -\partial_t^2+\nabla^2$ is 
\begin{equation}
    G \left( t, \vec{x} \right) = -\frac{1}{4 \pi \left|\vec{x} \right|} \delta \left(t-\left|\vec{x} \right| \right)\,,
\end{equation}
which indeed satisfies the distributional identity
\begin{equation}
    \Box G\left(t,\vec{x}\right) = \delta(t) \delta^{(3)} \left( \vec{x} \right) \,.
\end{equation}
The solution can then be written as
\begin{equation}
    h_{ij}^\mathrm{TT} \left(t,\vec{x} \right) = 4 \int \frac{\sigma_{ij} \left(t-\left|\vec{x}-\vec{x}' \right|,\vec{x}' \right)}{\left|\vec{x}-\vec{x}' \right|} \mathrm{d}^3 x'.
\end{equation}
In order to find the source $\sigma_{ij}$ from the stress-energy tensor, we have to invert the Eq.~\ref{eq:stressenergypoisson}. To this purpose we can formally define the projector
\begin{equation}
    P_{ij} = \delta_{ij}-\nabla^{-2} \partial_i \partial_j,
\end{equation}
and write
\begin{equation}
\label{eq:sigmafromstress}
    \sigma_{ij} = \left(P_i^{\ k} P_j^{\ l} -\frac{1}{2} P_{ij} P^{kl} \right) T_{kl}.
\end{equation}
Using the fact that partial derivatives commute and thus $\nabla^{-2}\partial_i=\partial_i \nabla^{-2}$, one can show that Eq.~\ref{eq:sigmafromstress} implies $\partial_i \sigma^{ij}=0$ and $\sigma^i_{\ i}=0$, i.e. Eq.~\ref{eq:sigmafromstress} correctly defines the transverse and traceless part of the matter stress energy tensor.
We can therefore write
\begin{equation}
   h_{ij}^\mathrm{TT} = -16 \pi \Box^{-1} \sigma_{ij} = -16 \pi \Box^{-1} \left(P_i^{\ k} P_j^{\ l} -\frac{1}{2} P_{ij} P^{kl} \right) T_{kl}.
\end{equation}
Because the flat d'Alembertian and partial derivatives commute, we can then proceed to write
\begin{equation}
\begin{aligned}
   & h_{ij}^\mathrm{TT} = -16 \pi \left(P_i^{\ k} P_j^{\ l} -\frac{1}{2} P_{ij} P^{kl} \right) \Box^{-1} T_{kl}= \\
   & = 4\left(P_i^{\ k} P_j^{\ l} -\frac{1}{2} P_{ij} P^{kl} \right) \int \frac{T_{kl} \left(t-\left|\vec{x}-\vec{x}' \right|,\vec{x}' \right)}{\left|\vec{x}-\vec{x}' \right|} \mathrm{d}^3 x'= \\
   & = \frac{4}{r} \left(P_i^{\ k} P_j^{\ l} -\frac{1}{2} P_{ij} P^{kl} \right) \int T_{kl} \left(t-\left|\vec{x}-\vec{x}' \right|,\vec{x}' \right)  \mathrm{d}^3 x' \left[1+\mathcal{O} \left( \frac{1}{r} \right)\right],
\end{aligned}    
\end{equation}
where in the last step, besides  approximating $\left|\vec{x}-\vec{x}' \right|$ with $r=|\vec{x}|$, we have also commuted $1/r$ with the projectors, which is appropriate at leading order in $1/r$.
Moreover, in the last step we can approximate, up to subleading terms in  $1/r$,
\begin{equation}
    P_{ij} \approx \delta_{ij}-n_i n_j,
\end{equation}
where $n_i = x_i/r$ is a unit vector in the direction of the observer. 
Defining $\mathcal{P}_{ij}^{\ \ kl} = P_i^{\ k} P_j^{\ l} -\frac{1}{2} P_{ij} P^{kl}$, one can write this ``Green'' formula in more compact form as
\begin{equation}
\label{eq:greenformula}
    h_{ij}^\mathrm{TT} = \frac{4}{r}\mathcal{P}_{ij}^{\ \ kl} \int T_{kl} \left(t-\left|\vec{x}-\vec{x}' \right|,\vec{x}' \right) \mathrm{d}^3 x'.
\end{equation}

To go from this equation to the quadrupole formula, one can note that from the conservation of the stress energy tensor (in flat space) it follows that
\begin{equation}
    \partial_t^2 \left( T^{tt}x^i x^j \right) = 2 T^{ij} + \partial_k \partial_l \left( T^{kl}x^i x^j \right)-2 \partial_k \left( T^{ik} x^j + T^{kj} x^i \right).
\end{equation}
Using this equation, and neglecting surface terms that vanish if the source is confined, Eq.~\ref{eq:greenformula} then becomes
\begin{equation}
    h_{ij}^\mathrm{TT} = \frac{2}{r}\mathcal{P}_{ij}^{\ \ kl}\int \partial_t^2 \left(T^{tt} x'^k x'^l \right) \mathrm{d}^3 x'.
\end{equation}
Defining the inertia tensor
\begin{equation}
    I_{ij} = \int \mathrm{d}^3 x' \rho\, x'^i x'^j
\end{equation}
and the quadrupole tensor
\begin{equation}
    Q_{ij} = I_{ij}-\frac{1}{3} I \delta_{ij},
\end{equation}
where $I=I^i_{\ i}$, 
we finally arrive at the ``quadrupole formula''
\begin{equation}
    h_{ij}^\mathrm{TT} = \frac{2G}{c^4 r} \mathcal{P}_{ij}^{\ \ kl} \Ddot{Q}_{kl}\,,
\end{equation}
where $\dot{\phantom{A}}=\mathrm{d}/{\mathrm{d} t}$ and we have reinstated $G$ and $c$ for physical clarity.

While this final result looks reasonable, two key assumptions were used to derive it, namely \textit{(i)} linear perturbation theory and \textit{(ii)} the conservation of the stress energy tensor on flat space. Neither of these assumptions is justified for a compact binary system, as \textit{(i)} the spacetime is not a perturbation of Minkowski space near black holes or neutron stars, and \textit{(ii)} $\partial_\mu T^{\mu\nu}=0$ implies the geodesic equation in flat space (cf. section~\ref{sec:prerequisites}), which in turn implies straight line motion (which clearly cannot describe quasicircular binary systems).
In fact, when applying the Green and quadrupole formulae to binary systems one gets into paradoxes such as that described in the next exercise. This shows that a better, more rigorous treatment of gravitational wave generation is needed, which will prompt us to go beyond linear theory in the next section.
\\
\\
{\bf Exercise 2:} {\it Consider an equal-mass, Keplerian binary (i.e. a binary with large separation, for which the laws of Newtonian mechanics are applicable) on a circular orbit on the $(x,y)$ plane, and an observer (far from the source) along the $z$ axis. Compute the gravitational-wave signal according to the ``Green formula'' of the lectures, and according to the quadrupole formula. Show that the amplitudes of the two predictions differ by a factor 2.}

\subsection{Dimensional analysis}\label{dimsec}
Let us try to derive the quadrupole formula from dimensional arguments. A matter source can be characterized by its multipole moments, e.g. the mass monopole $M=\int \rho\, \mathrm{d}^3 x$, the mass dipole $D^i= \int \rho\, x^i\, \mathrm{d}^3 x$, the mass quadrupole $Q_{ij}$, the angular momentum (i.e. the first moment of the mass current) $L_i=\int  \rho\,e_{ijk}\, x^j\, v^k\,\mathrm{d}^3 x $ etc. Since metric perturbations are dimensionless, one can try to write a monopole gravitational wave signal using dimensional analysis as $h\sim G M/(r c^2)$, a dipole signal as $h\sim G \dot{D}/(r c^3)\sim G P/(r c^3)$ (where $P$ is the linear momentum), an angular momentum term $h\sim G \dot{L}/(r c^4)$. These terms are zero (or static) because of conservation of mass, linear momentum and angular momentum. The quadrupole term, again by dimensional analysis, is instead $h\sim G \ddot{Q}/(r c^4)$. Note that radiation sourced by the mass monopole and dipole and by the angular momentum {\it can} be present beyond GR, because in that case the mass, linear momentum and angular momentum of matter may not be conserved (due to exchanges with additional gravitational degrees of freedom different from the tensor gravitons). Similarly, the static scalar and vector degrees of freedom of Eq.~\ref{pnfields} will generally become dynamical beyond GR.

\section{Post-Newtonian expansion}
\label{sec:postNewtonian}
In order to assess which of the two expressions for the generation of gravitational waves (the quadrupole formula or the ``Green
formula'') is correct, let us take a small detour. We will now study perturbations of flat space not by expanding in the perturbation amplitude (like we did previously), but in powers of $1/c$ (with $c\to\infty$). This is known as post-Newtonian (PN) expansion, and will allow us to re-derive the quadrupole formula in a more rigorous way.

\subsection{The motion of massive and masseless bodies}
Let us start by writing the following ansatz for the metric:
\begin{equation}
\begin{aligned}
\label{eq:ansatzpn}
    & g_{00} = -\left(1+\frac{2 \phi}{c^2}\right), \\
    & g_{0i} = \frac{\omega_i}{c^3},\\
    & g_{ij} = \left(1-\frac{2 \psi}{c^2}\right) \delta_{ij}+\frac{\chi_{ij}}{c^2},
\end{aligned}    
\end{equation}
where $\chi_{ij}$ is  traceless ($\chi^i_{\phantom{i}i}=0$) and we
have used Cartesian coordinates $x^\mu=(ct,x^i)$. Latin indices are meant to be raised and lowered with the flat spatial metric $\delta_{ij}$. Note that we have reinstated $c$ (which was set to 1 in the previous sections), as that will be our book-keeping parameter.
Before venturing into the actual calculation, let us note that the choice of powers of $c$ appearing in Eq.~\ref{eq:ansatzpn} is exactly the one that will be needed to consistently solve the Einstein equations (i.e. should we choose different powers, the Einstein equations would set the potentials to zero, or they would not allow for a consistent solution). However, it is possible to make sense of this ansatz also in a more physical way.

Let us consider a point particle moving in the geometry described by  Eq.~\ref{eq:ansatzpn}. From the point particle action (Eq.~\ref{eq:actionpointparticle}), one can obtain the Lagrangian (by recalling that by definition $S = \int L \mathrm{d}t$). By replacing
therefore  Eq.~\ref{eq:ansatzpn} into Eq.~\ref{eq:actionpointparticle}, one  obtains
\begin{align}
\label{eq:actionpn}
     S &= -mc^2 \int \frac{\mathrm{d}\tau}{\mathrm{d}t} \mathrm{d}t = -mc \int \sqrt{-g_{\mu \nu} \dot{x}^\mu \dot{x}^\nu} \mathrm{d}t\nonumber\\&= -mc \int \sqrt{c^2 + 2\phi-\frac{2\omega_i v^i}{c^2}-\left(1-\frac{2\psi}{c^2}\right)v^2-\frac{\chi_{ij}v^i v^j}{c^2}} \mathrm{d}t\nonumber\\&\approx
              -m c^2 \int \bigg(1+\frac{\phi}{c^2}-\frac{1}{2} \frac{v^2}{c^2}-\frac{\phi^2}{2 c^4}
              + \frac{\phi v^2}{2c^4}+ \frac{\psi v^2}{c^4}-\frac{v^4}{8 c^4}-\frac{v_i \omega^i}{c^4}-\frac{\chi_{ij}v^i v^j}{2c^4}
        \bigg) \mathrm{d}t,
\end{align}
where $v^i \equiv \dot{x}^i\equiv \mathrm{d} x^i/\mathrm{d}t$, $v^2 \equiv \delta_{ij}v^i v^j$, and in the last step we have Taylor expanded in $1/c$ (for $c\to\infty$). 
The Lagrangian for a point particle therefore reads
\begin{equation}
   L= -m c^2
   +m\left(\frac{v^2}{2}-\phi\right)
   +m \left(
   \frac{\phi^2}{2 c^2}
              - \frac{\phi v^2}{2c^2}- \frac{\psi v^2}{c^2}+\frac{v^4}{8 c^2}+\frac{v_i \omega^i}{c^2}+\frac{\chi_{ij}v^i v^j}{2c^2}\right)  +\ldots
\end{equation}
As can be seen, in the limit $c\to\infty$ the ansatz of Eq.~\ref{eq:ansatzpn}
leads to the correct Newtonian limit (note the appearance of the Newtonian Lagrangian after the irrelevant constant offset term), as well as to deviations from  the Newtonian Lagrangian that are suppressed by
${\cal O} (1/c^2)$ relative to the Newtonian dynamics. These are known as 1PN corrections (where $n$PN denotes terms that are suppressed by $1/c^{2n}$ relative to the leading order Newtonian term).

Note however that this counting is only applicable to massive particles, and not to photons (for which $v\sim c$). In the latter case, the terms $\psi v^2/c^2$, $\phi v^2/(2c^2)$  and ${\chi_{ij}v^i v^j}/(2 c^2)$, which are of order 1PN for a massive particle, are of order 0PN (i.e. Newtonian order).
In more detail, one can write the Lagrangian from photons as $L\propto\mathrm{d}\tau/\mathrm{d}t$, which using Eq.~\ref{eq:ansatzpn} leads to
\begin{align}
    L&\propto\sqrt{1-\beta^2 + 2\frac{\phi}{c^2}-\frac{2\omega_i \beta^i}{c^3}+2\frac{\psi\beta^2}{c^2}-\frac{\chi_{ij}\beta^i \beta^j}{c^2}}\nonumber\\
    &=\sqrt{1-\beta^2}+\frac{\gamma}{2 c^2} \left(2\phi+2\psi\beta^2-\chi_{ij}\beta^i \beta^j\right)+{\cal O}\left(\frac{1}{c^3}\right)\,,
\end{align}
with $\beta^i=v^i/c$ and $\gamma=1/\sqrt{1-\beta^2}$. 
In other words, the bending of light in GR is determined at leading order by both $g_{00}$ ($\phi$) and $g_{ij}$ ($\psi$ and $\chi_{ij}$), but not by $g_{0i}$ ($\omega_i$). The same can be seen, e.g., by using  Eq.~\ref{eq:ansatzpn} into the dispersion relation for a photon, $p^\mu p^\nu g_{\mu\nu}=0$, with $p^{\mu}=(E,p^i)$
the 4-momentum, and solving for the energy $E$ to derive the Hamiltonian describing the photon's motion.

\subsection{The Einstein equations}

Let us now compute the potential appearing in the metric from the Einstein equations. To do so, let us first perform the usual scalar-vector-tensor decomposition on the metric ansatz of Eq.~\ref{eq:ansatzpn}:
\begin{equation}
\begin{aligned}
    & \omega_i = \partial_i \omega + \omega_i^\mathrm{T}, \\
    &\chi_{ij} = \left( \partial_i \partial_j -\frac{1}{3} \delta_{ij} \nabla^2 \right) \chi + \partial_{(i} \chi_{j)}^\mathrm{T} + \chi_{ij}^\mathrm{TT}   ,
\end{aligned}    
\end{equation}
where the index ``T'' identifies divergenceless (i.e. transverse) vector fields and ``TT'' transverse and traceless tensors.  Let us first adopt the same gauge that we used in linear theory, i.e. the Poisson gauge, defined by
\begin{equation}
    0=\partial_i \omega^i = \partial_i \chi^{ij}\,,
\end{equation}
which yields $\omega=\chi=\chi_i^T=0$. Let us also describe matter
as a perfect fluid with stress energy tensor
\begin{equation}
T^{\mu\nu}= (p+\rho c^2) u^\mu u^\nu +p g^{\mu\nu}\,,
\end{equation}
with ${u^i}/{u^0}={v^i}/{c}$ [and therefore $u^0=1-(\phi-v^2/2)/c^2+{\cal O}(1/c^4)$ because of the 4-velocity normalization]. Using then the Einstein equations, in which we reinstate $c$ to obtain
$G_{\mu\nu}={8\pi} T_{\mu\nu}/c^4$, one gets the following equations for the potentials~\cite{Barausse:2013ysa}:
\begin{align}\label{phi_eq_psi}
&\psi=\phi+{\cal O}\left(\frac{1}{c^2}\right)\,,\\
&\nabla^2 \omega_T^i= 4 (4 \pi\rho v^i+\phi_{,ti})+{\cal O}\left(\frac{1}{c^2}\right)\,,\label{wT}\\
&\nabla^2 \phi=4\pi \left(3 \frac{p}{c^2} +\rho\right)+\frac{2}{c^2}{\phi}_{,i}{\phi}_{,i}
+8 \pi\rho \left(\frac{v}{c}\right)^2-\frac{3}{c^2}{\phi}_{,tt}+{\cal O}\left(\frac{1}{c^4}\right)\label{PNpoisson}\,,\\
&\nabla^2\chi^{TT}_{ij}={\cal O}\left(\frac{1}{c^2}\right)\label{eq:nabla2chi_ij}\,.
\end{align}
As a consistency check, note that by taking the divergence of Eq.~\ref{wT}, both sides evaluate to zero: the left hand side because $\omega_T^i$ is transverse, and the right hand side because of the continuity equation for the number density, which at leading 
 (Newtonian) order reads $\partial_t \rho+\partial_i (\rho v^i)=
\partial_t\nabla^2\phi/(4 \pi) +\partial_i (\rho v^i)=0$. (This is because the rest mass density and the energy density differ by the internal energy $p/[c^2(\Gamma-1)]$, with $\Gamma$ the adiabatic index; see e.g. \cite{Barausse:2013ysa}.)

One can then write
\begin{align}
    & \phi = \phi_\mathrm{N} + \frac{\phi_\mathrm{1PN}}{c^2}+\ldots,\\
    & \psi = \phi_\mathrm{N} + \frac{\psi_\mathrm{2PN}}{c^2}+\ldots ,\\
    & \omega_T^i = \omega^i_\mathrm{1PN} + \frac{\omega^i_\mathrm{2PN}}{c^2}+\ldots ,\\
    & \chi_{TT}^{ij} = \frac{\chi^{ij}_\mathrm{2PN}}{c^2} + \frac{\chi^{ij}_\mathrm{2.5PN}}{c^3}+\ldots , 
\end{align}    
where $\phi_\mathrm{N}$ is the Newtonian potential
(obtained by solving $\nabla^2 \phi_\mathrm{N}=4\pi\rho$), and we have left indicated the terms that appear at 1PN order and higher in the Lagrangian for massive particles derived in the previous section. These PN  terms can be obtained explicitly by solving Eqs.~\ref{phi_eq_psi}--\ref{eq:nabla2chi_ij} and their higher order generalizations. In particular, one can show explicitly that the leading order term for  $\chi_{TT}^{ij}$ appears at ${\cal O}(1/c^2)$ (2PN order in the Lagrangian for massive particles). This is a conservative term (as it is of even parity in $c$, i.e. it is left unchanged by a time reversal). However, a dissipative term appears at 
${\cal O}(1/c^3)$, i.e. at 2.5 PN order. This corresponds to the loss of energy and angular momentum to gravitational waves (see e.g. \cite{Barausse:2013ysa} for details).

One unsightly feature of the Poisson gauge is however apparent from Eq.~\ref{PNpoisson}, which features a double time derivative of $\phi$ on the right-hand side. That term corresponds to $\partial_t S$ in Eq.~\ref{laplPsi}, i.e.
one can re-express it in terms of the matter density by writing it as $-3 \phi_{,tt}/c^2=-12 \pi \nabla^{-2} \rho_{,tt}/c^2+{\cal O}(1/c^4)=-12 \pi S_{,t}/c^2+{\cal O}(1/c^4)$. That requires, however, solving a non-local equation to compute $\nabla^{-2} \rho$. A better option is to eliminate the term $-3 \phi_{,tt}/c^2$ by performing a gauge transformation with generator $\xi_0\propto \partial_t \mathbb{X}$,
where $\mathbb{X}=-2\nabla^{-2} \phi$ is the Newtonian ``superpotential''~\cite{Will:2014kxa} (see also the appendix of \cite{Bonetti:2015oda}). This leads to the ``standard PN gauge'', which is defined exactly as a gauge in which the 1PN spatial metric is isotropic (i.e. $\chi_{ij}$ is zero at 1PN, which we have seen to be already the case in our Poisson gauge) and in which no term proportional to $\phi_{,tt}$ appears at 1PN in the equation for $\nabla^2\phi$~\cite{Will:2014kxa,WillPoisson}. 

Even more simply, one can do the calculation starting directly in the standard PN gauge, which satisfies the gauge conditions~\cite{Will:2014kxa}
\begin{align}
    &\partial_\mu h_i^\mu-\frac12 \partial_i h^\mu_\mu=0\,,\\
    &\partial_\mu h_0^\mu-\frac12 \partial_0 h^\mu_\mu=-\frac12 \partial_0 h_{00}\,,
\end{align}
where $g_{\mu\nu}=\eta_{\mu\nu}+h_{\mu\nu}$ and the indices of $h_{\mu\nu}$ are raised and lowered with the Minkowski metric. In this gauge, the Einstein equations at 1PN order become
\begin{align}
&\nabla^2 \left(\phi - \frac{\phi^2}{c^2}+4 \frac{\Phi_2}{c^2}\right)=4\pi \left(\rho+2 \rho \frac{v^2}{c^2}+2\rho \frac{\phi}{c^2}+3 \frac{p}{c^2}\right)\,,\\
&\nabla^2 \psi=4\pi \rho\,,\\
&\nabla^2 \omega_j=16\pi\rho v_j+\partial_t\partial_i\phi
\end{align}
with $\nabla^2\Phi_{2} = 4\pi \rho \phi$. Solving these equations by using the Green function of the flat Laplacian
one gets the 1PN metric in the standard PN gauge as~\cite{Will:2014kxa}
\begin{align}\label{1pnG}
&g_{00} = -1 -2\dfrac{\phi_\mathrm{N}}{c^2} -2\dfrac{\phi_\mathrm{N}^2}{c^4} + 4\dfrac{\Phi_1}{c^4} + 4\dfrac{\Phi_2}{c^4} + 6\dfrac{\Phi_4}{c^4}\\
& g_{0i} =-\dfrac{7}{2}\dfrac{V_i}{c^3} - \dfrac{1}{2}\dfrac{W_i}{c^3}\\
& g_{ij} = \left(1-2\dfrac{\phi_\mathrm{N}}{c^2}\right)\delta_{ij}
\end{align}
in terms of the PN potentials
\begin{align}
 &V_{i}=\int d^{3}x' \ \frac{\rho(\vec{x} ', t) v_{i}(\vec{x} ', t)}{|\vec{x}-\vec{x} '|},
\\
 &W_{i}=\int d^{3}x' \ \frac{\rho(\vec{x} ', t) [ \vec{v}(\vec{x} ', t) \cdot (\vec{x}-\vec{x} ') ] (x-x')_{i}}{|\vec{x}-\vec{x}'|^3}, \\ 
&\Phi_{1}=\int d^{3}x' \frac{\rho(\vec{x} ', t) v(\vec{x} ', t)^{2}}{|\vec{x}-\vec{x}'|}, \\&\Phi_{2}=-\int d^{3}x' \frac{\rho(\vec{x} ', t) \phi_\mathrm{N}(\vec{x} ', t)}{|\vec{x}-\vec{x} '|}, \\ &\Phi_{4}=\int d^{3}x' \frac{p(\vec{x} ', t)}{|\vec{x}-\vec{x}'|}.
\end{align}
Note that to obtain this result we have used the relation~\cite{Will:2014kxa}  $\partial_t\partial_i \mathbb{X} = W_i - V_i$ (where $\nabla^2 \mathbb{X} = -2\phi_N$, as defined previously), which follows from the explicit expression for the Newtonian superpotential, $\mathbb{X} =   \int d^3 x' \rho(\vec{x} ', t) \ |\vec{x} - \vec{x}'|$, and from the continuity of the number density [$\partial_t \rho=-\partial_i (\rho v^i)$ at  Newtonian order]. While the choice of the standard PN gauge bears no physical significance (the Poisson gauge has the same physical validity, since observables in general relativity are gauge invariant), Eq.~\ref{1pnG} is the metric usually adopted to describe tests of general relativity in the solar system (e.g. periastron precession, light bending, Shapiro time delay, lunar laser ranging, frame dragging, etc; see~\cite{Will:2014kxa} for more details).

\subsection{A more rigorous derivation of the quadrupole formula}
\label{rig_quad}
By using now the PN expansion in place of the linear approximation, let us revisit the generation of gravitational waves from binary systems. This will lead us to re-derive the quadrupole formula in a more rigorous fashion, which will in turn shed light on the 
discrepancy between quadrupole formula and ``Green formula'', which we discovered in Exercise 2. As previously mentioned, the problem with the linear theory derivation of the quadrupole formula is two-fold: the assumption of ``weak gravity'' ($h_{\mu \nu}\ll 1$) and the use of the stress energy tensor conservation in flat space. Here, we will fix both of these shortcomings.

 To drop the weak gravity assumption, let us start from the
full Einstein equations, which we write in ``relaxed form''
in the harmonic gauge, defined by
\begin{equation}
\label{eq:harmonic}
    \Box x^\alpha = 0\,.
\end{equation}
It is important to keep in mind that the coordinates $x^\alpha$, despite the space-time indices, are not vectors but scalars, as can be seen from their transformation properties under diffeomorphisms.
Using this fact, it is straightforward to see that the 
 condition \ref{eq:harmonic} can be rewritten in terms of the
pseudo-tensor\footnote{A pseudo-tensor is an object that transforms as a tensor  under linear transformations, but not under more general coordinate transformations.}:
\begin{equation}
    \bar{H}^{\mu \nu} = \eta^{\mu \nu}-\sqrt{-g}g^{\mu \nu}.
\end{equation}
Expanding then
\begin{equation}
\begin{aligned}
    & \Box x^\alpha = \frac{1}{\sqrt{-g}} \partial_\mu \left(\sqrt{-g}g^{\mu \nu} \partial_\nu x^\alpha \right) = \frac{1}{\sqrt{-g}} \partial_\mu \left(\sqrt{-g}g^{\mu \nu} \delta_\nu^{\ \alpha} \right) = \\
    & = \frac{1}{\sqrt{-g}} \partial_\mu \left(\sqrt{-g}g^{\mu \alpha} \right)  \propto \partial_\mu \bar{H}^{\mu \nu}\,,
\end{aligned}    
\end{equation}
the condition \ref{eq:harmonic} turns out to be equivalent to
\begin{equation}
\label{eq:alternativeharmonic}
    \partial_\mu \bar{H}^{\mu \nu}=0\,.
\end{equation}
Note that the quantity $\bar{H}^{\mu \nu}$ 
becomes the trace reversed metric perturbation 
at linear order. Indeed, if $g^{\mu \nu} = \eta^{\mu \nu}-h^{\mu \nu}+\mathcal{O}(h^2)$, then $\delta g =
-h+\mathcal{O}(h^2)$, and 
$ \bar{H}^{\mu \nu} = h^{\mu \nu} - \frac{1}{2} h^{\mu \nu} + \mathcal{O}(h^2) = \bar{h}^{\mu \nu}+\mathcal{O}(h^2)$. As a result, at linear order
the harmonic gauge condition \ref{eq:alternativeharmonic} coincides with the Lorenz gauge condition \ref{eq:lorenz}.

In the harmonic gauge,  the {\it fully non-linear} Einstein equations take the form 
\begin{equation}
\label{eq:relaxedeinstein}
    \Box_\eta \bar{H}^{\mu \nu} = -16 \pi \tau^{\mu \nu},
\end{equation}
where $\Box_\eta= \eta^{\mu\nu}\partial_\mu\partial_\nu$ is the {\it flat} space d'Alembertian operator and
\begin{equation}
    \tau^{\mu \nu} = (-g) T^{\mu \nu} +\frac{\Lambda^{\mu \nu}}{16 \pi},
\end{equation}
with
\begin{equation}\label{eqLambda}
    \Lambda^{\mu \nu} = 16 \pi (-g) t_\mathrm{LL}^{\mu \nu} + \left(\partial_\beta \bar{H}^{\mu \alpha} \partial_\alpha \bar{H}^{\nu \beta} - \partial_\alpha \partial_\beta \bar{H}^{\mu \nu} \bar{H}^{\alpha \beta}\right).
\end{equation}
Here, $t_\mathrm{LL}^{\alpha \beta}$ is the Landau-Lifshitz pseudo-tensor,
\begin{eqnarray}
&& 16 \pi (-{ g})t_{_{\rm LL}}^{\alpha \beta } \equiv
 { g}_{\lambda\mu}{ g}^{\nu\rho}{ \bar{H}^{\alpha\lambda}}_{,\nu}{ \bar{H}^{\beta\mu}}_{,\rho}\label{LL}\\
&&+\frac{1}{2}
 { g}_{\lambda\mu}{ g}^{\alpha\beta}{ \bar{H}^{\lambda\nu}}_{,\rho}{ \bar{H}^{\rho\mu}}_{,\nu}
- 2{ g}_{\mu\nu}{ g}^{\lambda (\alpha}{ \bar{H}^{\beta )\nu}}_{,\rho}{ \bar{H}^{\rho\mu}}
_{,\lambda}\nonumber
\\
&&+ \frac{1}{8}
(2{ g}^{\alpha\lambda}{ g}^{\beta\mu}-{ g}^{\alpha\beta}{ g}^{\lambda\mu})
(2{ g}_{\nu\rho}{ g}_{\sigma\tau}-{ g}_{\rho\sigma}{ g}_{\nu\tau})
{ \bar{H}^{\nu\tau}}_{,\lambda}{ \bar{H}^{\rho\sigma}}_{,\mu}  \;,\nonumber
\label{landau}
\end{eqnarray}
which describes the stress-energy of the gravitational field.
Because it is a pseudo-tensor, it can be non-zero in a set of coordinates but vanishing in a different one. This is simply a consequence of the well known fact that in general relativity the gravitational field can be locally set to zero (by choosing Riemann normal coordinates where
 the metric is locally $\eta_{\mu \nu}$ and the Christoffel symbols vanish, c.f. section \ref{sec:RNC}). Indeed, there is no way of defining a local energy density for the gravitational field: only the global energy (or mass) of an asymptotically flat spacetime is well defined in general relativity.
 
Taking now a (partial derivative) divergence
of Eq.~\ref{eq:relaxedeinstein} and using the condition \ref{eq:alternativeharmonic}, one obtains the conservation law $\partial_\mu \tau^{\mu \nu}=0$, which is equivalent to the equations of motion of matter (c.f. section~\ref{sec:prerequisites}).
Note that because we have not made any approximations thus far, these are the fully nonlinear equations of motion of matter, i.e. unlike in the case of linear theory, we are {\it not} assuming straight-line motion or $\partial_\mu T^{\mu\nu}=0$\footnote{The fact that $\partial_\mu \tau^{\mu \nu}=0$ does not imply straight-line motion can be tracked back to the presence of second derivatives of $\bar{H}^{\mu\nu}$ in Eq.~\ref{eqLambda}: as a result, even though the left hand side of the relaxed Einstein equations is written in terms of the flat wave operator, wavefronts do not follow straight lines in the eikonal approximation.}. 
We can then follow the same procedure of section~\ref{sec:quadrupolewrong} to derive the quadrupole formula, i.e. we can invert
Eq.~\ref{eq:relaxedeinstein} as
\begin{equation}
    \bar{H}^{\mu \nu} = -16 \pi \Box^{-1} \tau^{\mu \nu}\,,
\end{equation}
which gives in particular the ``Green formula''
\begin{equation}\label{green2}
    \bar{H}^{ij}\left(t,\vec{x} \right) \approx \frac{4}{r} \int \tau^{ij} \left(t-\frac{r}{c},\vec{x}' \right) \mathrm{d}^3 x'\,.
\end{equation}
Using the conservation law $\partial_\mu\tau^{\mu \nu}=0$ like in section~\ref{sec:quadrupolewrong}, one can write
\begin{equation}
    \bar{H}^{\mu \nu} \approx \frac{2}{r} \Ddot{Q}_{ij},
\end{equation}
where now
\begin{equation}\label{eq:q2}
    Q_{ij} = \int \tau^{tt} x'^i x'^j \mathrm{d}^3 x'.
\end{equation}
Let us now examine the relation between $\tau^{\mu \nu}$ and $T^{\mu \nu}$. Reinstating the appropriate powers of $c$, and using the PN expanded metric of Eq.~\ref{eq:ansatzpn} and the stress energy tensor for a system of two point particles [cf. Eq.~\ref{eq:pointparticlestressenergy}], one finds that~\cite{bonetti}
\begin{equation}\label{tau}
\begin{aligned}
    & \tau^{tt} = T^{tt} \left[1+\mathcal{O}\left(\frac{1}{c^2}\right)\right], \\
    & \tau^{ti} = T^{ti} \left[1+\mathcal{O}\left(\frac{1}{c^2}\right)\right], \\
    & \tau^{ij} = \left[T^{ij}+\frac{1}{4\pi}\left(\partial^i \phi \partial^j \phi-\frac{1}{2} \delta^{ij} \partial_k \phi \partial^k \phi \right) \right]\left[1+\mathcal{O}\left(\frac{1}{c^2}\right)\right]. \\
\end{aligned}    
\end{equation}
Therefore, at leading PN order $\tau^{tt} \approx T^{tt}$, and Eq.~\ref{eq:q2} reduces to the quadrupole formula derived in linear theory; however, $\tau^{ij}$ {\it cannot} be approximated by $T^{ij}$ at  leading PN order, i.e. the Green formula \ref{green2} 
does {\it not} reduce to the Green formula of linear theory. It follows that if one applies the formulae derived in linear theory, only the quadrupole formula is correct.
\\
\\
{\bf Exercise 3:} {\it Show that the additional terms contributing to $\tau^{ij}$ in Eq.~\ref{tau} solve the factor 2 discrepancy between the Green and quadrupole formula found in Exercise 2. [Hint: use the fact that the solution to 
$\nabla^2 g(\boldsymbol{x},\boldsymbol{y}',\boldsymbol{y}'') = |\boldsymbol{x}-\boldsymbol{y}'|^{-1}
|\boldsymbol{x}-\boldsymbol{y}''|^{-1}$ (with $\nabla^2$ the Laplacian with 
respect  to $\boldsymbol{x}$)
in the sense of distributions is
$g=\ln (|\boldsymbol{x}-\boldsymbol{y}'|+|\boldsymbol{x}-\boldsymbol{y}''|+|\boldsymbol{y}'-\boldsymbol{y}''|)+{\rm constant}$.]}

\section{Local flatness and the equivalence principle}
\label{sec:RNC}
In the previous section, when discussing the Landau-Lifshitz pseudotensor, we recalled that in general relativity the gravitational field can always be locally set to zero, i.e. it is 
always possible to choose a ``Local Inertial Frame'' where the gravitational force vanishes (i.e. where the metric is locally flat and the Christoffel symbols vanish). This can be seen as a manifestation of the equivalence principle of general relativity. In this section, we will provide a proof of this statement, which will clarify issues such as that of the non-existence of a covariant stress-energy tensor for the gravitational field. The coordinates that we introduce will also be useful when deriving the response of a gravitational wave detector in the following.

\subsection{The local flatness theorem and Riemann normal coordinates}
The local flatness theorem states that at any given event (i.e. space-time point) $P$, there exists a coordinate system such that $g_{\mu \nu}|_P = \eta_{\mu \nu}$ and $\Gamma^\alpha_{\ \mu \nu}|_P=0$ (or equivalently $\partial_\alpha g_{\mu \nu}|_P=0$).
We will now provide two proofs of this theorem.

\textit{Algebraic proof:} Let us start with a system of coordinates $\{ x^\alpha\}$ such that $P$ corresponds to $x^\alpha=0$. Let us then perform a coordinate transformation to some new coordinates $\{ x^{\alpha'}\}$ also centered in $P$:
\begin{equation}
    x^{\alpha'} = A^{\alpha'}_{\ \beta} x^\beta + \mathcal{O}(x^2) \quad \Leftrightarrow \quad x^{\alpha} = A^{\alpha}_{\ \beta'} x^{\beta'} + \mathcal{O}(x'^2),
\end{equation}
where $A^{\alpha'}_{\ \beta}$ and $A^{\alpha}_{\ \beta'}$ are constant matrices (the Jacobian of the transformation and its inverse) that satisfy
\begin{equation}
    A^{\alpha'}_{\ \mu} A^\mu_{\ \beta'} = \delta^{\alpha'}_{\ \beta'} \quad \text{and} \quad  A^{\alpha}_{\ \mu'} A^{\mu'}_{\ \beta} = \delta^{\alpha}_{\ \beta}\,.
\end{equation}
The metric at $P$ transforms as
\begin{equation}
    g_{\alpha' \beta'}|_P = g_{\alpha \beta} \frac{\partial x^\alpha}{\partial x^{\alpha'}}\frac{\partial x^\beta}{\partial x^{\beta'}} \bigg|_P = g_{\alpha \beta} A^\alpha_{\ \alpha'} A^\beta_{\ \beta'}.
\end{equation}
The matrix $A$ has 16 coefficients, 10 of which can be chosen to set $g_{\alpha' \beta'}|_P = \eta_{\alpha' \beta'}$. The remaining 6 degrees of freedom  correspond to the 6  generators of the Lorentz transformations, which are isometries of the Minkowski metric.

In order to show that the Christoffel symbols vanish, let us expand the transformation to second order:
\begin{equation}
    x^{\alpha'} = A^{\alpha'}_{\ \beta} x^\beta + \frac{1}{2} B^{\alpha'}_{\ \beta \gamma} x^\beta x^\gamma + \mathcal{O}(x^3),
\end{equation}
where $B^{\alpha'}_{\ \beta\gamma}$ is the Hessian of the transformation. Note that  $B^{\alpha'}_{\ \beta\gamma}$ has the same symmetries as the Christoffel symbols, 
i.e. it is symmetric under 
the exchange $\beta \leftrightarrow \gamma$. As well known, the Christoffel symbols are {\it not} tensors under generic coordinate transformations (otherwise it would not be possible to set all of them  to zero with a choice of coordinates, which is what we are trying to prove), but they transform according to 
\begin{equation}
\begin{aligned}
   & \Gamma^{\alpha'}_{\ \beta' \gamma'} = {\Gamma^\alpha_{\ \beta \gamma} \frac{\partial x^\beta}{\partial x^{\beta'}}\frac{\partial x^\gamma}{\partial x^{\gamma'}}\frac{\partial x^{\alpha'}}{\partial x^{\alpha}}}- \frac{\partial x^{\alpha'}}{\partial x^{\beta} \partial x^\gamma} \frac{\partial x^\beta}{\partial x^{\beta'}}\frac{\partial x^\gamma}{\partial x^{\gamma'}} = \\
   & = A^{\alpha'}_{\ \alpha} A^{\beta}_{\ \beta'} A^{\gamma}_{\ \gamma'} \Gamma^\alpha_{\ \beta \gamma}- B^{\alpha'}_{\ \beta \gamma}A^{\beta}_{\ \beta'} A^{\gamma}_{\ \gamma'}.
\end{aligned}    
\end{equation}
We can therefore impose $\Gamma^{\alpha'}_{\ \beta' \gamma'} {=}0$ by solving for $B^{\alpha'}_{\ \beta \gamma}$. This equation has a unique solution, because $B$ and $\Gamma$ share the same symmetries.

This concludes our first proof of the local flatness theorem. The coordinates where the latter holds are known as ``Riemann Normal Coordinates'' (RNCs). We will now give a more geometric proof of the theorem, which includes a procedure to explicitly construct these coordinates.

\textit{Geometric proof:} Let us consider a spacetime endowed with coordinates $x^\mu$, and explicitly construct RNCs $x^{\mu'}$ around an event $P$.
To assign coordinates to a neighboring point $P'$, let us consider the unique geodesic connecting  $P$ and $P'$, and the vector $v$ tangent to this geodesic in $P$. Let us decompose this vector onto its components on a tetrad centered in $P$, i.e. on a basis of four orthogonal unit-norm vectors $\{e_{(\alpha)}\}_{\alpha=1,\ldots ,4}$:
\begin{equation}
    v^\mu = \Omega^{(\alpha)} e_{(\alpha)}^\mu.
\end{equation}
By definition, the tetrad vectors satisfy  the orthonormality and completeness relations
\begin{equation}
\begin{aligned}
    & e_{(\alpha)} \cdot e_{(\beta)} \equiv g_{\mu \nu} e^\mu_{(\alpha)} e^\nu_{(\beta)} = \eta_{\alpha\beta} \,,\\
    & e_{(\alpha)}^\mu e^{(\alpha)}_\mu = \delta^\mu_{\ \nu} \,,
\end{aligned}    
\end{equation}
where the tetrad (bracketed) indices are raised and lowered with the Minkowski metric $\eta_{\alpha\beta}$, whereas the space-time indices are raised and lowered with the space-time metric $g_{\mu \nu}$. 

As a working hypothesis, let us choose the new coordinates of the point $P'$ to be 
\begin{equation}\label{rncdef}
    x^{\alpha'} = \Omega^{(\alpha')} \Delta \lambda\,,
\end{equation}
with $\Delta \lambda = \lambda_{P'}-\lambda_P$, where $\lambda$ is the affine parameter of the  geodesic connecting $P$ and $P'$.
First, let us check that this definition is invariant under a re-parametrization of the geodesic. We know that the geodesic equation is invariant under affine transformations of the parameter, $\lambda'=a\lambda+b$. Under this transformation, $\Delta \lambda' = a \Delta \lambda$, and
\begin{equation}
    v^\mu\vert_{\lambda'} = \frac{\mathrm{d}x^\mu}{\mathrm{d}\lambda'} = \frac{\mathrm{d}x^\mu}{\mathrm{d}\lambda} \frac{\mathrm{d}\lambda}{\mathrm{d}\lambda'} = \frac{v^\mu}{a}.
\end{equation}
Thus, the coordinates of $P'$ remain unchanged:
\begin{equation}
    x^{\alpha'}|_{\lambda'} = a \Delta \lambda \,\frac{\Omega^{(\alpha')}}{a} = x^{\alpha'}|_{\lambda}.
\end{equation}

To see that the coordinates that we constructed are indeed RNCs, 
let us first check that the metric at the event $P$ is given by the Minkowski metric in the new coordinates. To this purpose let us first compute the Jacobian of the transformation from the old to the new coordinates evaluated at the point $P$. From Eq.~\ref{rncdef} one has
\begin{equation}
 \frac{\mathrm{d} x^{\alpha'}}{\mathrm{d}\lambda}\bigg|_P  =\Omega^{(\alpha')}\,.  
\end{equation}
Therefore, one also has
\begin{equation}
    \frac{\mathrm{d} x^{\alpha}}{\mathrm{d}\lambda}\bigg|_P=v^\alpha=\Omega^{(\mu)} e_{(\mu)}^\alpha=\frac{\partial x^{\alpha}}{\partial x^{\alpha'}}\bigg|_P \frac{\mathrm{d} x^{\alpha'}}{\mathrm{d}\lambda}\bigg|_P= \frac{\partial x^{\alpha}}{\partial x^{\alpha'}}\bigg|_P \Omega^{(\alpha')}\,.
\end{equation}
From the arbitrariness of the components $\Omega^{(\mu)}$ one then gets
\begin{equation}
    \frac{\partial x^{\alpha}}{\partial x^{\alpha'}}\bigg|_P = e^\alpha_{(\alpha')}\,,
\end{equation}
and therefore
\begin{equation}
   g_{\alpha'\beta'}\vert_P=  g_{\alpha\beta}\vert_P \frac{\partial x^{\alpha}}{\partial x^{\alpha'}}\bigg|_P \frac{\partial x^{\beta}}{\partial x^{\beta'}}\bigg|_P = g_{\alpha\beta} \vert_P\, e^\alpha_{(\alpha')} e^\beta_{(\beta')}=\eta_{\alpha'\beta'}\,.
\end{equation}

To compute instead the Christoffel symbols at $P$, let us consider a one parameter family of events along the geodesic connecting
$P$ and $P'$. This family has coordinates growing linearly with $\lambda$, i.e. $x^{\alpha'} \propto \lambda$, which implies ${\mathrm{d}^2 x^{\alpha'}}/{\mathrm{d}\lambda^2}=0$. Since 
this one-parameter family is (by definition) a geodesic with affine parameter $\lambda$, one must have
\begin{equation}
    \frac{\mathrm{d}^2 x^{\alpha'}}{\mathrm{d}\lambda^2} \bigg|_P = -\Gamma^{\alpha'}_{\ \mu' \nu'} \big|_P \frac{\mathrm{d} x^{\mu'}}{\mathrm{d}\lambda} \bigg|_P \frac{\mathrm{d} x^{\nu'}}{\mathrm{d}\lambda} \bigg|_P = -\Gamma^{\alpha'}_{\ \mu' \nu'}  \big|_P \Omega^{(\mu')} \Omega^{(\nu')}=0.
\end{equation}
Since the procedure can be repeated for all geodesics originating from $P$, the components $\Omega^{(\mu')}$ are arbitrary, from which it follows that
$\Gamma^{\alpha'}_{\ \mu' \nu'}\big|_P=0$, which completes the proof.

We can therefore conclude that around the event $P$, the metric in RNCs is $g_{\mu'\nu'}=\eta_{\mu\nu}+{\cal O}(x')^2$. It is possible to prove~\cite{Poisson_book} that the quadratic terms ${\cal O}(x')^2$
are proportional to components of the Riemann tensor, i.e. those terms, being dimensionless, scale as $(x'/L)^2$, with $L$ the curvature radius of the spacetime (defined from the Riemann tensor).

\subsection{Fermi Normal Coordinates}
Let us now slightly modify the idea behind the geometric construction of RNCs to build a set of coordinates describing the reference frame of an observer in motion along a generic timelike worldline $\gamma$ with 4-acceleration $a^\mu$. The coordinates, usually referred to as ``Fermi normal coordinates'' (FNCs) are defined in a worldtube surrounding the worldline $\gamma$.

Let us consider a tetrad $e_{(\alpha)}$ (with $\alpha=1,\ldots 4$) attached to the wordline $\gamma$, with $e_{(0)}^\mu=u^\mu$
(i.e. the ``time'' leg of the tetrad coincides with the 4-velocity of the worldline). The tetrad is assumed to be Fermi-Walker transported along $\gamma$.\footnote{A vector
$\omega^\mu$ is said to be Fermi-Walker transported along a worldline with 4-velocity $u^\mu$, proper time $\tau$ and acceleration $a^\mu$ if
\begin{equation}
    \frac{\mathrm{D} \omega^\mu}{\mathrm{d} \tau} = - \omega_\nu \left(a^\mu u^\nu - a^\nu u^\mu \right),
\end{equation}
with $\mathrm{D}/{\mathrm{d} \tau}$ the covariant derivative along the four velocity. This transport preserves angles and internal products,  and reduces to  parallel transport when $a^\mu =0$. Moreover, it is easy to check that the 4-velocity is always Fermi-Walker transported, which explains why we can always choose the time leg of our Fermi-Walker transported tetrad to be the 4-velocity.}
For any point $P'$ within a worldtube surrounding $\gamma$, let us consider the unique (spacelike) geodesic that passes through $P'$ and intersects orthogonally the trajectory $\gamma$. The intersection point will be denoted by $P$. Let us assign to $P'$ a new time coordinate 
\begin{equation}
    x^{0'}=\tau\,,
\end{equation}
 where $\tau$ is the proper time of the worldline at $P$. As for the spatial coordinates, let us consider, at the point $P$, the vector $v$ tangent to the spacelike geodesic linking $P$ and $P'$, and decompose it on the spatial ``triad'' $e_{(i)}$. (Note that the projection on $e_{(0)}$  vanishes, because the spacelike geodesic is constructed to be orthogonal to $\gamma$.) Denoting by $s$ the proper length along this spacelike geodesic, let us therefore write
\begin{equation}
    v^{\mu} = \Omega^{(i)} e^{\mu}_{(i)}\,,
\end{equation}
where the $\Omega^{(i)}$ are the projections on the triad, and define the new spatial coordinates of the point $P'$ as 
\begin{equation}
    x^{i'}=\Delta s\, \Omega^{(i')}\,.
\end{equation}
Proceeding like in the case of RNCs, it is easy to check that this definition is invariant under affine reparametrizations of the geodesic connecting $P$ and $P'$. 

To explore the consequences of these definitions, let us  compute the Jacobian of the transformation between the old coordinates $x^\mu$ and the new FNCs $x^{\mu'}$, on the worldline $\gamma$.
First, note that
\begin{equation}
\label{eq:jacobian0}
    \frac{\partial x^{\alpha}}{\partial x^{0'}}\bigg|_\gamma = u^{\alpha} = e_{(0)}^{\alpha}\,
\end{equation}
simply because $x^{0'}=\tau$.
Again because of how we defined FNCs, we have
\begin{equation}
    \frac{\mathrm{d} x^{\alpha}}{\mathrm{d}s}\bigg|_\gamma = v^{\alpha} = \Omega^{(i)} e_{(i)}^{\alpha},
\end{equation}
but also
\begin{equation}
    \frac{\mathrm{d} x^{\alpha}}{\mathrm{d}s}\bigg|_\gamma = \frac{\partial x^{\alpha}}{\partial x^{i'}} \bigg|_\gamma\frac{\mathrm{d}x^{i'}}{\mathrm{d}s}\bigg|_\gamma = \frac{\partial x^{\alpha}}{\partial x^{i'}}\bigg|_\gamma \Omega^{(i')}.
\end{equation}
Comparing the two, we then obtain
\begin{equation}
\label{eq:jacobiani}
   \frac{\partial x^{\alpha}}{\partial x^{i'}} \bigg|_\gamma =  e_{(i')}^{\alpha}\,.
\end{equation}
Equations \ref{eq:jacobian0} and \ref{eq:jacobiani} can 
then be expressed concisely as
\begin{equation}\label{jacFNC}
    \frac{\partial x^{\mu}}{\partial x^{\mu'}} \bigg|_\gamma =  e_{(\mu')}^{\mu}.
\end{equation}
The metric at point $P$ in FNCs is then
\begin{equation}
    g_{\mu' \nu'} |_\gamma = \frac{\partial x^{\mu}}{\partial x^{\mu'}} \bigg|_\gamma\frac{\partial x^{\nu}}{\partial x^{\nu'}}\bigg|_\gamma g_{\mu \nu}|_\gamma = e_{(\mu')}^{\mu} e_{(\nu')}^{\nu} g_{\mu \nu}|_\gamma  = \eta_{\mu'\nu'}\,.
\end{equation}
Therefore, FNCs 
ensure that the metric is Minkowskian on the {\it whole} worldline $\gamma$.

Let us now see what happens for the Christoffel symbols on $\gamma$. Since the spacelike geodesic connecting $P$ and $P'$ 
is parametrized by $x^{0'}=\,$ const and $x^{i'}= \Delta s\, \Omega^{(i')}$ in FNCs, one must have
\begin{equation}
\label{eq:geodesic}
    \frac{\mathrm{d}^2 x^{\mu'}}{\mathrm{d}s^2}\bigg|_\gamma + \Gamma^{\mu'}_{\ {\alpha'} {\beta'}}\big|_\gamma\frac{\mathrm{d} x^{\alpha'}}{\mathrm{d}s} \bigg|_\gamma\frac{\mathrm{d} x^{\beta'}}{\mathrm{d}s}\bigg|_\gamma=\Gamma^{\mu'}_{\ {i'} {j'}}\big|_\gamma \Omega^{(i')} \Omega^{(j')} =0\,,
\end{equation}
and since the components $\Omega^{(i')}$ are generic, $\Gamma^{\mu'}_{\ i'j'}|_\gamma=0$.

Let us then parametrize the worldline $\gamma$ in FNCs, i.e. $x^{0'}=\tau$ and $x^{i'}=0$. Since $\gamma$ is not necessarily a geodesic, one has
\begin{equation}
\label{eq:worldline}
  a^{\mu'} = \frac{\mathrm{d}^2 x^{\mu'}}{\mathrm{d}\tau^2}\bigg|_\gamma + \Gamma^{\mu'}_{\ {\alpha'} {\beta'}}\big|_\gamma\frac{\mathrm{d} x^{\alpha'}}{\mathrm{d}\tau} \bigg|_\gamma\frac{\mathrm{d} x^{\beta'}}{\mathrm{d}\tau}\bigg|_\gamma
= \Gamma^{\mu'}_{\ {0'} {0'}}\big|_\gamma\,.
  \end{equation}
Moreover, by inverting Eq.~\ref{jacFNC}
one finds
\begin{equation}\label{jacFNC2}
    \frac{\partial x^{\mu'}}{\partial x^{\mu}} \bigg|_\gamma =  e^{(\mu')}_{\mu}\,,
\end{equation}
which allows for computing   $a^{\mu'}$:
\begin{equation}
    a^{\mu'}=\frac{\partial x^{\mu'}}{\partial x^{\mu}} \bigg|_\gamma a^\mu=e^{(\mu')}_{\mu}a^\mu\,,
\end{equation}
which yields $a^{0'}=0$ (since the time leg of the tetrad is the 4-velocity, which is orthogonal to the acceleration because of the unit-norm condition) and $a^{i'}=a^{(i')}=e^{(i')}_{\mu} a^\mu$. From Eq. \ref{eq:worldline} one therefore obtains
$\Gamma^{0'}_{0'0'}|_\gamma=0$ and $\Gamma^{i'}_{0'0'}|_\gamma=a^{(i')}$.

Finally,  let us consider a space-like unit-norm vector
$\omega$ orthogonal to the worldline, i.e. in FNCs $\omega^{\mu'}=(0,\bar{\Omega}^{(i)})$, with $\bar{\Omega}^{(i)}=\,$const such that $\delta_{ij} \bar{\Omega}^{(i)}\bar{\Omega}^{(j)}=1$.
 Its components in the original coordinates are
\begin{equation}
    \omega^{\mu} = \frac{\partial x^{\mu}}{\partial x^{\mu'}} \bigg|_\gamma\omega^{\mu'} = e^{\mu}_{(\mu')} \omega^{\mu'} = e^{\mu}_{(i')} \bar{\Omega}^{(i)}.
\end{equation}
Because the components $\bar{\Omega}^{(i)}$
are constant and because the triad $e_{(i)}$ 
is Fermi-Walker transported along $\gamma$,
this vector is also
 Fermi-Walker transported along $\gamma$, and thus
\begin{equation}\label{dw1}
    \frac{\mathrm{D} \omega^{\mu'}}{\mathrm{d} \tau} = -(a^{\mu'} u^{\nu'}-a^{\nu'} u^{\mu'}) \omega_{\nu'} = a_{\nu'} \omega^{\nu'} u^{\mu'} = a_{(i')} \bar{\Omega}^{(i')} \delta^{\mu'}_{0'}\,,
\end{equation}
where we have used the fact that $u^{\mu'}=\delta^{\mu'}_{0'}$. However, by definition one also has
\begin{equation}
\frac{\mathrm{D} \omega^{\mu'}}{\mathrm{d} \tau}= \frac{\mathrm{d}\omega^{\mu'}}{\mathrm{d}\tau} + \Gamma^{\mu'}_{\alpha' \beta'} u^{\alpha'} \omega^{\beta'} = \Gamma^{\mu'}_{0' i'}  \bar{\Omega}^{(i')}\,, 
\end{equation}
from which, by  comparing to Eq.~\ref{dw1}, one obtains
\begin{equation}
    \Gamma^{0'}_{0'i'} = a_{(i')} \quad \text{and} \quad \Gamma^{j'}_{ 0'i'} = 0.
\end{equation}
In summary, collecting our findings, the Christoffel symbols on $\gamma$ in FNCs are
\begin{equation}
    \Gamma^{\mu'}_{i'j'}=0, \quad \Gamma^{0'}_{0'0'} =0, \quad \Gamma^{i'}_{0'0'} = a^{(i')}, \quad \Gamma^{0'}_{0'i'} = a_{(i')}, \quad  \Gamma^{j'}_{0'i'} = 0.
\end{equation}
These conditions  completely determine  $\partial_{\mu'} g_{\alpha' \beta'}$ on the worldline, from which one can expand the metric in the worldtube up to  second order in $x^{i'}$ to obtain
\begin{equation}\label{FNCmetric}
\begin{aligned}
    & g_{0'0'} = -1 -2a_{(i')}x^{i'} + \mathcal{O}({x'})^2, \\
    & g_{0'i'} = \mathcal{O}({x'})^2,\\
    & g_{i'j'} = \delta_{i'j'}+\mathcal{O}({x'})^2\,.
\end{aligned}    
\end{equation}
This is the metric ``felt'' by an observer on the trajectory $\gamma$. Like in the case of RNCs, the quadratic remainders 
$\mathcal{O}({x'})^2$ are actually proportional to $1/L^2$, with $L$ the curvature radius~\cite{Poisson_book}.

As can be seen, the metric is not locally flat unless the worldline is geodesic, i.e. unless the observer is in free fall. If that is not the case, the acceleration enters the metric, and specifically $g_{0'0'}$, as an apparent Newtonian potential. This potential encodes the apparent forces due to the non-inertial motion of the observer. On the other hand, if the acceleration vanishes, FNCs generalize RNCs to the entire worldtube.
Finally, note that the form of the metric in FNCs would change if we were to use a different transport law for the triad $e_{(i)}$, which can be interpreted as the spatial frame of the observer's laboratory. If the triad were not Fermi-Walker transported, additional terms would appear in the metric, corresponding to more general apparent forces (centrifugal forces, Coriolis forces, etc).

\section{The stress energy tensor of gravitational waves}

Let us now revisit the problem of defining the stress energy tensor of the gravitational field in GR. As clear from the local flatness theorem, it is always possible to choose a coordinate chart in which the effect of the gravitational field locally vanishes. In these local coordinates, the metric is therefore flat
and local experiments obey the laws of physics without gravity. Choosing these coordinates, as should be clear from the geometric construction of FNCs provided in the previous section, corresponds to adopting a reference frame attached to a free falling observer, e.g. the frame of a free falling elevator. Their existence is thus a manifestation of the well-known equivalence principle of GR.

Because the effect of the gravitational field can always be made to vanish with a proper choice of coordinates, it is clear that it is not possible to define a stress energy {\it tensor} for the gravitational field, because if a tensor is non-zero in a set of coordinates, it is non-zero in any other set of coordinates connected to the first by a non-singular transformation. The most we can aspire to is therefore to build a stress energy {\it pseudotensor} for the gravitational field in GR.
We have already encountered this pseudotensor, the Landau-Lifschitz
pseudotensor, when deriving the relaxed form of the Einstein equations in section~\ref{rig_quad}. An alternative derivation exploits the fact that for a field theory defined by a Lagrangian density ${\cal L}(\psi_n,\partial_\mu \psi_n)$, where $\psi_n$ are fields, the stress energy tensor in flat space can be defined as
\begin{equation}\label{bonifante}
    T^{\mu}_{\phantom{\mu}\nu} =\sum_n \partial_\nu \psi_n \frac{\partial {\cal L}}{\partial(\partial_\mu \psi_n)}-\delta^{\mu}_{\phantom{\mu}\nu}\, {\cal L}\,,
\end{equation}
which satisfies the conservation equation $\partial_\mu T^{\mu}_{\phantom{\mu}\nu}=0$ because of the Lagrange equations.
This stress energy tensor is in general not symmetric in its indices (when the latter are both covariant or both contravariant), and therefore needs to be redefined as $  T^{\mu}_{\phantom{\mu}\nu}\to    T^{\mu}_{\phantom{\mu}\nu}+\partial_\alpha S^{\alpha \mu}_{\phantom{\alpha \mu}\nu}$, with a suitable tensor  $S^{\alpha \mu}_{\phantom{\alpha \mu}\nu}$ satisfying $S^{[\alpha \mu]\nu}=0$. As a result of this symmetry, the redefined tensor still satisfies
$\partial_\mu T^{\mu}_{\phantom{\mu}\nu}=0$.

Because this construction only works for a Lagrangian depending on first derivatives of the fields, for GR let us start from the $\Gamma$-$\Gamma$ (or Schr\"odinger) action
\begin{align}
    &S=-\frac{1}{16 \pi} \int \mathrm{d}x^4 {\cal L}\,,\\
&{\cal L}=-\sqrt{-g} g^{\alpha\beta} \left(\Gamma^{\mu}_{\alpha\beta} \Gamma^{\nu}_{\mu\nu}-
    \Gamma^{\nu}_{\alpha\mu} \Gamma^{\mu}_{\beta\nu}    
    \right)\,,
\end{align}
which differs from the usual Einstein-Hilbert action by a surface term. One can then define the Einstein-Schr\"odinger complex 
\begin{equation}
    t^{\mu}_{\phantom{\mu}\nu} =-\frac{1}{16 \pi \sqrt{-g}} \left[ 
    \partial_\nu g_{\alpha\beta} \frac{\partial {\cal L}}{\partial(\partial_\mu g_{\alpha\beta})}-\delta^{\mu}_{\phantom{\mu}\nu} {\cal L}\right]\,,
\end{equation}
which satisfies $\partial_\mu [\sqrt{-g}(t^{\mu}_{\phantom{\mu}\nu}+T^{\mu}_{\phantom{\mu}\nu})]$, where
$T^{\mu}_{\phantom{\mu}\nu}$ is the matter stress energy tensor.
Not only is this object a pseudotensor and  not a tensor, but it is also not symmetric in its two indices, when the lower index is raised. The idea is therefore to redefine it as explained above to achieve symmetry between the two indices.

In more detail, one can then show~\cite{defelice,bertschinger} that
\begin{equation}
16 \pi  \sqrt{-g}(t^{\mu}_{\phantom{\mu}\nu}+T^{\mu}_{\phantom{\mu}\nu})=\partial_\alpha S^{\mu\alpha}_{\phantom{\alpha \mu}\nu}
\end{equation}
where long algebraic manipulations give $S^{\mu\alpha}_{\phantom{\alpha \mu}\nu}=U^{\mu\alpha}_{\phantom{\alpha \mu}\nu}+\partial_\sigma W^{\mu\alpha\sigma}_{\phantom{\alpha \mu\delta}\nu}$, with $W^{\mu[\alpha\sigma]}_{\phantom{\alpha \mu\delta\,\,\,\,}\nu}=0$ and
\begin{equation}
U^{\mu\alpha}_{\phantom{\alpha \mu}\nu}=U^{[\mu\alpha]}_{\phantom{[\alpha \mu]}\nu}=\frac{1}{\sqrt{-g}} g_{\nu\beta} \partial_\sigma U^{\mu\alpha\beta\sigma}\,,
\end{equation}
where $U^{\mu\alpha\beta\sigma}=(-g) (g^{\mu\beta} g^{\alpha\sigma}-g^{\alpha\beta} g^{\mu\sigma})$ is the Landau-Lifschitz superpotential. 
As a result of the symmetries of $W^{\mu\alpha\sigma}_{\phantom{\alpha \mu\delta}\nu}$, one then has
\begin{equation}
16 \pi  \sqrt{-g}(t^{\mu}_{\phantom{\mu}\nu}+T^{\mu}_{\phantom{\mu}\nu})=\partial_\alpha U^{\mu\alpha}_{\phantom{\alpha \mu}\nu}\,.
\end{equation}
By manipulating this equation, one can then define a new complex, the Landau-Lifschitz pseudotensor
$\tau_{LL}^{\mu\nu}$, which is symmetric in the two indices and which
satisfies
\begin{equation}
    16 \pi (-g) (T^{\mu\nu}+\tau_{LL}^{\mu\nu})=\partial_\alpha\partial_\beta U^{\mu\alpha\nu\beta}\,.
\end{equation}
Since the Landau-Lifschitz superpotential is antisymmetric under exchanges of the first and second indices and under exchanges of the third and fourth indices, one finally obtains the conservation equation
\begin{equation}
    \partial_\mu \left[  (-g) (T^{\mu\nu}+\tau_{LL}^{\mu\nu})\right]=0\,.
\end{equation}

Evaluating the Landau-Lifschitz pseudotensor for a flat spacetime with a linear transverse traceless perturbation ($\partial_\mu h_{\mathrm{TT}}^{\mu\nu}=h_{\mathrm{TT}\,\mu}^{\mu}=0$) gives 
\begin{equation}\label{SETgw}
    \tau_{\alpha \beta}^{LL}= \frac{1}{32 \pi} \partial_{\alpha}h^{\mathrm{TT}}_{\rho \sigma} \partial_{\beta} h_{\mathrm{TT}}^{\rho \sigma}\,,
\end{equation}
which can be interpreted as encoding the stress energy of gravitational waves. The same result can be obtained in an even simpler way by starting from the action for a linear transverse traceless perturbations on flat space. The latter is derived by replacing $g_{\mu\nu}\approx\eta_{\mu\nu}+h^{\mathrm{TT}}_{\mu\nu}$ in the Einstein-Hilbert or Schr\"odinger action and then expanding at quadratic order:
\begin{equation}
    S=-\frac{1}{64\pi}\int \mathrm{d}^4 x \left(\partial_{\alpha}h^{\mathrm{TT}}_{\rho \sigma} \eta^{\alpha\beta}\partial_{\beta} h_{\mathrm{TT}}^{\rho \sigma}\right)\,.
\end{equation}
One can then apply the procedure of Eq.~\ref{bonifante} to get to 
Eq.~\ref{SETgw}.

This derivation of a ``stress-energy tensor'' for gravitational waves, while correct, can be confusing. We have indeed started by noticing that defining a (local) stress-energy tensor for the gravitational field is impossible (because of the equivalence principle and the local flatness theorem) and we have ended up deriving one for gravitational waves. The only possible explanation is that the ``stress-energy tensor'' for gravitational waves is inherently a {\it non-local} object. This will become clear from the alternative derivation that we will now undertake.

\begin{figure}[h]
\centering
\includegraphics[scale=.25]{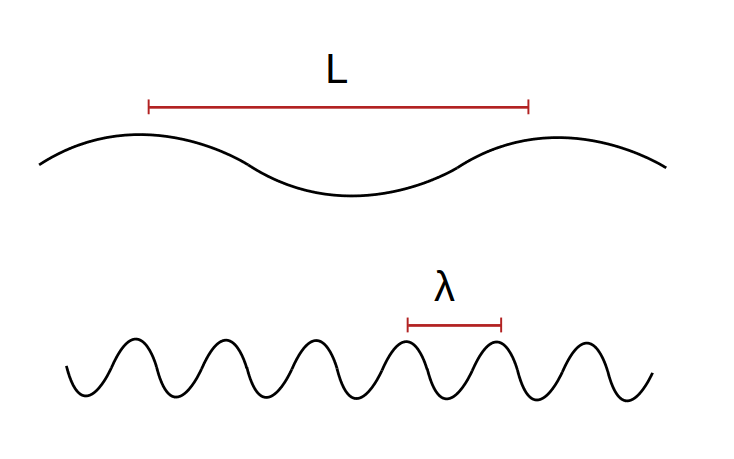}
\caption{\footnotesize Sketch of a perturbed spacetime in the geometric-optics regime. $L$ is the characteristic length of the background and $\lambda$ is the wavelength of the perturbation.}
\label{fig:st_background}
\end{figure}

Let us start by considering a 
background spacetime with a small perturbation. Let us also assume that the perturbation changes on a characteristic time and length scale $\lambda$
much smaller than the background's curvature radius $L$. This situation is usually referred to as geometric-optics regime and is shown schematically in Fig.~\ref{fig:st_background}.
Let us also define an average  $\langle \dots \rangle$ over 
lengths and times $\gg \lambda$ and $\ll L$.

Let us split the spacetime metric as
\begin{equation}
 g_{\alpha \beta}=g^{B}_{\alpha \beta} + \epsilon h_{\alpha \beta} + \epsilon^2 j_{\alpha \beta} +\mathcal{O}(\epsilon^3),
\end{equation}
where $g^{B}_{\alpha \beta}$ is unperturbed background metric,
while $h_{\alpha \beta}$ and $j_{\alpha \beta}$ are the first and second order perturbations ($\epsilon$ being a small parameter).
Let us then consider the vacuum Einstein equations
$ G_{\alpha \beta}= 0$, and expand them in $\epsilon$ as
\begin{equation}\label{2ndOrderEq}
    0= G_{\alpha \beta} [g^{B}] + \epsilon G^{(1)}_{\alpha \beta}[h,g^{B}] + \epsilon^2\Big(G^{(1)}_{\alpha\beta} [j,g^B] + G^{(2)}_{\alpha \beta}[h,g^B] \Big).
\end{equation}
Here, the first term is the
 Einstein tensor computed with the background metric, the second one gives the equations of motion for the first order perturbations (c.f. section~\ref{curved_space_pert}), while the $\epsilon^2$ term gives the equations of motion for the second order perturbations (which are in turn comprised of terms quadratic in $h$, denoted by $G_{\alpha\beta}^{(2)}[h,g^B]$, and  terms linear in $j$, denoted by $G^{(1)}_{\alpha\beta} [j,g^B]$).
 
 Taking now an average of this equation, one can note that since
 $G^{(1)}[\ldots,g^B]$ is a linear operator (the linearized Einstein tensor on the background metric),
 the average commutes with it, giving therefore
 $\langle G^{(1)}_{\alpha\beta} [h,g^B]\rangle=
  G^{(1)}_{\alpha\beta} [\langle h \rangle,g^B]=0$
(because the first order perturbations are oscillatory and thus average to zero, cf. section~\ref{curved_space_pert}). Similarly, $\langle G^{(1)}_{\alpha\beta} [j,g^B]\rangle=
  G^{(1)}_{\alpha\beta} [\langle j \rangle,g^B]$,
  where we are allowing for $\langle j \rangle\neq0$. In fact,  the second order equations of motion have the cartoon form $\Box j_{\alpha\beta}={\cal O}(h)^2 $, i.e. second order perturbations are sourced by products of first order ones. As such, the average of the second order perturbations cannot be zero as $\langle {\cal O}(h)^2\rangle$ will not vanish, in general.
  The average of Eq.~\ref{2ndOrderEq} thus yields
\begin{equation}
    0= G_{\alpha \beta} [g^{B}] +\epsilon^2\Big(G^{(1)}_{\alpha\beta} [\langle j\rangle,g^B] + \langle G^{(2)}_{\alpha \beta}[h,g^B]\rangle \Big)\,,
\end{equation}
where we have used the fact that $\langle G_{\alpha \beta} [g^{B}]\rangle= G_{\alpha \beta} [g^{B}]$ since the background metric varies on scales much larger than those on which the average is performed.
This equation can then be written, by resumming the Taylor expansion, as 
\begin{equation}
    G_{\mu\nu} [g^B +\epsilon^2\langle j \rangle]=8\pi G T_{\mu \nu}^{\rm GW},
\end{equation}
with 
\begin{equation}
    T^{\rm GW}_{\mu \nu}=-\frac{1}{8\pi } \langle G_{\mu\nu}^{(2)}[h,g^B] \rangle ,
\end{equation}
To compute this average,  one can write $G_{\mu\nu}^{(2)}[h,g^B] $ explicitly, use the fact that 
covariant derivatives commute up to terms depending on the Riemann tensor, and show that these terms are subleading in the geometric-optics regime $\lambda\ll L$. A more detailed calculation can be found in \cite{Flanagan:2005yc} and \cite{misner1973gravitation} and  and yields (for transverse traceless perturbations)
\begin{equation}\label{SETgw2}
    T_{\alpha \beta}^{\rm GW}= \frac{1}{32 \pi} \langle\nabla_{\alpha}h^{\mathrm{TT}}_{\rho \sigma} \nabla_{\beta} h_{\mathrm{TT}}^{\rho \sigma}\rangle.
\end{equation}
Remarkably, this is the same result that we derived previously, except for the average, which shows explicitly that the stress-energy tensor of gravitational waves is a non-local object
(i.e. it makes no sense to define the stress-energy tensor of gravitational waves pointwise, but only on scales larger than the wavelength). Just as 
the gravitational 
force disappears for an  observer in a free-falling  elevator
 if the elevator is much smaller than the Earth (otherwise non-local tidal effects appear, cf. Eq.~\ref{FNCmetric}), the gravitational wave perturbation can be set to zero by going to RNCs locally, but only on scales smaller than the spacetime curvature radius (which is given by $\lambda$ for the perturbed spacetime represented in Fig.~\ref{fig:st_background}).

\subsection{The gravitational contribution to the mass of a compact star}

In this section we will take a short detour and investigate another example showing that the gravitational field, although it can be set to zero locally, provides a finite contribution to the energy of the system on non-local scales. We will consider indeed a compact spherically symmetric star, and compute the contribution of the gravitational field to its mass.

Let us start by modeling the metric as
\begin{equation}
    ds^2 = -B(r)dt^2 + A(r) dr^2 + r^2 d\Omega^2,
\end{equation}
and the matter by a perfect fluid. The Einstein equations and the conservation of the fluid's stress energy tensor then yield (assuming asymptotic flatness) the famous
Tolman-Oppenheimer-Volkoff equations~\cite{Shapiro:1983du}:
\begin{equation}\label{eq:ein_tensor_gw}
\begin{aligned}
    & \frac{dp}{dr} = -\rho \frac{m(r)}{r^2} \Bigg( 1+ \frac{p}{\rho}\Bigg) \Bigg( 1+ \frac{4\pi r^3 p}{m(r)}\Bigg) \Bigg( 1 - \frac{2m(r)}{r}\Bigg)^{-1} ,\\
    & \frac{1}{2B}\frac{dB}{dr} = - \frac{1}{p+\rho} \frac{dp}{dr},\\
    & A(r) = \frac{1}{1- \frac{2m(r)}{r}},
\end{aligned}
\end{equation}
where
\begin{align*}
    m(r) = \int^r_0 \rho(r') 4 \pi r'^2 dr'.
\end{align*}
    Clearly, the first equation is the relativistic Euler equation, where $\frac{p}{\rho}$, $\frac{4\pi r^3 p}{m(r)}$ and $\frac{2m(r)}{r}$ are the relativistic corrections, which disappear in the Newtonian limit $c \rightarrow \infty$ (if $c$ is reinstated). As can be seen, the solution to these equations reduces to the Schwarzschild metric in vacuum (and thus in the exterior of the star). In particular, the metric in the exterior is given by the Schwarzschild metric with mass $m(\infty)=\int^R_0 \rho(r') 4 \pi r'^2 dr'$, where $R$ is the radius of the star. This mass can be interpreted as the star's gravitational mass, as measured by an observer that were to fly satellites far from the star and interpret their motion with Kepler's law. More formally, one can show that this mass matches the Arnowitt-Deser-Misner mass of the spacetime (which in turn matches its Komar mass), see e.g.~\cite{townsend,wald}.

The total baryonic mass of the star can instead  be obtained from the continuity equation for the baryonic current $j^\mu=m_b n u^\mu$,
where $m_b$ is the average baryon mass and $n$ is the baryon number density. The baryonic mass is then
\begin{equation}
    M_b=\int\mathrm{d}x^3 \sqrt{-g} j^t =m_b \int 4\pi r^2 \sqrt{A(r)} n(r) dr\,.
\end{equation}
This mass corresponds to the sum of the rest masses of all the baryons of the system, but does not include the internal energy. Since the internal energy density of a fluid is given by the difference $\rho-m_b n$, we would expect the total mass of the system to be given simply by
\begin{equation}
    M_\star= \int 4\pi r^2 \sqrt{A(r)} \rho(r) dr\,.
\end{equation}
This mass, however, still differs from $m(\infty)$. To understand this discrepancy one may compute the difference between the two, reinstate factors of $G$ and $c$ (using e.g. dimensional analysis) and expand it in orders of $1/c^2$.
Equivalently, one can expand the difference in the weak gravity limit $m(r)/r\ll 1$. By using the fact that
$A(r)=[1-2 Gm(r)/(r c^2)]^{-1}$, one then finds
\begin{align}
    &m(\infty)  -M_\star= - \int \mathrm{d}r \frac{4 \pi r^2 \rho(r) Gm(r)}{rc^2}= U_{\rm self}/c^2 \,,\\
    & U_{\rm self}=- G\int \mathrm{d}m(r) \frac{m(r)}{r}\,.
\end{align}
This shows \textit{(i)} that the gravitational mass $m(\infty)$ is always smaller than the expected value $M_\star$, and \textit{(ii)} that in the Newtonian limit the difference is given exactly by the contribution of the (Newtonian) gravitational self-energy,  i.e. (in absolute value) the work that one would need to perform against the gravitational force to destroy the star. Once again, we have found that even though it can be locally set to zero, the gravitational field does contribute to the mass of an extended object. This contribution is of the order of $10-20\%$ for neutron stars.

\section{The inspiral and merger of binary systems of compact objects}
In this section we will use the results that we have derived
to gain some semi-quantitative understanding of the physics of binary systems of compact objects (black holes and neutron stars).

Let us first apply the quadrupole formula to a system of two compact objects with masses $m_1$ and $m_2$ on a circular orbit of radius $r$ and orbital frequency $\Omega$, which at lowest (i.e. Newtonian) order is given by Kepler's law
\begin{equation}\label{eq:kepler}
    \Omega= \sqrt{\frac{M}{r^3}}\,,
\end{equation}
with $M=m_1+m_2$. The  gravitational wave signal predicted by the quadrupole formula, for an observer at distance $D$ and angle $\iota$
with respect to the direction orthogonal to the orbital plane, then reads
\begin{equation}\label{hTT}
h^{ \rm TT}_{ij} = h
 \left[\begin{matrix}
\frac{1+\cos^2\iota}{2}\cos2\Omega t  & \cos\iota\sin2\Omega t & 0 \cr
\cos\iota\sin2\Omega t &-\frac{1+\cos^2\iota}{2}\cos2\Omega t & 0 \cr
0 & 0 & 0 \cr
\end{matrix}\right]
\end{equation}
with the amplitude (also referred to as ``strain'') being
\begin{equation}\label{hQ}
h = \frac{4\mu\Omega^2 r^2}{D} = \frac{4\mu M^{2/3}\Omega^{2/3}}{D}\;,
\end{equation}
where we have also introduced the reduced mass $\mu=m_1 m_2/M$.

Several comments are worth making here. First, the frequency of the gravitational wave signal is twice the orbital frequency. This is due to the tensor nature of gravitational waves. By reinstating $G$ and $c$ (recalling that $h$ must be dimensionless) and computing explicitly the  amplitude, one finds e.g. a strain of $\sim 10^{-22}$
for a system of two neutron stars of masses $m_1=m_2=1.4 M_\odot$, orbital period of 10 ms and distance of 50 Mpc. Similarly, a strain
$h\sim 10^{-21}$ can be obtained e.g. for an equal mass binary of black holes of 30 $M_\odot$ each, at a distance of 400 Mpc and with the same orbital period of 10 ms. Clearly, these sources have gravitational wave frequencies $\sim 100$ Hz, and as we will see they are detectable by current ground based interferometers (LIGO-Virgo-KAGRA).
Similarly, typical sources in the bands of LISA (mHz) and pulsar-timing arrays (nHz) are e.g. a binary of  massive black holes of $10^7 M_\odot$ each, with period of a few hours and distance of few Gpc ($h\sim 10^{-16}$) or a   binary of supermassive black holes of $10^9 M_\odot$ each, with period of one year and distance $\sim 1$ Gpc ($h\sim 10^{-15}$), respectively. Note how small the amplitude of these metric perturbations is compared to the amplitude of the metric perturbation at the surface of the Sun, $h\sim G M_\odot/(R_\odot c^2) \sim 10^{-6}$. Another important observation is that the strain decays as $1/D$. This is why gravitational wave observations allow for exploring the Universe up to high redshift. Not only do gravitational waves interact very weakly with matter (since the interaction is only gravitational), but  interferometers detect directly the gravitational wave strain $h$, which decays more slowly than the electromagnetic fluxes ($\propto 1/d^2$) collected e.g. by optical telescopes.

Using now Eq.~\ref{hQ} in the stress energy tensor of gravitational waves, one can get the gravitational wave flux $T^{\rm GW}_{ti}$. Like all fluxes, this decays as $1/D^2$, but we stress again that interferometers observe $h$ directly, and not the flux. Integrating the flux on a sphere far from the source, one 
finds that the gravitational wave luminosity of a binary (i.e. the 
 energy carried away by gravitational waves per unit time)
 takes the simple form
\begin{equation}\label{Lgw}
    \dot{E}_{\rm GW}= \frac{32}{5}\frac{G}{c^3}\Bigg(\frac{Gm_1 m_2}{r^2} \Bigg)^2 \Bigg( \frac{v}{c} \Bigg)^2\,,
\end{equation}
where we have explicitly reinstated $G$ and $c$.
The energy that gravitational waves remove from the source must of course come from the system's kinetic and potential energy. For a circular Keplerian binary, the sum of kinetic and potential energy is simply given, in the center of mass frame, by
\begin{equation}\label{eq:en_binary}
    E_{\rm tot} = \frac{1}{2} \mu v^2 - \frac{GM\mu}{r} = -\frac{1}{2}\frac{GM\mu}{r}\,.
\end{equation}
By requiring energy conservation ($\dot{E}_{\rm tot}= -\dot{E}_{\rm GW}$), one can then obtain an expression of the rate of change of the separation, $\dot{r}$, due to gravitational wave emission. Clearly $\dot{r}<0$, i.e. the binary slowly spirals in under the backreaction of gravitational waves. This can of course be interpreted as a PN contribution to the acceleration of the system, i.e.
\begin{equation}
    \vec{a}_{12}= \vec{a}_N \Big( 1 + \mathcal{O}\Big(\frac{1}{c^2}\Big) +  \mathcal{O} \Big(\frac{1}{c^4}\Big) + \mathcal{O} \Big(\frac{1}{c^5}\Big)      \Big),
\end{equation}
where $\vec{a}_N$ is the Newtonian acceleration, $\mathcal{O} (\frac{1}{c^2})$ is the 1PN (conservative) correction, $\mathcal{O}(\frac{1}{c^4})$ is the 2PN (conservative) correction and $\mathcal{O}(\frac{1}{c^5})$ is the 2.5PN (dissipative) backreaction of gravitational waves. Note that the $\mathcal{O}(\frac{1}{c^5})$  scaling of the last term follows from the factors of $1/c$ in Eq.~\ref{Lgw}.

Using Kepler's law, $\dot{r}$ can be recast into the rate of change of the orbital angular frequency, $\Omega$. Using then the relation between gravitational wave frequency and $\Omega$, $f=2 f_{\rm orb}=\Omega/\pi$, one finally obtains 
\begin{equation}\label{fdot}
    \begin{aligned}
    & \dot{f}=\frac{96}{5} \frac{1}{c^5} \pi^{8/3}(GM_c)^{5/3} f^{11/3}\,,
    \end{aligned}
\end{equation}
where we have introduced the \emph{chirp mass}
\begin{equation}\label{Mc}
    M_c= M \eta^{3/5},
\end{equation}
\noindent
where $\eta =\frac{\mu}{M}$ is the symmetric mass ratio.
The chirp mass is indeed the quantity that can be most easily estimated from the gravitational wave signal from inspiraling binaries:  as gravitational waves remove
 energy and angular momentum, the binary spirals in, the  
 separation decreases, and the frequency of gravitational waves increases depending on $M_c$ alone (at leading PN order).
 This expression can also be recast into an equation for the rate of change of the orbital period. The latter is the quantity monitored in binary pulsar systems, i.e. systems at least one component of which is a millisecond pulsar. The presence of the pulsar  allows for tracking the period of the binary system with exquisite accuracy,  historically providing for the first time evidence for the existence of gravitational waves~\cite{Taylor:1979zz}.

In order to gain more qualitative understanding of the inspiral phase beyond the leading PN order, we will now make a short detour and recall the most salient features of geodesic orbits in Schwarzschild and Kerr spacetimes. While this is only applicable to binaries with very small mass ratio $q=m_2/m_1\ll1 $, the qualitative features that we will discover (e.g. the effect of spins, the plunge) will survive even at  mass ratios $q\approx 1$. This is somewhat expected from Newtonian mechanics, where one can map a binary with arbitrary masses into a particle with the reduced mass $\mu$ around a particle with the total mass $M$, but it is not at all obvious in GR. Only recently has evidence started accumulating that a similar mapping between arbitrary binaries and the test-particle limit may exist even in PN theory, although approximately. This approximate mapping goes under the name of `effective-one-body' model~\cite{eob}.

\subsection{Geodesics in Schwarzschild and Kerr}\label{geo}

Let us start by studying geodesics in the Schwarzschild spacetime, 
whose line element we write in areal coordinates in the usual form
\begin{equation}\label{schw}
    \mathrm{d}s^2=-\left(1-2 \frac{M}{r}\right)\mathrm{d}t^2+\frac{\mathrm{d}r^2}{1-2 M/r}+r^2\mathrm{d}\Omega^2\,.
\end{equation}
 Since the metric is  static and spherically symmetric, we can look at equatorial geodesics without loss of generality (i.e. the coordinates can always be chosen to be such that orbits have $\theta=\frac{\pi}{2}$). 

Let us start with particles having non-zero mass (timelike geodesics).
From the existence of the two Killing vectors $\partial_t$ and $\partial_\phi$, it follows that 
the specific~\footnote{By ``specific'', we mean ``normalized by the particle's mass''.}
energy and angular momentum observed at infinity, i.e. $E=-u_t$ and $L=u_\phi$ with $u^\mu=\mathrm{d} x^\mu/\mathrm{d}\tau$ the four-velocity, are conserved. One can then obtain the first-order equations
\begin{equation}\label{eq: schw_en_ang_mom}
\begin{aligned}
& \left(1-\frac{2 M}{r}\right) \frac{\mathrm{d} t}{\mathrm{d} \tau}=E, \\
& r^{2} \frac{\mathrm{d} \phi}{\mathrm{d} \tau}=L\,.
\end{aligned}
\end{equation}
Moreover, by using these equations in 
 the conservation of the norm $u^\mu u_\mu=-1$, 
 one can obtain a first-order equation for the radial motion:
\begin{equation}\label{eq:rad_ev}
\frac{1}{2}\left(\frac{\mathrm{d} r}{\mathrm{d} \tau}\right)^{2}+V(r)=\frac{1}{2} E^{2},
\end{equation}
with the effective potential being given by
\begin{equation}
V(r)=\frac{1}{2} - \frac{M}{r}+\frac{L^{2}}{2 r^{2}}-\frac{M L^{2}}{r^{3}}\,.\label{Veff}
\end{equation}
Apart from the first (constant and thus irrelevant) term, this
potential includes the Newtonian potential, the usual Newtonian centrifugal term, but the last term does not appear in Newtonian mechanics. In fact, it is a 1PN term, as can be seen by reinstating $c$ by dimensional analysis.

As a consequence of this term, the behavior of the potential at small $r$ drastically differs from the Newtonian one. Unlike the latter, which predicts the existence of stable circular orbits down to arbitrarily small radii, Eq.~\ref{Veff} 
predicts the existence of an innermost stable circular orbit (ISCO)
at $r=6 M$. This can be seen by solving the equations defining circular orbits, $E^2/2-V(r)=V'(r)=0$, and checking the sign of $V''(r)$ to assess stability. As can also be understood by plotting $V$, for $E>1$ only unstable circular orbits exist. However, for each value of $E<1$ (corresponding to bound orbits) two circular orbits exist, with the one at larger radius being stable and the other being unstable. These orbits exist only for $L\geq 2\sqrt{3} M$ and coincide for $L=2\sqrt{3} M$, which corresponds to the ISCO ($r=6M$ and $E=2 \sqrt{2}/3$). The unstable circular orbits lie instead at $r<6M$, but they always have $r>3M$ (they only approach $r=3M$ in the ultra-relativistic limit $E,L\to \infty$).

By redoing the same analysis for null orbits, one can prove that the radial motion obeys 
\begin{equation}
\frac{1}{E^{2}}\left(\frac{d r}{d \lambda}\right)^{2}=-V_{\rm ph}=1-\frac{b^{2}}{r^{2}}\left(1-\frac{2 M}{r}\right),\label{Vph}
\end{equation}
where $\lambda$ is an affine parameter and 
$b=L/E$ is usually referred to as impact parameter. An analysis of the effective potential $V_{\rm ph}$ shows that a circular orbit exists only for a critical value of $b$, namely $b=3 \sqrt{3} M$. This circular null orbit (also known as light ring) lies at a radius of $r=3 M$ and is unstable. Its existence and properties are not only important for the interpretation of the observations by the Event Horizon Telescope (EHT)~\cite{eht}, but also for the physics of black hole quasinormal modes, as we will see in the following. 

The case of geodesics in a Kerr spacetime is slightly more involved, because one can no longer assume equatorial motion due to the absence of spherical symmetry. In Boyer-Lindquist coordinates the metric reads
\begin{eqnarray} 
\label{Kerr}
ds^2 &=& 
- \left( 1-\frac{2Mr}{\Sigma} \right) ~dt^2
+ \frac{\Sigma}{\Delta}~dr^2
+ \Sigma~d\theta^2 \nonumber \\
&&+ \left( r^2+a^2 + \frac{2Ma^2r}{\Sigma}\sin^2\theta \right)\sin^2\theta~d\phi^2 - \frac{4Mar}{\Sigma}\sin^2\theta~dt~d\phi,
\end{eqnarray}
where $a=S/M$ (with $S$ the spin), $M$ is the mass and
\begin{equation}
 \Sigma = r^2 + a^2\cos^2\theta, \quad \Delta = r^2 - 2Mr + a^2. 
\end{equation}
The Killing vectors $\partial_t$ and $\partial_\phi$ still exist, and imply that the specific energy $E=-u_t$ and angular momentum (in the spin direction) $L=u_\phi$ must be conserved. However, one more conservation equation (besides the unit norm condition) is needed to reduce the equations of motion to first order. Fortunately, the Kerr geometry has a `hidden' symmetry, which can be described by a Killing-Yano tensor. This symmetry implies the existence of an additional constant of motion, the Carter constant $Q$, which allows for writing the equations for timelike geodesics (with particle mass $\mu$) as
\begin{align}
&\left(\frac{dr}{d\lambda}\right)^2 = V_r(r), &
&\frac{dt}{d\lambda} = V_t(r,\theta),&
\nonumber \\
&\left(\frac{d\theta}{d\lambda}\right)^2 = V_\theta(\theta), &
&\frac{d\phi}{d\lambda} = V_\phi(r,\theta)\;,&
 \label{geodesics}
\end{align}
with 
\begin{subequations} \label{detailed geodesics}
\begin{align}
&V_t(r,\theta)
  \equiv E \left( \frac{\varpi^4}{\Delta} - a^2\sin^2\theta \right)
     + aL \left( 1 - \frac{\varpi^2}{\Delta} \right),&
\label{tdot} \\
&V_r(r)
  \equiv \left( E\varpi^2 - a L \right)^2
  - \Delta\left[ r^2 + (L - a E)^2 + Q\right],&
\label{rdot}\\
&V_\theta(\theta) \equiv Q - L^2 \cot^2\theta - a^2(1 - E^2)\cos^2\theta,&
\label{thetadot}\\
&V_\phi(r,\theta)
  \equiv L \csc^2\theta + aE\left(\frac{\varpi^2}{\Delta} - 1\right) - \frac{a^2L}{\Delta},&
\label{phidot}
\end{align}
\end{subequations}
where we have defined 
\begin{equation}
\varpi^2 \equiv r^2 + a^2
\end{equation}
and the ``Carter time'' $\lambda$ by
\begin{equation}
\frac{d\tau}{ d\lambda}\equiv\Sigma\;.
\end{equation}

\begin{figure} 
\centering
    \begin{subfigure}[b]{0.48\textwidth}
        \includegraphics[width=\textwidth]{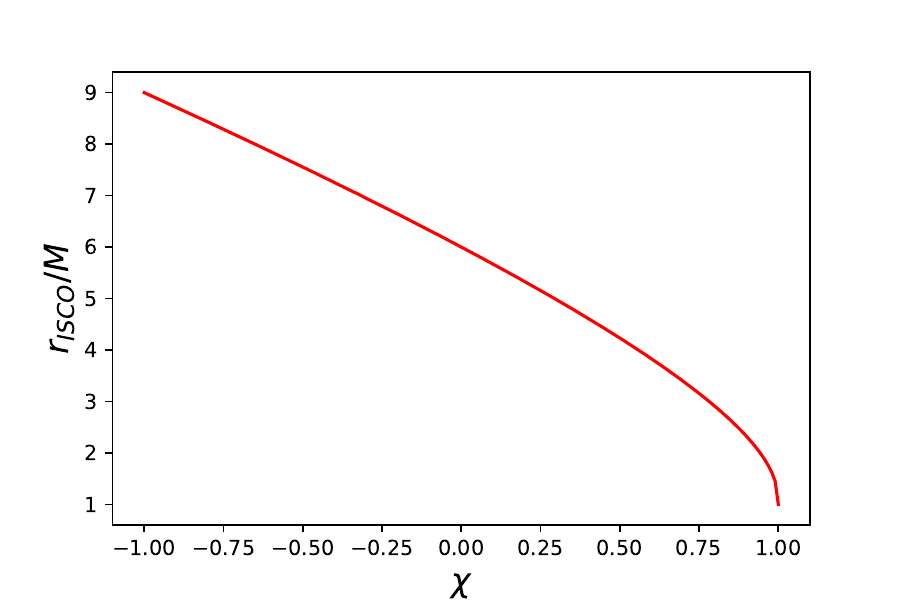}
    \end{subfigure}
    \begin{subfigure}[b]{0.48\textwidth}
        \includegraphics[width=\textwidth]{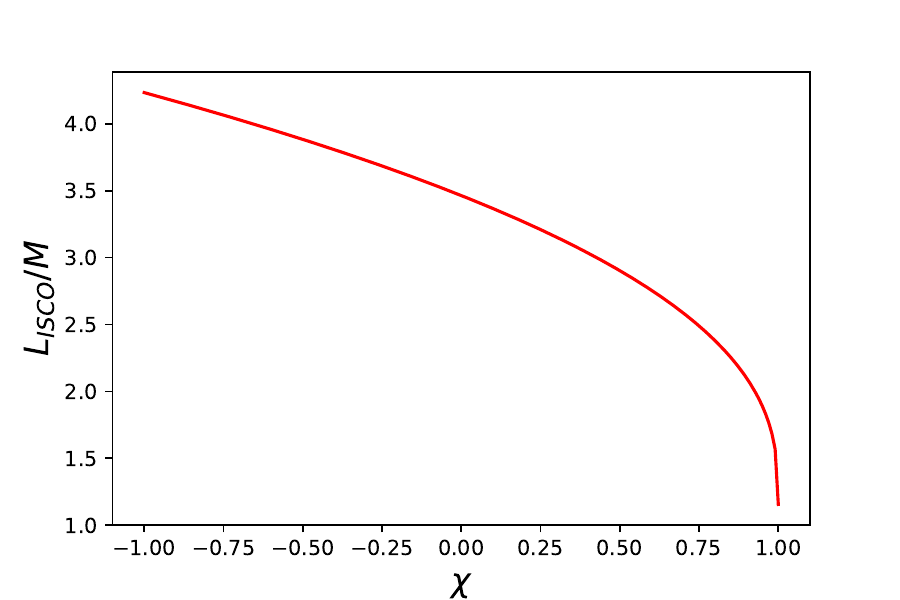}
    \end{subfigure}
        \begin{subfigure}[b]{0.5\textwidth}
        \includegraphics[width=\textwidth]{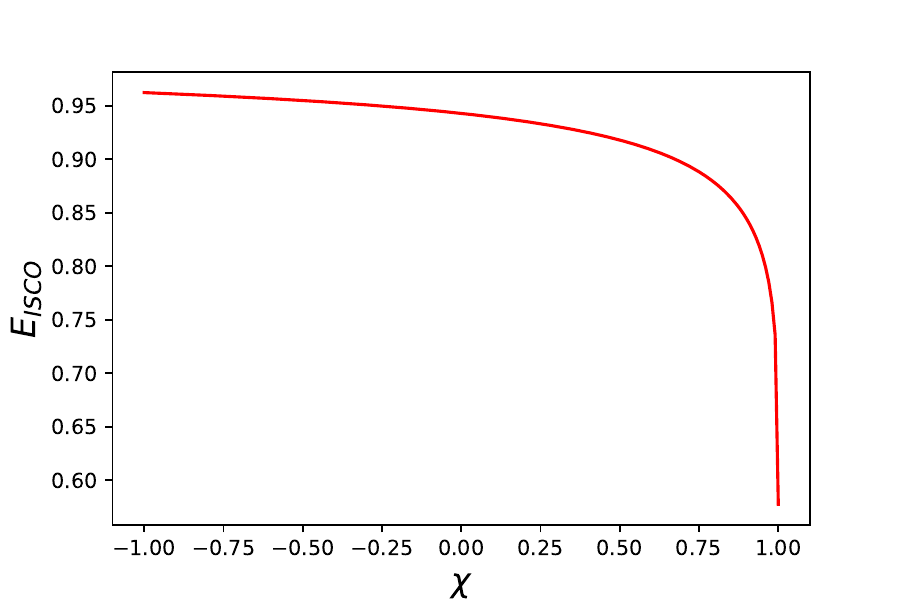}
    \end{subfigure}
    \caption{\footnotesize ISCO radius, specific angular momentum and specific energy for equatorial orbits in Kerr.}
    \label{fig:isco}
\end{figure}{}

The most striking difference from the Schwarzschild case is the presence of a non-trivial equation for the angular variable $\theta$, which 
describes the {\it precession} of the orbital angular momentum of the particle around the spin of the Kerr geometry. Moreover, the spin parameter $a$ also enters the effective potential for the radial motion. This has profound implications for the particle's motion. Focusing for instance on non-precessing (i.e. equatorial) orbits, we can set $\theta=\pi/2$ and $Q=0$, and we can obtain the radius, specific energy and specific angular momentum of circular orbits by solving $V_r(r)=V'_r(r)=0$. Like in the Schwarzschild case,
this analysis shows the existence of an ISCO, whose properties depend on the spin as~\cite{Bardeen:1972fi}
\begin{eqnarray}
\label{eisco} 
&&{E}_{_{\rm ISCO}}(\chi)=\sqrt{1-\frac{2}{3{r}_{_{\rm ISCO}}(\chi)}}\,,\label{eisco_eq}\\ 
&&{L}_{_{\rm ISCO}}(\chi)=\frac{2}{3\sqrt{3}}\left[1+2\sqrt{3{r}_{_{\rm ISCO}}(\chi)-2}\right]\,,
\label{lisco} 
\end{eqnarray}
\begin{eqnarray} 
&&{r}_{_{\rm ISCO}}(\chi)=3+Z_{2}-\frac{\chi}{|\chi|}\sqrt{(3-Z_{1})(3+Z_{1}+2Z_{2})}\,,\\ 
&&Z_{1}=1+(1-\chi^2)^{1/3}\left[(1+\chi)^{1/3}+(1-{\chi})^{1/3}\right]\,,\\ 
&&Z_{2}=\sqrt{3{\chi}^{2}+Z_{1}^{2}}\,,\label{z2}
\end{eqnarray}
where $\chi=a/M$. The parameter $\chi$ ranges from -1 to 1, with positive values corresponding to orbits co-rotating with the Kerr black hole, while negative values correspond to counter-rotating orbits. In the limit $\chi\to0$, these expressions reduce to those for the Schwarzschild ISCO. 

The dependence of these expressions on the spin is a special case of a more general phenomenon appearing in GR, the dragging of inertial frames or Lense-Thirring effect. Unlike what happens in Newtonian theory, the spin has a clear impact on the dynamics, ``dragging'' matter into rotation. In accordance with this, Eqs.~\ref{eisco}--\ref{z2} predict that as the spin increases in magnitude, prograde  orbits (i.e. ones co-rotating with the Kerr black hole) have smaller and smaller ISCO radii, down to ${r}_{_{\rm ISCO}}=M$
in the extremal limit $\chi=1$.~\footnote{Note that although the ISCO
seems to coincide with the event horizon in this limit, this is simply an artifact of the Boyer-Lindquist coordinates becoming singular in the extreme limit, as can be seen by computing the proper distance between the ISCO and the event horizon~\cite{Bardeen:1972fi}.}
As a result, the values of ${E}_{_{\rm ISCO}}$ and ${L}_{_{\rm ISCO}}$ 
decrease  as $\chi$ grows from 0 to 1.
For retrograde orbits ($\chi<0$) the ISCO instead moves to larger radii as the spin magnitude increases, up to $r=9M$ in the extreme limit. As a consequence, the values of ${E}_{_{\rm ISCO}}$ and ${L}_{_{\rm ISCO}}$ increase  as $\chi$ goes from 0 to -1. These behaviors are shown in Fig.~\ref{fig:isco}.

\begin{figure} 
\centering
    \begin{subfigure}[b]{0.48\textwidth}
        \includegraphics[width=\textwidth]{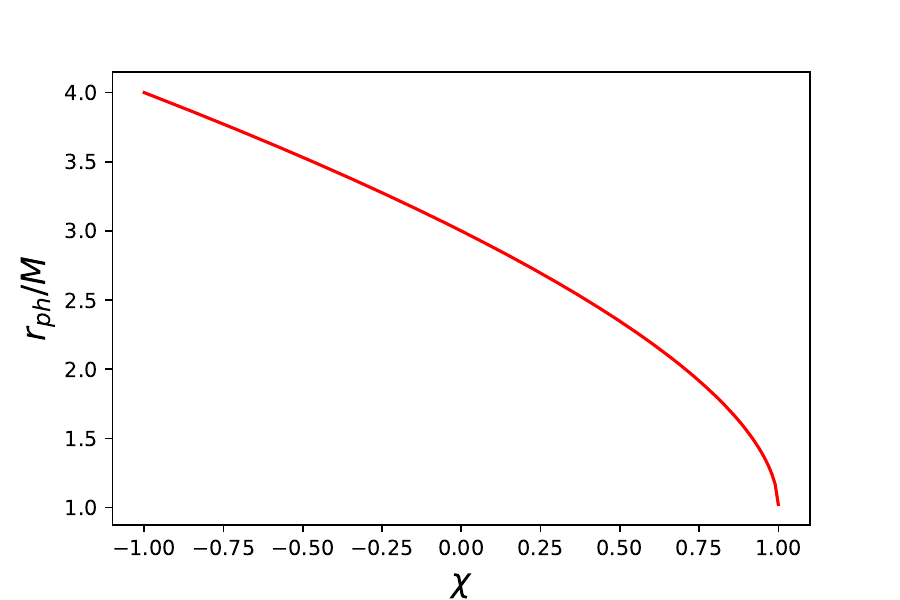}
    \end{subfigure}
    \begin{subfigure}[b]{0.48\textwidth}
        \includegraphics[width=\textwidth]{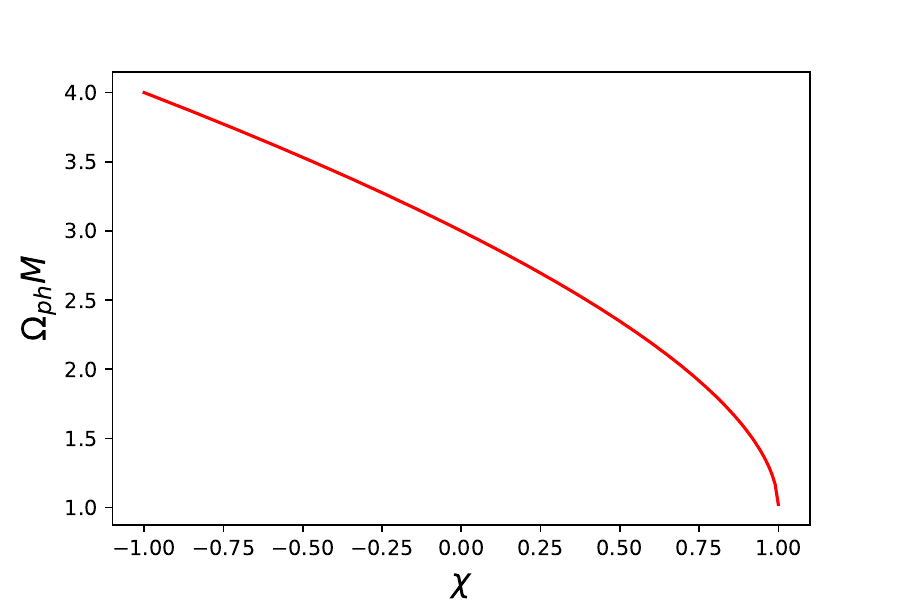}
     \end{subfigure}
    \caption{\footnotesize Circular photon orbit's radius and orbital frequency for equatorial orbits in Kerr.}
    \label{ph_orbit}
\end{figure}

For a photon in the Kerr geometry, the geodesics equations are instead~\cite{Bardeen:1972fi}
\begin{align}
&\left(\frac{dr}{d\tilde{\lambda}}\right)^2 = V_r(r), &
&\frac{dt}{d\widetilde{\lambda}} = V_t(r,\theta),&
\nonumber \\
&\left(\frac{d\theta}{d\widetilde{\lambda}}\right)^2 = V_\theta(\theta), &
&\frac{d\phi}{d\widetilde{\lambda}} = V_\phi(r,\theta)\;,&
 \label{geodesics2}
\end{align}
with $\widetilde{\lambda}$ a ``time'' parameter,  $b=L/E$,
$q=Q/E^2$, and
\begin{subequations} \label{detailed geodesics ph}
\begin{align}
&\widetilde{V}_t(r,\theta)
  \equiv  \frac{-a^2 \Delta  \sin ^2\theta +a b (\Delta -\varpi^2)+\varpi^4}{\Delta }
\label{tdot ph} \\
&\widetilde{V}_r(r)
  \equiv  (\varpi^2-a b)^2-\Delta  \left[(a-b)^2+q\right],&
\label{rdot_ph}\\
&\widetilde{V}_\theta(\theta) \equiv 
a^2 \cos ^2\theta -b^2 \cot ^2\theta+q,&
\label{thetadot_ph}\\
&\widetilde{V}_\phi(r,\theta)
  \equiv \frac{b}{\sin^2\theta}-\frac{a (a b+\Delta -\varpi^2)}{\Delta }\,.&
\label{phidot_ph}
\end{align}
\end{subequations}
Specializing again to equatorial orbits ($q=0$, $\theta=\pi/2$), one finds, like in Schwarzschild, that there exists an unstable circular photon orbit, whose coordinate radius reads~\cite{Bardeen:1972fi}
\begin{equation}
r_{\rm ph}(\chi)=2M \{1+\cos[2/3 \arccos(- \chi)]\}\,.
\end{equation} 
This is plotted in Fig.~\ref{ph_orbit}, together with the orbital frequency $\Omega_{\rm ph}=\mathrm{d}\phi/\mathrm{d}t$ (obtained from Eq.~\ref{detailed geodesics ph}). As can be seen, the frame dragging once again makes the photon orbit radius decrease  with $\chi$, with $r_{\rm ph}\to M$ as $\chi\to 1$.~\footnote{As for the ISCO, this is due to the coordinates becoming singular in the extreme limit. The proper distance between the circular photon orbit and the event horizon remains non-zero~\cite{Bardeen:1972fi}.}
Like in the Schwarzschild case, not only are circular photon orbits relevant for EHT observations, but also for the physics of quasinormal modes, as we will see below.

\subsection{A qualitative description of the inspiral and merger}

\begin{figure}
    \centering
    \includegraphics[width=\textwidth]{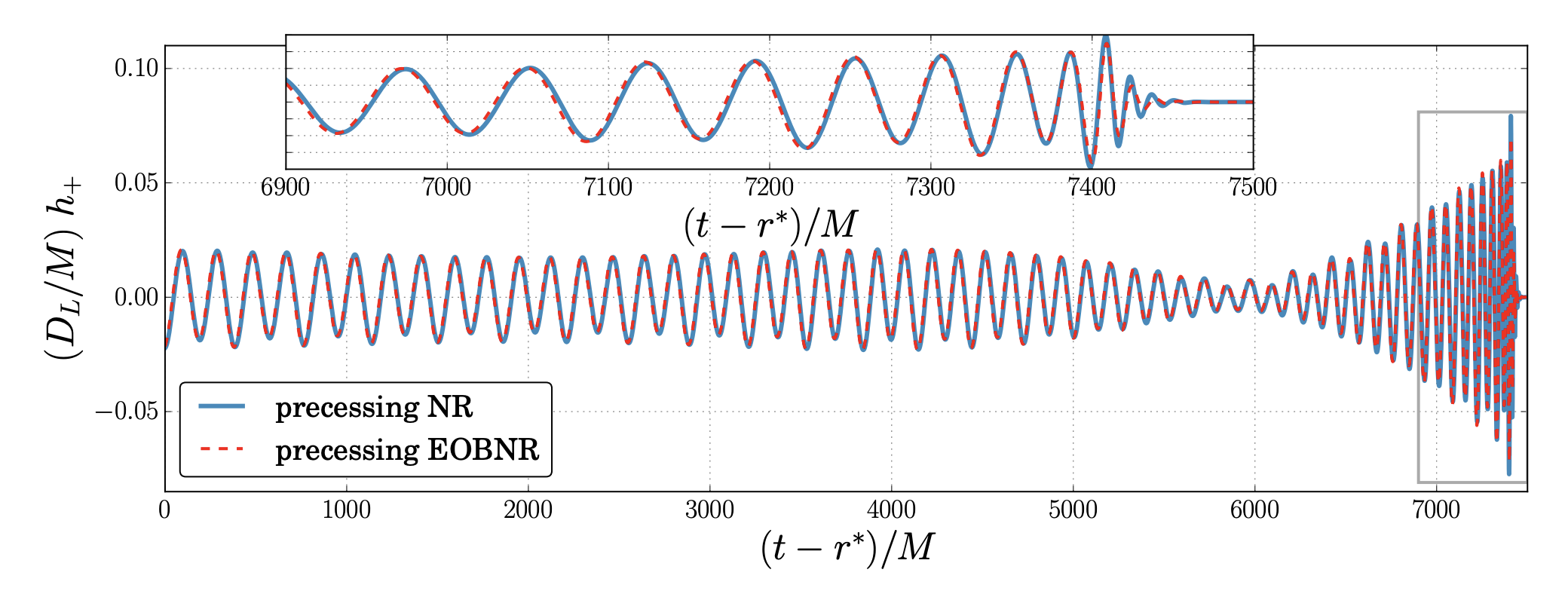}
    \caption{\footnotesize 
Two waveform approximants for
the signal $h_+$ (normalized by the luminosity distance $D_L$ and the system's total mass $M$)
produced by a black hole binary with mass ratio $1:5$, spin parameters
0.5 and 0 (for the heavier and lighter object respectively, and with the non-vanishing spin 
initially in the
orbital plane) as function of retarded time.
The observer is located 60 degrees away from the orbital angular momentum axis. Taken from Ref.~\cite{eobprec}
}
    \label{precession}
\end{figure}

Let us now utilize what we have learned so far to gain some qualitative understanding of the evolution of a quasi-circular binary of compact objects (black holes or neutron stars). As have seen, gravitational waves are emitted and carry energy and angular momentum away from  the system. As a result, the separation of the binary decreases and the orbital frequency increases. The rates of change of these quantities, as we have seen, only depend on a combination of the two masses, the chirp mass, at leading PN order. However, at higher PN orders  the gravitational wave fluxes  also depend on the individual masses and spins. 
PN corrections  also appear  
in the conservative sector, changing e.g.  the relation between the orbital frequency and the system parameters (which is only given by Kepler's law at Newtonian order, cf. sections~\ref{sec:postNewtonian} and~\ref{geo}), giving rise to precession (if at least one spin is non-zero and misaligned with the orbital angular momentum, cf. section~\ref{geo}), etc.
Precession (spin-spin and spin-orbit) introduces modulations in the gravitational waveforms (both in amplitude and phase), as shown for instance in Fig.~\ref{precession}. This
 makes measurements of the spin directions possible (at least in principle) with gravitational wave detectors.

As the binary's separation shrinks, the system  transitions from one circular orbit to the next until it either reaches the ISCO or the two bodies touch. We have seen that an ISCO exists in the test-particle limit (i.e. for geodesics), but a similar transition to unstable circular orbits occurs also for comparable-mass binaries of black holes~\cite{LeTiec:2011dp} (neutron stars touch and interact before they reach this effective ISCO). When the separation reaches the effective ISCO, 
or when the two bodies touch (in the case of neutron stars), the binary plunges and merges. The merger phase can only be studied via numerical-relativity simulations, but at least for  black holes the post-merger phase can be understood analytically in terms of quasi-normal modes, as we will see in the next section. As for neutron stars, the post-merger phase depends critically on the microphysics (e.g. on the equation of state of nuclear matter) and can only be predicted via numerical simulations.

\begin{figure} 
\centering
    \hskip0.9cm \begin{subfigure}[t]{0.8\textwidth}
       \includegraphics[width=\textwidth]{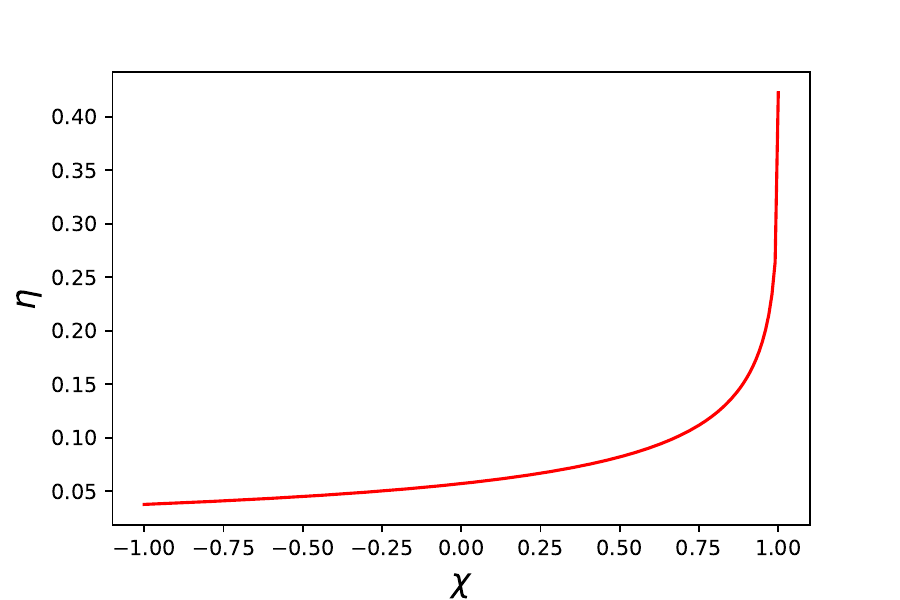}
    \end{subfigure}
    \vskip 0.5cm
    \begin{subfigure}[b]{0.85\textwidth}
        \includegraphics[width=\textwidth]{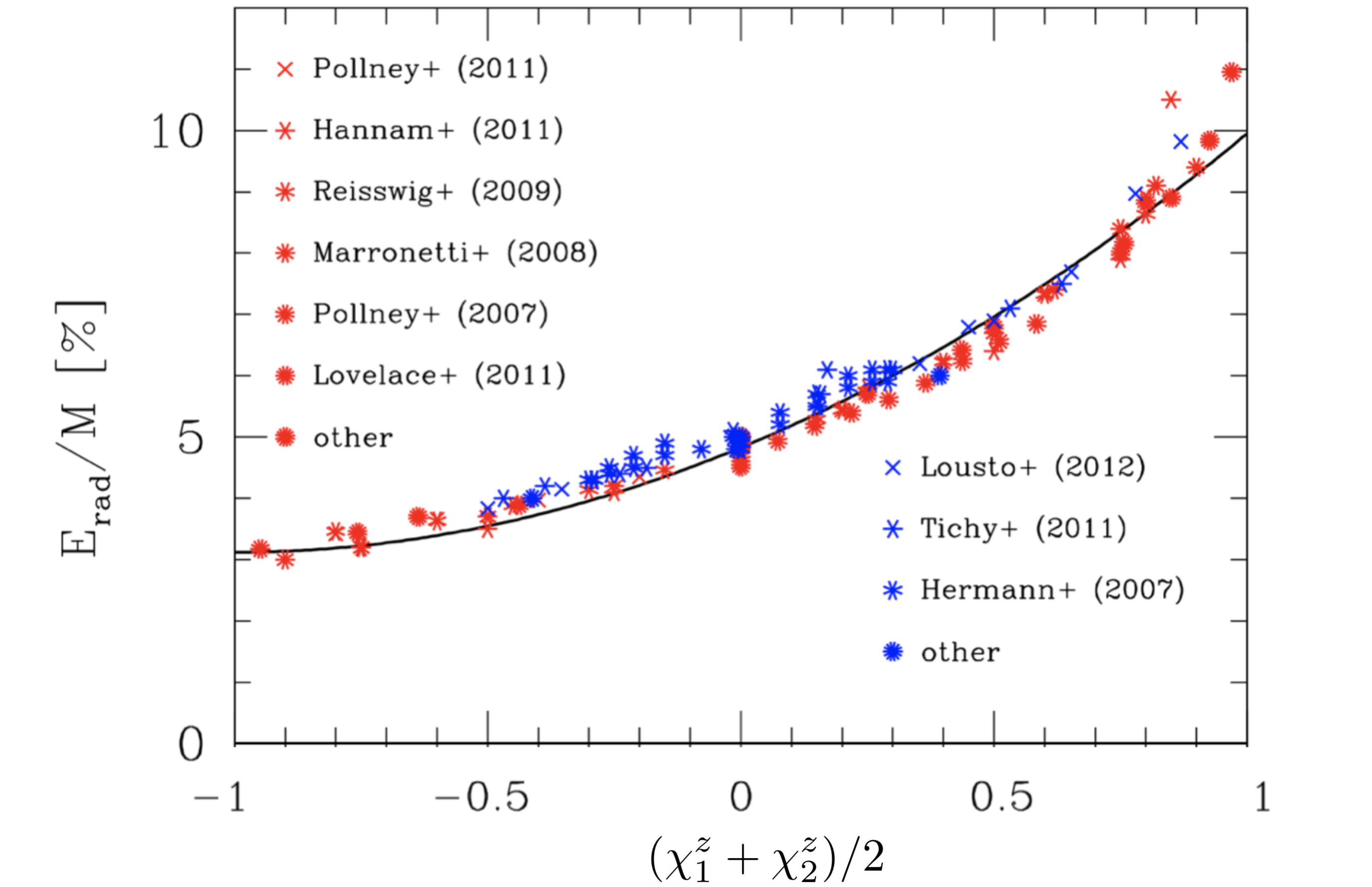}
     \end{subfigure}
    \caption{\footnotesize Top: The radiative efficiency of a geometrically thin, optically thick accretion disk around a Kerr black hole with spin parameter $\chi$. Note that the maximum spin expected for  black holes surrounded by such accretion disks is $\chi=0.998$~\cite{thorne}, for which the efficiency is $\approx 32\%$. Bottom: the fraction of the total mass $M=m_1+m_2$ emitted by a black hole binary system in gravitational waves, as a function of the average of the projections of the two spins on the orbital angular momentum axis, $(\chi_1+\chi_2)/2$. (Adapted from Ref.~\cite{morozova}.)}
    \label{eta}
\end{figure}

\begin{figure}
\centering
\includegraphics[width=1\textwidth]{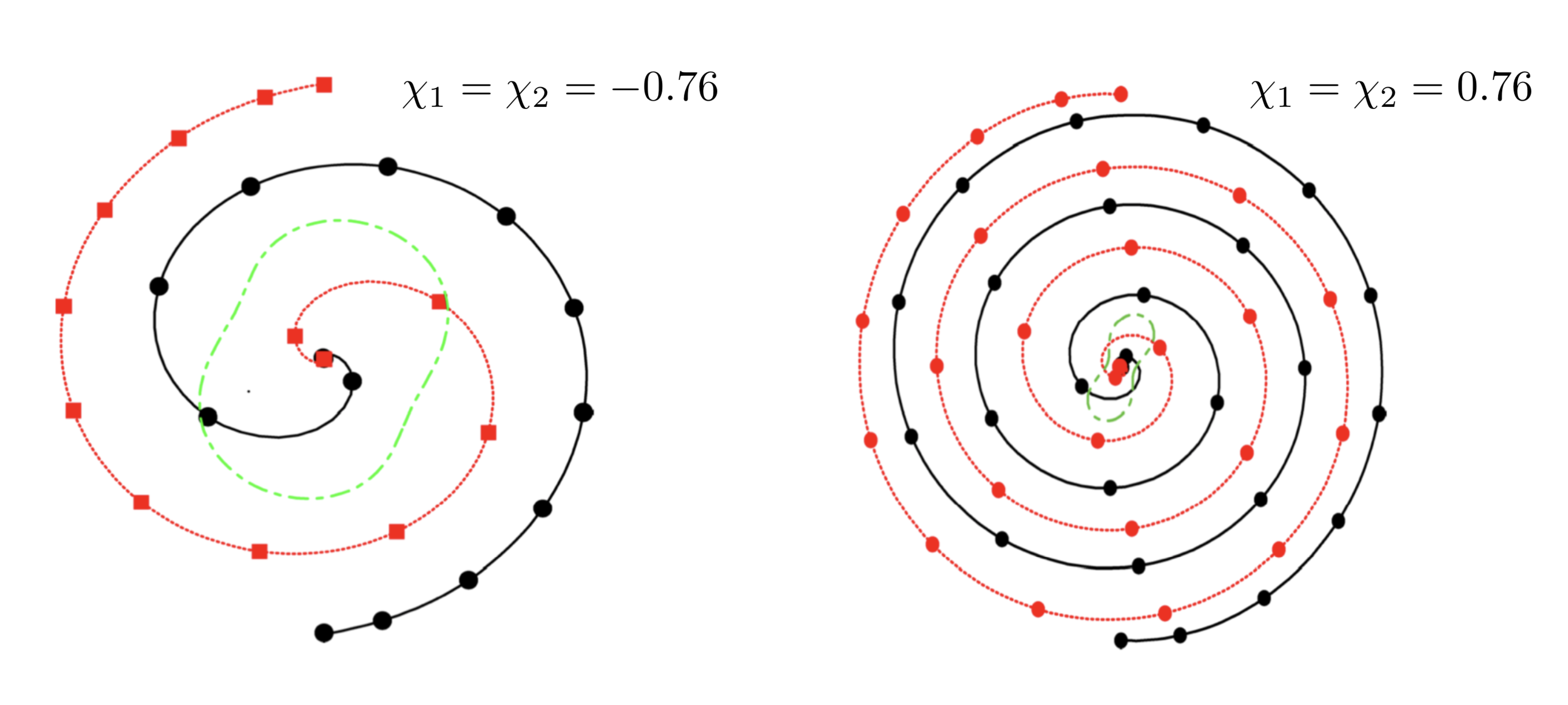}
\caption{\footnotesize Numerical relativity simulations of binary black holes with spins $|\chi_1|=|\chi_2|=0.76$ either antialigned (left) or aligned (right) with the orbital angular momentum. The squares and circles represent the puncture positions every 10 $M$ (with $M=m_1+m_2$) of evolution, and the lines joining them can therefore be thought as the trajectories of the black holes. The dashed green circles represent the first common apparent horizon.}
\label{rit}
\end{figure}

The position of the effective ISCO, just like in the test-particle limit, depends critically on the spins of the two objects. The larger the spins, the more the effective ISCO moves inwards and the longer the binary emits gravitational waves before plunging. This is exactly the same effect that takes place, in the electromagnetic case, for geometrically thin, optically thick accretion disks. In the latter, the gas spirals in on quasi-circular orbits, as it loses energy because of friction with the neighboring gas elements. The gas potential and kinetic energy is thus converted into heat, and eventually radiated away (e.g. in the optical, infrared, UV and/or X-ray bands). This process can only continue, however, until the gas reaches the ISCO of the central black onto which it is accreting. For a gas element with unit mass starting at rest at infinity, the energy when it reaches the ISCO is $E_{_{\rm ISCO}}$ (as given by Eq.~\ref{eisco_eq}). By energy conservation, the radiative efficiency of an accretion disk is therefore $\eta=1-E_{_{\rm ISCO}}$, which is a strong function of spin, as can be seen from Fig.~\ref{eta} (top panel).

\begin{figure}
\centering
\includegraphics[trim=0.cm 7cm 0cm 10,width=1\textwidth]{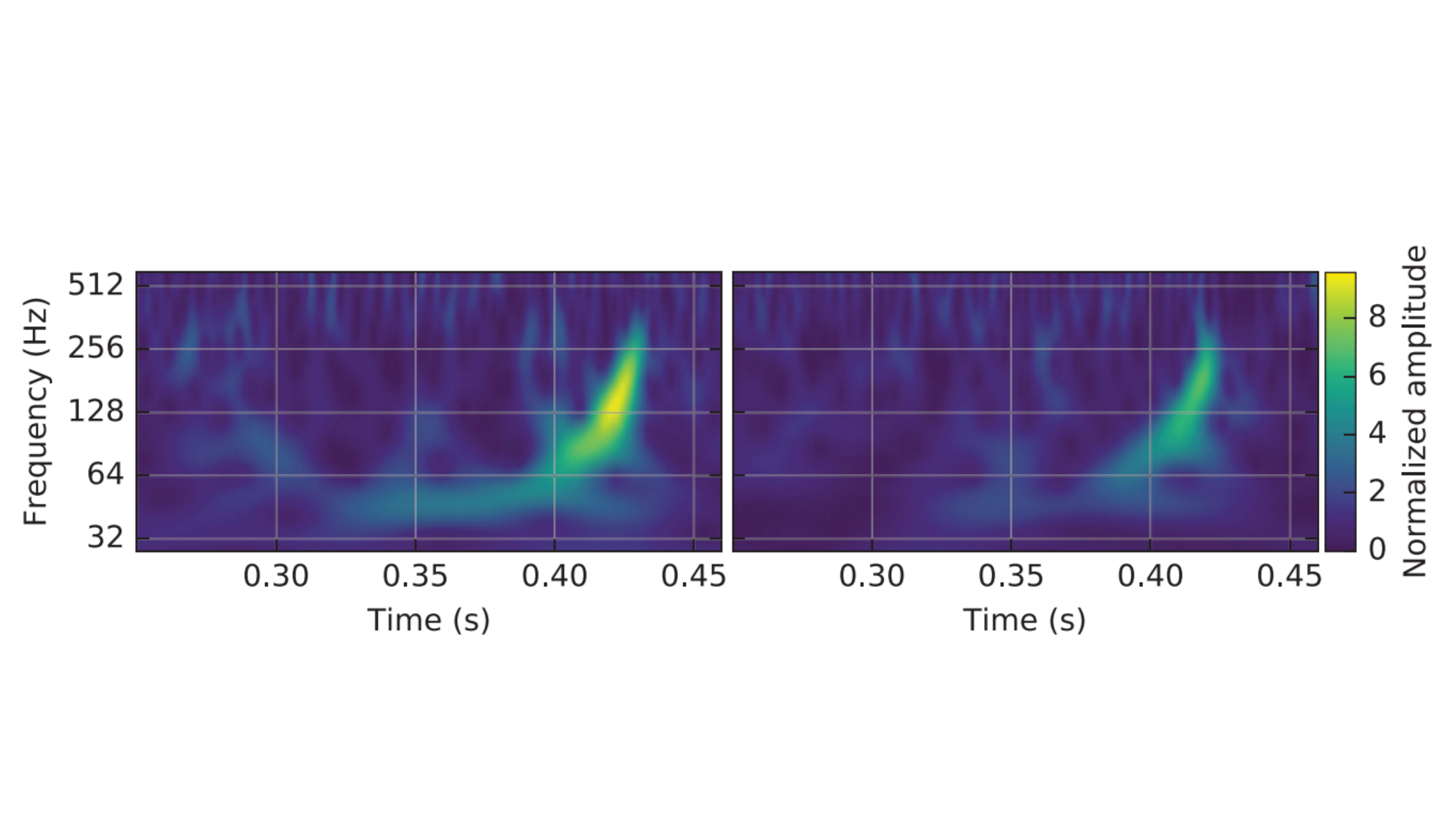}
\caption{\footnotesize A spectrogram of the GW150914 event  observed by  LIGO Hanford (left) and Livingston (right). The color code represents the normalized strain amplitude. Figure adapted from Ref.~\cite{GW150914}.}
\label{fig:LIGO}
\end{figure}

Computing a similar efficiency for gravitational waves from binary systems is not easy away from the test-particle limit, as the effective ISCO energy is not known analytically for comparable masses. However, one can perform numerical relativity simulations, which show in fact exactly the same effect (qualitatively). Fig.~\ref{rit}, taken from Ref.~\cite{campanelli}, shows  the ``trajectories'' (in some sense) of two black holes with spins respectively aligned (right) and antialigned (left) with the orbital angular momentum. It can be visually seen that the orbit with aligned spins (corresponding to prograde Kerr geodesics) reaches smaller separations and performs more cycles than that for antialigned spins (corresponding to retrograde   Kerr geodesics). This effect is known in the literature as ``orbital hang up'', but it is really a manifestation of the frame dragging of GR. The same effect can be seen at play in Fig.~\ref{eta} (bottom panel), adapted from Ref.~\cite{morozova}, which collected the energy emitted in gravitational waves in various black hole binaries simulated in the literature, and plotted it as function of a combination of the two spins (projected on the orbital angular momentum axis). Although the emission efficiency tops at about 10\% at high spins, thus remaining lower than the electromagnetic efficiency shown in the top panel of Fig.~\ref{eta}, the behavior is qualitatively the same as the latter.

As a final application, let us show that the knowledge that we have gained so far, albeit qualitative, allows for interpreting the data of GW150914~\cite{GW150914}, the first direct gravitational wave detection, and for concluding that the components of this binary system must be black holes. A spectrogram of this event is shown in Fig.~\ref{fig:LIGO}: the color code represents the
gravitational wave strain amplitude in a given bin of time (x-axis) and frequency (y-axis). One can clearly recognize a ``chirp'', i.e. an increase of the gravitational wave frequency with time, which one can fit with Eq.~\ref{fdot} to obtain $M_c\approx 30 M_\odot$. This in turn implies, through Eq.~\ref{Mc}, that the total mass must be $M\gtrsim 70 M_\odot$. The power's peak, which one expects to coincide with the plunge/merger phase, lies at $f\sim 150$ Hz. Translating that into an orbital frequency (dividing by a factor 2), and using Kepler's law to convert to a separation (assuming for simplicity roughly equal masses), one finds that the plunge/merger takes place when the binary separation decreases to just 350 km. This is very close to the sum of the Schwarzschild radii of the two objects, $GM/c^2\gtrsim 210$ km. Therefore, the separation at which the plunge happens is comparable to the effective ISCO radius, which seems to favor the hypothesis that the two objects are black holes. In fact, if the objects were stars, they would touch, interact and plunge way before reaching the effective ISCO separation, i.e. the two objects must be very compact. Among compact objects -- i.e. ones with $Gm/(Rc^2)={\cal O}(1)$, with $m$ and $R$ the object's mass and radius -- the only options (in GR) are black holes and neutron stars. Neutron stars, however, are excluded because they cannot be more massive than $2-3 M_\odot$. This leaves black holes as the only possibility.

\section{The post-merger signal}
\begin{figure}
\centering
\includegraphics[width=1\textwidth]{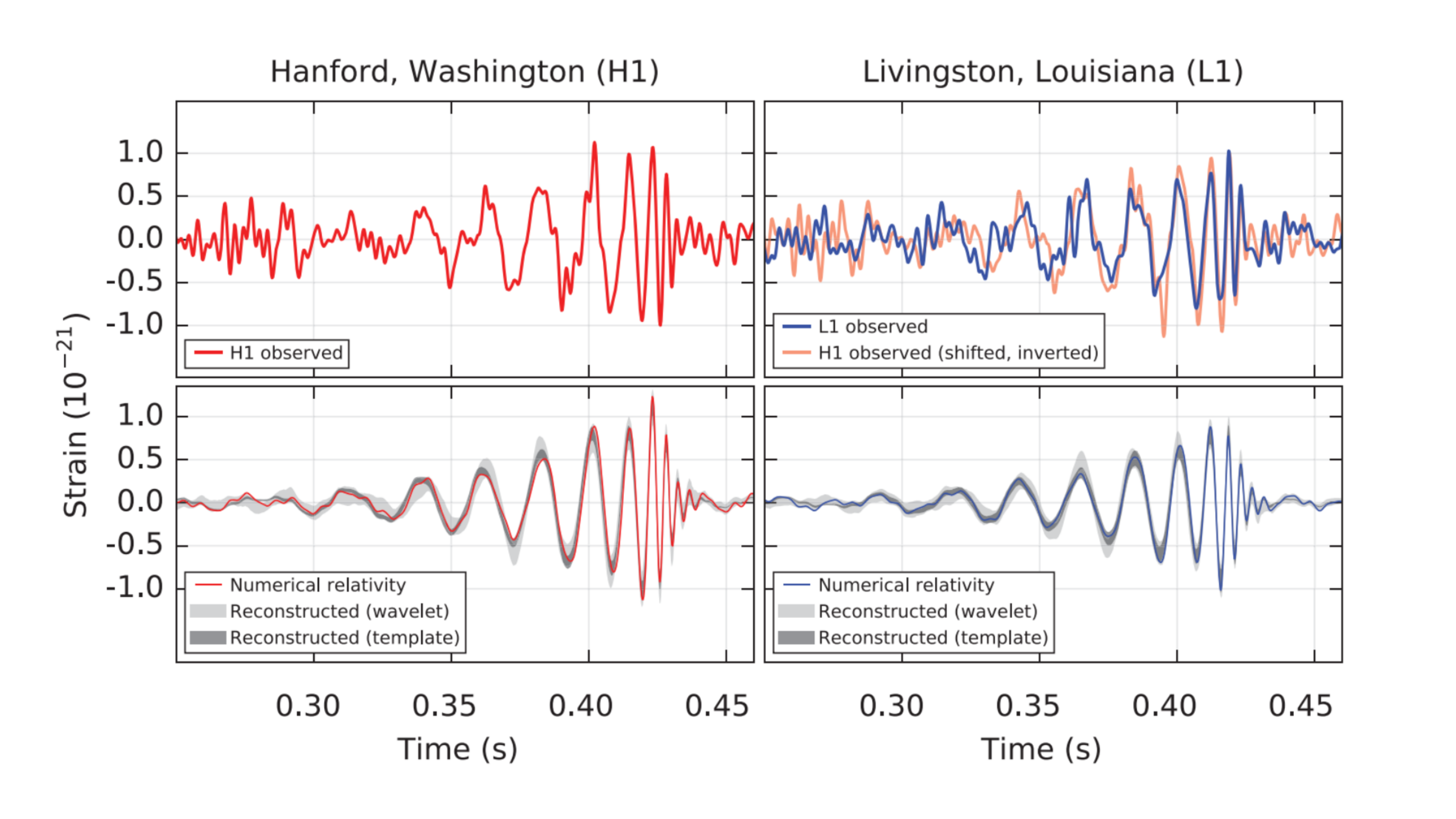}
\caption{\footnotesize Top: the GW150914 data, as observed by the Hanford and Livingstone LIGO detectors. Bottom: the de-noised signal, reconstructed with template and wavelet techniques, alongside the prediction from numerical relativity simulations. Figure adapted from Ref.~\cite{GW150914}}
\label{strain}
\end{figure}

As can be seen even in real strain data (c.f. Fig.~\ref{strain} for the GW150914 event), after the amplitude peaks at the merger, the gravitational wave signal from a black hole binary seems to be well described by one (or more) damped sinusoids. This is in fact what happens, as can be understood rather easily by employing linear perturbation theory on a Schwarzschild or Kerr background, as we will do in this section. We will start with the simple toy problem of a test Klein-Gordon field on a Schwarzschild background, and we will then move to the case of gravitational perturbations on the same geometry. We will finally generalize the treatment to the Kerr case, which will allow for concluding that indeed the post-merger signal is well described by a linear superposition of {\it quasi-normal} modes of the final (spinning) black hole resulting from the merger. For more details,
we refer the reader to the extensive review~\cite{Berti:2009kk}.

\subsection{Scalar perturbations of non-spinning black holes}

Let us start by considering the toy problem of a free scalar field on a Schwarzschild background, i.e. let us consider the Klein-Gordon equation
\begin{equation}g^{\mu \nu} \nabla_{\mu} \nabla_{\nu} \varphi=0,\label{KG}\end{equation}
with
$g^{\mu \nu}$ the contravariant components of the Schwarzschild metric given by Eq.~\ref{schw}. Since the metric is static and spherically symmetric, it is natural to decompose the scalar field in spherical harmonics ($Y_{\ell m}$) and Fourier modes ($e^{-i \omega t}$) as
\begin{equation}\varphi=\sum_{\ell, m} \frac{R_{\ell m}(r) }{r} Y_{\ell m} (\theta, \phi) e^{-i \omega t},\label{decomposition}\end{equation}
where $R_{\ell m}$ characterizes the radial profile of the scalar field. 
By replacing this ansatz in the Klein-Gordon equation, that reduces 
to a single ordinary differential equation in the radial coordinate,
\begin{equation}
\frac{\mathrm{d}^{2} R_{\ell m}}{\mathrm{d} r_{*}^{2}}+\left(\omega^{2}-V_{\ell}\right) R_{\ell m}=0,\label{1d}\end{equation}
where \begin{equation}r_{*} \equiv r+2 M\ln \left(\frac{r}{2 M}-1\right),\end{equation}
is the tortoise coordinate and 
the potential  $V_\ell$ the potential
is given by
\begin{equation}\label{eq:pot}
V_{\ell}(r) \equiv\left(1-\frac{2 M}{r}\right)\left[\frac{\ell(\ell+1)}{r^{2}}+\frac{2M}{r^3}\right].\end{equation}
As can be seen, in the geometric optics (eikonal) limit $\ell\gg 1$ this potential reduces to the effective potential for the radial motion of photons, Eq.~\ref{Vph}, if one identifies the 
impact parameter $b=L/E$
of photons  with $\ell$. This is  expected, since in the eikonal limit the wavefronts of a scalar field satisfying the Klein-Gordon equation move along null geodesics of the metric.
We also stress that going from the partial differential equation \ref{KG} to the single ordinary differential equation \ref{1d} is highly non-trivial, and depends critically on the choice of the ansatz of Eq.~\ref{decomposition}.

To solve Eq.~\ref{1d}, we need to impose suitable boundary conditions. As $r_*\to\infty$, in order for nothing to enter the system, we need to impose outgoing boundary conditions $R_{\ell m}\sim  \exp{(i\omega r_*)}$. Conversely, since nothing can escape the horizon (which corresponds to
$r_*\to-\infty$), we need to impose
$R_{\ell m}\sim  \exp{(-i\omega r_*)}$ there (ingoing boundary conditions).
Solving this boundary-value problem, one obtains a discrete spectrum of {\it complex} frequencies $\omega$. The corresponding excitations are referred to as {\it quasinormal} modes (as opposed to normal modes, which have real frequencies). Moreover, one can check that  all frequencies in the spectrum have negative imaginary part, which corresponds to damped modes (c.f. Eq.~\ref{decomposition}). This shows that a test scalar field is (linearly) stable on the Schwarzschild geometry.
\\
\\
{\bf Exercise 4: }  {\it Plot the effective potential of Eq.~\ref{eq:pot} and approximate it qualitatively with a rectangular potential. Solve Eq.~\ref{1d} in the three regions in which this rectangular potential is constant, and impose appropriate junction conditions at the transition radii and ingoing/outgoing boundary conditions at the horizon and at infinity. By counting the integration constants, show that the spectrum is discrete and complex. Solve numerically for a few frequencies in the spectrum.}

\subsection{Tensor perturbations of non-spinning black holes}

A similar analysis can be performed for the metric perturbation $h_{\mu\nu}$ of a Schwarschild spacetime. To exploit again the fact that the Schwarzschild geometry is static and spherically symmetric, we can decompose the time dependence in Fourier modes and the angular dependence in scalar, vector and tensor harmonics. In more detail, $h_{tt}$
$h_{tr}$, $h_{rr}$ are scalars on the two-sphere (i.e. under rotations), and can thus be expanded in the usual (scalar) spherical harmonics $Y_{\ell m}$. The cross terms $h_{tA}$ and  $h_{rA}$ (with capital Latin letters
spanning the two angles $\theta,\phi$)
are instead vectors on the two sphere. A basis for vectors on the two-sphere can be obtained by taking gradients of the scalar harmonics, 
\begin{equation}
 Y^{{\cal E},\ell m}_A=\partial_A Y_{\ell m} \,,  
\end{equation}
or exterior derivatives of these gradients,
\begin{equation}
Y^{{\cal B},\ell m}_A=\epsilon_A^{\phantom{B}B} \partial_B Y_{\ell m}\,,
\end{equation}
where $\epsilon_{AB}$ is the Levi-Civita tensor on the two-sphere and the angular indices are raised and lowered with the metric of the two-sphere, $\gamma_{AB}$:
\begin{gather}
    \gamma_{AB} \mathrm{d} X^A\mathrm{d} X^B= \mathrm{d} \theta^2+ \sin^2\theta\, \mathrm{d} \phi^2\,,\\
    \epsilon_{AB}=\sqrt{\gamma}\, e_{AB}\,,
\end{gather}
with $\gamma={\rm det}(\gamma_{AB})=\sin\theta$ and $e_{\theta\phi}=-e_{\phi\theta}=1$, $e_{\theta\theta}=e_{\phi\phi}=0$.
Because scalar harmonics have parity $(-1)^\ell$, $Y^{{\cal E}}_A$ and $Y^{{\cal B}}_A$ have respectively parity $(-1)^\ell$ and $(-1)^{\ell+1}$.
Similarly, the components $h_{AB}$ transform as a tensor on the two-sphere, and can thus be decomposed in the following basis
\begin{equation}
    \begin{aligned} 
    &Y^{{\cal E},\ell m}_{AB}=Y^{\ell m}_{;A ;B}-\frac12 \gamma_{AB}Y^{;C}_{\ell m;C}  , \\
   &Y^{{\cal T},\ell m}_{AB}=Y_{\ell m} \gamma_{AB}\,, \\ 
    &Y^{{\cal B},\ell m}_{AB}=
    Y^{{\cal B},\ell m}_{(A;B)}
        =\frac12\left(\epsilon_{A}^{\phantom{B}C} Y^{{\cal E},\ell m}_{CB}+\epsilon_{B}^{\phantom{B}C} Y^{{\cal E},\ell m}_{CA}\right)  \,,\end{aligned}
\end{equation}
where the semicolon denotes covariant derivatives on the two-sphere. Because  of the parity of the scalar harmonics, $Y^{{\cal E}}_{AB}$ and $Y^{{\cal T}}_{AB}$ have parity $(-1)^\ell$, while 
$Y^{{\cal B}}_{AB}$ has parity $(-1)^{\ell+1}$. Note also that 
by using the explicit expressions for the scalar spherical harmonics, the vector harmonics $Y^{{\cal E},\ell m}_{A}$ and
$Y^{{\cal B},\ell m}_{A}$ start at $\ell=1$ (i.e. $Y^{{\cal E},00}_{A}=Y^{{\cal B},00}_{A}=0$), while $Y^{{\cal E},\ell m}_{AB}$ and $Y^{{\cal B},\ell m}_{AB}$ start at $\ell=2$ (i.e.
$Y^{{\cal E},\ell m}_{AB}=Y^{{\cal B},\ell m}_{AB}=0$ for $\ell\leq1$).

Expanding the metric perturbation in these scalar, vector and tensor harmonics one can then set
\begin{gather}
    h_{tt}=\sum_{\ell m} \left(1-\frac{2M}{r}\right) H_0(r) Y_{\ell m} (\theta, \phi) e^{-i \omega t}\\
    h_{tr}=\sum_{\ell m} H_1(r) Y_{\ell m} (\theta, \phi) e^{-i \omega t}\\
    h_{rr}=\sum_{\ell m} \left(1-\frac{2M}{r}\right)^{-1} H_2(r) Y_{\ell m} (\theta, \phi) e^{-i \omega t}\\
    h_{tA}= \sum_{\ell m} [-h_0(r) Y^{{\cal B},\ell m}_A (\theta, \phi)+{\cal H}_0(r) Y^{{\cal E},\ell m}_A (\theta, \phi)] e^{-i \omega t}\\
      h_{rA}= \sum_{\ell m} [-h_1(r) Y^{{\cal B},\ell m}_A (\theta, \phi)+{\cal H}_1(r) Y^{{\cal E},\ell m}_A (\theta, \phi)] e^{-i \omega t}\\
      h_{AB}=\sum_{\ell m} [r^2 K(r) Y^{{\cal T},\ell m}_{AB} (\theta, \phi) + r^2 G(r) Y^{{\cal E},\ell m}_{AB} (\theta, \phi)+ h_2(r) Y^{{\cal B},\ell m}_{AB} (\theta, \phi)] e^{-i \omega t}\,,
\end{gather}
where $H_0$, $H_1$, $H_2$, $h_0$, $h_1$, $\mathcal{H}_0$, $\mathcal{H}_1$, $K$, $G$, $h_2$ are free radial functions.

By similarly expanding the generator $\xi^\mu$ of gauge transformations in scalar ($\xi^t$ and $\xi^r$) and vector ($\xi^A$) harmonics, one can set $h_2={\cal H}_0={\cal H}_1=G=0$ (Regge-Wheeler gauge~\cite{RW}).
Moreover, without loss of generality one can set $m=0$ when studying perturbations of Schwarzschild. In fact, because of spherical symmetry the modes with $m\neq0$ can be set to zero by rotation $\phi\to\phi\,+\,$const. Note that is also true for the scalar field considered above (indeed $m$ does not appear in the potential of Eq.~\ref{eq:pot}).
In the odd sector the metric perturbation then becomes
\begin{eqnarray}
{h}^{\rm odd}_{\mu \nu}= \left(
 \begin{array}{cccc}
 0 & 0 &0 & h_0(r)
\\ 0 & 0 &0 & h_1(r)
\\ 0 & 0 &0 & 0
\\ h_0(r) & h_1(r) &0 &0
\end{array}\right)
\left(\sin\theta\frac{\partial}{\partial\theta}\right)
Y_{l0}(\theta)e^{-i \omega t}\,, \label{oddgauge}
\end{eqnarray}
whereas in the even sector one has
\begin{eqnarray}
{h}^{\rm even}_{\mu \nu}= \left(
 \begin{array}{cccc}
 H_0(r) \left(1-\frac{2M}{r}\right) & H_1(r) &0 & 0
\\ H_1(r) & H_2(r) \left(1-\frac{2M}{r}\right)^{-1}  &0 & 0
\\ 0 & 0 &r^2K(r) & 0
\\ 0 & 0 &0 & r^2K(r)\sin^2\theta
\end{array}\right) \, Y_{l0}(\theta)e^{-i \omega t}\,\, \label{evengauge}
\end{eqnarray}
with $x^\mu=(t,r,\theta,\phi)$. Because of their different parity, the even and odd parity perturbations decouple when this ansatz is replaced into the Einstein equations.
By using the latter, in the odd sector one obtains the famous Regge-Wheeler equation~\cite{RW}
\begin{equation}
     \label{waveeq}
\frac{d^2 \Psi}{dr_*^2} + \left(\omega^2-V\right)\Psi=0\,.
\end{equation}
with
\begin{equation}
V=V^{-}=  \left(1-\frac{2 M}{r}\right)
\left\lbrack\frac{\ell(\ell+1)}{r^2}-\frac{6M}{r^3}\right\rbrack\,
\label{vodd}
\end{equation}
and
\begin{equation}
\Psi=\Psi^-=  \frac{h_1(r)}{r} \left(1-\frac{2 M}{r}\right)\,,\qquad
h_0=\frac{i}{\omega}\frac{d}{dr_*}\left(r\Psi^-\right)\,.
\label{qodd}
\end{equation}
Similarly, the even sector is described by the same 
Eq.~\ref{waveeq}, but with the Zerilli~\cite{zerilli} potential
\begin{equation}
V=V^{+}= \frac{2}{r^3}\left(1-\frac{2M}{r}\right)
\frac{3\lambda^2Mr^2+\lambda^2\left(1+\lambda\right)r^3+9M^2\left(M+\lambda
r\right)} {\left(3M+\lambda
r\right)^2} \,,\label{veven}
\end{equation}
where $\lambda\equiv (\ell-1)(\ell+2)/2$, 
and the Zerilli variable $\Psi=\Psi^+$ is defined implicitly by
\begin{eqnarray}
K&=& \frac{6M^2+\lambda \left(1+\lambda\right)r^2+3M \lambda
r} {r^2\left(3M+\lambda r\right)}
\Psi^+
+\frac{d\Psi^+}{dr_*} \,, \\
H_1&=& \frac{i\omega\left(3M^2+3\lambda Mr-\lambda r^2\right)} {r \left(3M+\lambda r\right)
(1-2 M/r)}\Psi^+ -
\frac{i \omega r}{1-2 M/r}\frac{d\Psi^+}{dr_*}\,. \label{4.11}
\end{eqnarray}
Note that $H_0$ can also be obtained from the
algebraic relation
\begin{eqnarray}
&&\left [(\ell-1)(\ell+2)+\frac{6M}{r}\right ]H_0+\left[i\frac{\ell(\ell+1)}{\omega\,r^2}M-2i\omega\,r\right ]H_1\,\nonumber\\
&-&\left [(\ell-1)(\ell+2)+\frac{2M}{r}-\frac{2\omega^2r^2+2 M^2/r^2}{1-2M/r}\right ]K=0\,,
\end{eqnarray}
which also follows from the Einstein equations.

A few comments are in order at this stage. First, even though the Regge-Wheeler and Zerilli potentials are different, they give rise to the same spectrum of quasinormal mode frequencies when ingoing boundary conditions are imposed at the horizon and outgoing boundary conditions are imposed at infinity, i.e. the two potentials are isospectral. The resulting frequencies have negative imaginary part, which is consistent with the Schwarzschild geometry being linearly stable to gravitational perturbations. Moreover, both the   
Regge-Wheeler and Zerilli potentials coincide, in the eikonal limit $\ell\gg 1$, with the scalar potential of Eq.~\ref{eq:pot} and the photon potential of Eq.~\ref{Vph}. As discussed above, this is expected and simply amounts to the fact that gravitational wavefronts travel along null geodesics, but it is also of practical importance. In fact, in the eikonal limit
the peak of the quasinormal mode potential must coincide with the peak of the photon potential, which lies at the
location of the (unstable) circular photon orbit.
The real part of the quasinormal mode frequencies can then be shown to be 
simply proportional to the orbital frequency of the circular photon orbit, while the imaginary part turns out to be related 
to the Lyapunov exponent of null
geodesics near the circular photon orbit, which in turn
depends on the curvature of the photon effective  potential near its peak~\cite{eikonal1,eikonal2}.
Intuitively, this means that quasinormal modes can be thought of as being
generated at
the circular photon orbit, after which they slowly leak outwards
(because the circular photon orbit 
is unstable to radial perturbations).

\subsection{Tensor perturbations of spinning black holes}
The calculation of the quasinormal modes of spinning black holes
is considerably more complicated. In fact, already for a test scalar field in 
the  Kerr geometry it is not obvious at all that the Klein-Gordon equation can be reduced to a one-dimensional Schr\"odinger-like equation. This is definitely not the case if the scalar is decomposed in spherical harmonics (as in Eq.~\ref{decomposition}). Still, the existence of a Killing-Yano tensor for the Kerr geometry, which allows for separating the equations for geodesic motion (which also regulate the motion of scalar and tensor wavefronts in the eikonal limit), suggests that the equations for both scalar and tensor perturbations may similarly separate under an appropriate choice of basis on the two-sphere.

In fact, one can try to solve the Klein-Gordon equation on the Kerr metric in Boyer-Lindquist coordinates (Eq.~\ref{Kerr}) by separation of variables, i.e.
\begin{equation}
    \varphi=R(r)\Theta(\theta) e^{i m\phi} e^{-i \omega t}\,,
\end{equation}
which yields~\cite{brill}
\begin{align}
& \Delta \frac{\partial}{\partial r}\left(\Delta \frac{\partial R}{\partial r}\right)+\left[a^{2} m^{2}-4 M r a m \omega+\left(r^{2}+a^{2}\right)^{2} \omega^{2}\right] R=\left(Q+m^{2}+\omega^{2} a^{2}\right) \Delta R, \\\label{ang_eq}
& \frac{1}{\sin \theta} \frac{\partial}{\partial \theta}\left(\sin \theta \frac{\partial \Theta}{\partial \theta}\right)+\left(a^{2}\omega^{2}\cos ^{2} \theta-\frac{m^{2}}{\sin ^{2} \theta}\right) \Theta=-\left(Q+m^{2}\right) \Theta \,,
\end{align}
where $Q$ is a separation constant (defined to reduce to the Carter constant in the eikonal limit).
As can be seen, for $a=0$ the second equation reduces to the equation defining associated Legendre polynomials, i.e. for $a=0$, $\Theta(\theta) e^{i m\phi}$ reduces to spherical harmonics. For $a\neq0$, Eq.~\ref{ang_eq} defines instead the scalar spheroidal harmonics.

A similar calculation is possible, albeit much more involved, for metric perturbations in Kerr. We will provide no proof here, but just state the result. Let us first introduce the
Newman-Penrose scalars
\begin{gather}
{\bf \Psi}_0=-C_{\mu\nu\lambda\sigma}l^\mu m^\nu l^\lambda m^\sigma\,,\\
{\bf \Psi}_4 = -C_{\mu\nu\lambda\sigma}n^\mu m^{*\nu} n^\lambda m^{*\sigma}\,,
\end{gather}
with $C_{\mu\nu\lambda\sigma}$  the Weyl curvature tensor, and $l\,,n\,,m\,,m^*$  a (complex) null tetrad defined at each spacetime point.
Note that ${\bf \Psi}_0$ and ${\bf \Psi}_4$ can be thought of as describing ingoing and outgoing gravitational wave signals. If one  defines the tensor spheroidal harmonics ${}_sS_{lm}$~\cite{teuk}
\begin{equation}
      {\biggl[}\frac{\partial}{\partial u} (1-u^2)\frac{\partial}{\partial u}{\biggr]}\,{}_sS_{lm}
+{\biggl[}a^2\omega^2u^2-2a\omega s u +s+\, _sA_{lm}
-\frac{(m+s u)^2}{1-u^2}  {\biggr]}\,{}_sS_{lm}=0\,,\label{angular}
\end{equation}
with $s=\pm 2$ and $u= \cos\theta$, one can then decompose
\begin{equation}\psi(t\,,r\,,\theta\,,\phi) =\frac{1}{2\pi}\int e^{-i\omega t}
\sum_{l=|s|}^{\infty}\sum_{m=-l}^{l}
e^{im\phi}\,{}_sS_{lm}(\theta)R_{lm}(r)d\omega\,,
\label{psiteuk}\end{equation}
where $\psi$ stands for either
${\bf \Psi}_0$ (in which case $s=2$) or $\rho^{-4}{\bf \Psi}_4$ [with $\rho \equiv -1/(r-ia \cos\theta)$ and $s=-2$]. 
Here, $A_{\ell  m}$ is a separation constant, which for $a=0$ can be computed analytically to be $A_{\ell  m}(a=0)=\ell (\ell +1)-s(s+1)$.
This ansatz allows for solving the linearized Einstein equations by separation of variables, and lead to the master radial equation~\cite{teuk}
\begin{equation}
    \Delta \partial^2_r
R_{lm}+2(s+1)(r-M)\partial_rR_{lm}+VR_{lm}=0\,. \label{radial}
\end{equation}
with
\begin{equation}
    V=2is\omega r -a^2\omega^2-\, _sA_{lm}+\frac{1}{\Delta} {\biggl[}
(r^2+a^2)^2\omega^2- 4M am\omega r +a^2m^2+2is\left(am(r-M)-M\omega(r^2-a^2)        \right ) {\biggr]} \,.
\end{equation}
 Again, this equation can be solved as a boundary value problem with  ingoing boundary conditions at the event horizon and outgoing boundary conditions at infinity. This yields a discrete spectrum of complex quasinormal mode frequencies, whose imaginary part is again negative for $|a|\neq M$, pointing at linear stability. In the extreme limit $a\to \pm M$, the imaginary part goes to zero for some modes, which suggests that extreme Kerr is only marginally stable.

As can be seen, the master equation defining the quasinormal mode spectrum only depends on the mass and spin of the black hole, $M$ and $a$. This is a manifestation of the no-hair theorem of GR, and can be used to perform consistency tests of the latter~\cite{detweiler}. Indeed, if two modes are observed, one can use the real and imaginary part of the first to obtain $M$ and $a$, and then  use them to predict the real and imaginary part of the second mode. These predictions can then be compared to the measurements.

\section{The detection of gravitational waves}\label{sec:det}

\begin{figure} 
\centering
    \begin{subfigure}[b]{0.48\textwidth}
        \includegraphics[width=\textwidth]{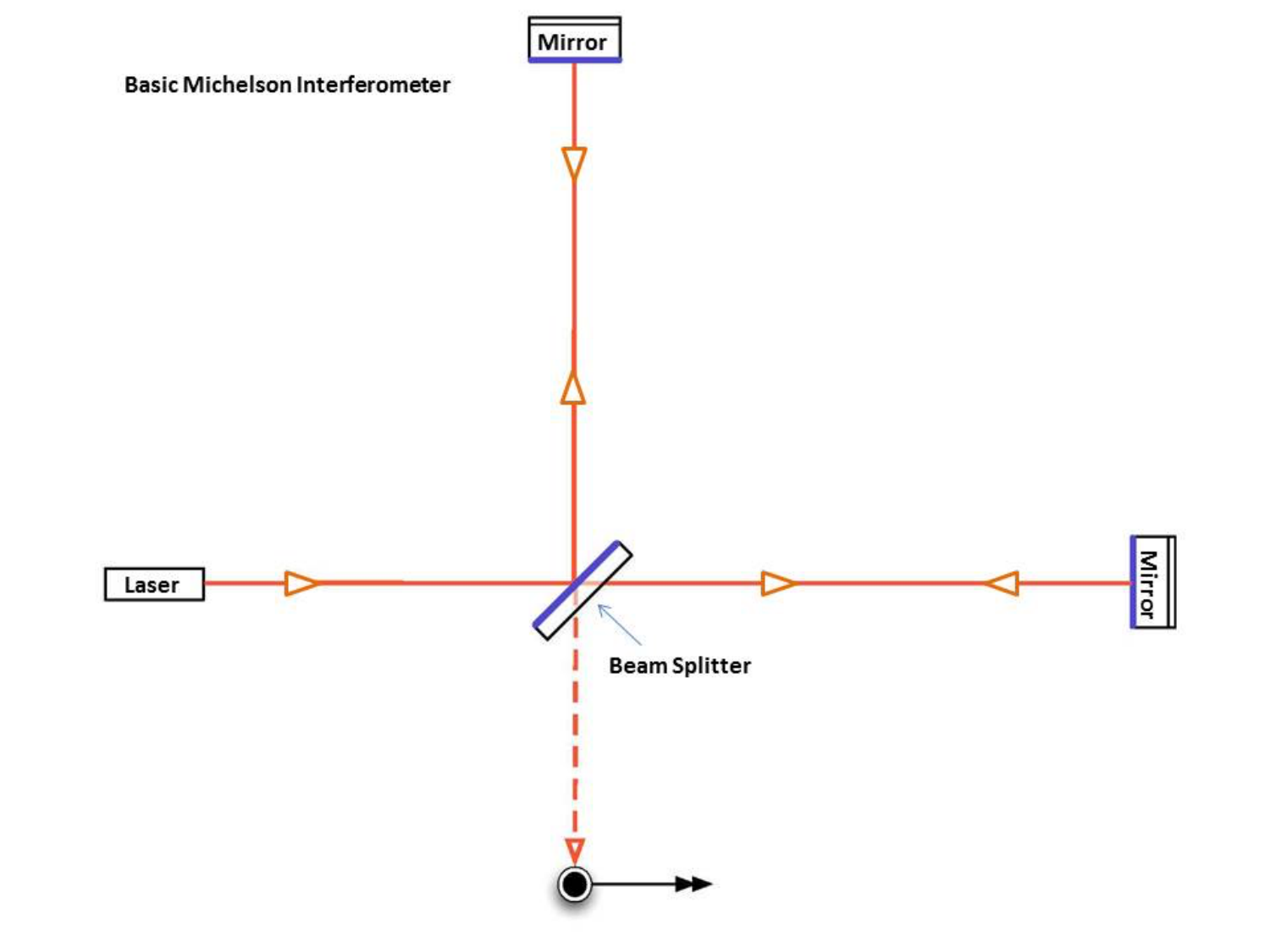}{}
    \end{subfigure}
    \begin{subfigure}[b]{0.5\textwidth}
        \includegraphics[width=\textwidth]{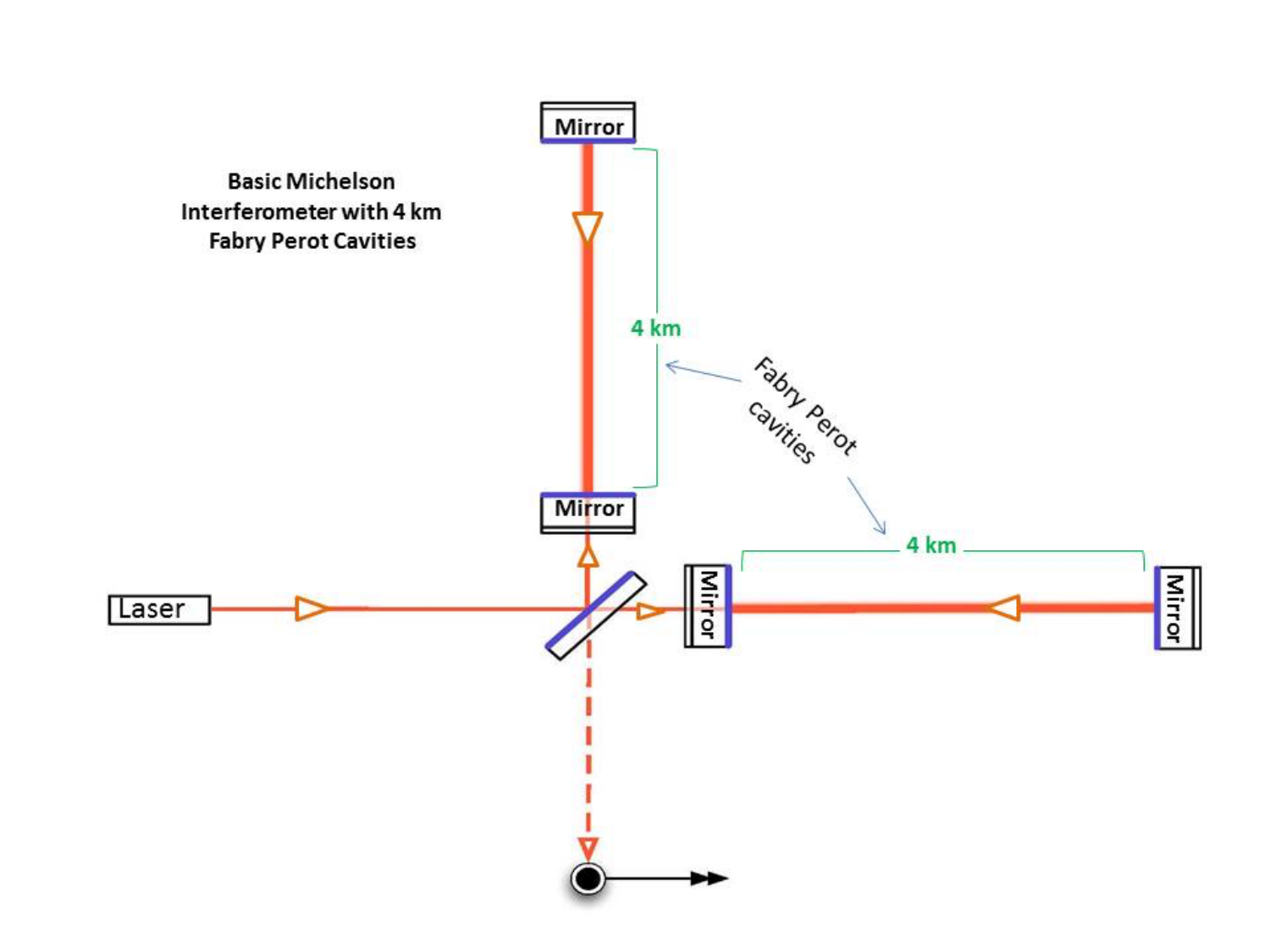}
    \end{subfigure}
    \caption{\footnotesize Schematic representation of a Michelson  (left) and a Fabry-Perot  (right) interferometer (courtesy Caltech/MIT/LIGO Laboratory). In both, the 45-degrees slab represents a beam splitter (which reflects 50\% of the photons and lets 50\% through) and the photodetector is represented by a dot. The two mirrors close to the beam splitter in the Fabry-Perot design can be thought of as semi-transparent, e.g. a photon will go back and forth many times along each arm before reaching the photodetector. This ``Fabry-Perot cavity'' thus increases the effective length of the arms (by a factor $\sim 300$ for LIGO). Note that e.g. LISA will instead be a Michelson interferometer.}
    \label{fig:interf}
\end{figure}

In this section, we will review the principles behind the experimental detection of gravitational waves. The effort to observe these signals  started in the 60s with Weber~\cite{weber}, who pioneered the use of resonant bars. The idea behind this setup is that gravitational waves change the size of
the bar (or any object for that matter)
in a periodic fashion (with frequency given by the wave's frequency $f$). If the latter coincides with one of the characteristic frequencies of the bar, the system can resonate and the detector's response is therefore amplified. A similar idea is behind the more modern interferometric detectors, in which lasers travel back and forth between mirrors in two or more arms, before being collected at the same point (a photodetector; c.f. schematic setup in Fig.~\ref{fig:interf}). Because gravitational waves have a non-trivial angular dependence, the lengths of the two arms  change in different fashions. This time-varying length difference gives rise to  a phase difference between the lasers, which can be measured by observing  their interference pattern at the photodetector. Interferometric detectors include for instance the ground-based LIGO, VIRGO and KAGRA experiments; the next generation ground-based detectors Cosmic Explorer and Einstein Telescope; and the future space-borne interferometer LISA. In the following, we will derive in detail the response of an interferometer to a gravitational signal.

\subsection{The response of a gravitational wave detector: the low frequency limit}\label{subsec:7_1}

{}

Let us  consider a laser traveling back and forth between two mirrors in free fall.~\footnote{For ground detectors such as LIGO and Virgo, the mirrors cannot be thought of as strictly speaking in free fall since the experiment takes place on Earth. However, the mirrors are isolated from the Earth's motion and vibrations by a sophisticated suspension system, and can thus be thought of as effectively in free fall, at least at high frequencies.} More precisely, let us 
assume that the mirrors move 
 along geodesics in a flat spacetime perturbed by a gravitational wave $h^\mathrm{TT}_{\mu\nu}$:
\begin{equation}\label{eq:tt_gauge}
\begin{aligned}
   & g_{\mu \nu}= \eta_{\mu \nu} + h ^{\mathrm{TT}}_{\mu\nu},\\
    &\frac{\mathrm{d}^2x^{\mu}}{\mathrm{d}\tau^2} + \Gamma^{\mu}_{\alpha \beta} \frac{\mathrm{d}x^{\alpha}}{\mathrm{d}\tau} \frac{\mathrm{d}x^{\beta}}{\mathrm{d}\tau}=0\,.
\end{aligned}
\end{equation}
Changing variable from the proper time $\tau$ to the coordinate time $t$,  one can compute the mirror's coordinate acceleration as
\begin{equation*}
    \frac{\mathrm{d}^2x^{\mu}}{\mathrm{d}t^2}= \frac{\mathrm{d}}{\mathrm{d}t}\Bigg( \frac{\mathrm{d}x^{\mu}}{\mathrm{d}\tau}  \frac{\mathrm{d}\tau}{\mathrm{d}t}  \Bigg) 
    = \frac{\mathrm{d}}{\mathrm{d}\tau}\Bigg[\frac{\mathrm{d}x^{\mu}}{\mathrm{d}\tau}  \left(  
    \frac{\mathrm{d}t}{\mathrm{d}\tau}
    \right)^{-1}  
    \Bigg]
    \frac{\mathrm{d}\tau}{\mathrm{d}t}
    =
    \Bigg[ \frac{\mathrm{d}^2x^{\mu}}{\mathrm{d}\tau^2}\left(  
    \frac{\mathrm{d}t}{\mathrm{d}\tau}
    \right)^{-1}  - \left(  
    \frac{\mathrm{d}t}{\mathrm{d}\tau}
    \right)^{-2}  \frac{\mathrm{d}x^\mu}{\mathrm{d}\tau} \frac{\mathrm{d}^2t}{\mathrm{d}\tau^2}\Bigg]\frac{\mathrm{d}\tau}{\mathrm{d}t}.
\end{equation*}
Expressing $ \frac{\mathrm{d}^2x^{\mu}}{\mathrm{d}\tau^2}$ and $\frac{ \mathrm{d}^2t}{\mathrm{d}\tau^2}$ by using the geodesic equations, one obtains
\begin{equation*}
     \frac{\mathrm{d}^2x^{\mu}}{\mathrm{d}t^2}= - {\Gamma^{\mu}_{\alpha \beta} \dot{x}^{\alpha}\dot{x}^{\beta}} +  \dot{x}^{\mu} \Gamma^0_{\alpha \beta} \dot{x}^{\alpha}\dot{x}^{\beta}.
\end{equation*}
where the dot denotes $\mathrm{d}/\mathrm{d}t$.
Since $\dot{x}^t=1$, the spatial acceleration is
\begin{equation}\frac{\mathrm{d}^{2} x^{i}}{\mathrm{d} t^{2}}=-\left(\Gamma_{t t}^{i}+2 \Gamma_{t j}^{i} v^{j}+\Gamma_{j k}^{i} v^{j} v^{k}\right)+v^{i}\left(\Gamma_{t t}^{t}+2 \Gamma_{t j}^{t} v^{j}+\Gamma_{j k}^{t} v^{j} v^{k}\right),\end{equation}
where $v^i\equiv \dot{x}^i$. Assuming that the mirror moves at $v\ll 1$, one finally obtains
\begin{equation}\frac{\mathrm{d}^{2} x^{i}}{\mathrm{d} t^{2}}\approx-\Gamma_{t t}^{i}.\end{equation}
It is now easy to see that 
\begin{equation}\Gamma_{t t}^{ i}=\frac{1}{2} \delta^{i j} \left(2 \partial_{t} h_{j t}^{ \mathrm{TT}}-\partial_{j} h_{t t}^{ \mathrm{TT}}\right)=0,\end{equation}
for a gravitational perturbation in the transverse traceless gauge (c.f. 
section~\ref{flat_prop}). We have thus reached the apparently paradoxical result that the mirrors do not move under the effect of a passing gravitational wave (if they start at rest). 

One must not forget, however, that coordinates (and therefore also the coordinate acceleration) have no physical meaning in GR. What is physically relevant is not the coordinate distance between the free falling mirrors, but their proper distance (which is independent of the coordinates). It is indeed the proper distance that determines the light travel time between   mirrors, and thus in turn the phase difference (the laser ``fringes'') at the photodetector.

Let us assume for simplicity that the mirrors are on the $x$ axis. Their proper distance is then
\begin{equation}
 L^{\rm proper}=\int_{0}^{L} \mathrm{d} x \sqrt{g_{x x}}=\int_{0}^{L} \mathrm{d} x \sqrt{1+h^{\mathrm{TT}}_{xx}}\simeq L \sqrt{1+h_{x x}^{\mathrm{TT}}(t, x=0)}  \simeq L\left[1+\frac{1}{2} h_{x x}^{\mathrm{TT}}(t,x=0)\right],
\end{equation}
where $L$ is the coordinate distance, and
we have assumed not only that the metric perturbation is small, but also that 
it has wavelength much larger than $L$.
(This assumption is needed in the third step.)

The change $\delta L$ in proper distance is therefore
\begin{equation}\frac{\delta L}{L} \simeq \frac{1}{2} h_{x x}^{\mathrm{TT}}(t, z=0)= \frac{1}{2} h^{\mathrm{TT}}_{ij}u^iu^j\label{dL}\end{equation}
where in the last step we have introduced the unit-norm vector $u$ to denote the detector's arm direction. In this way, this equation is valid in a general reference frame. 

An alternative (and more instructive) derivation of the same result can be obtained by resorting to the geodesic deviation equation,
\begin{equation}\label{geodev}
\frac{\mathrm{D}^2 v^\mu}{\mathrm{d}\tau^2}=R^{\mu}_{\alpha\beta\gamma}u^\alpha u^\beta v^\gamma\,,    
\end{equation}
with $R^{\mu}_{\alpha\beta\gamma}$ the Riemann tensor, $\mathrm{D}/\mathrm{d}\tau$ the covariant derivative along the four velocity,
and $v^\mu=\delta x^\mu$ the separation vector
between two neighboring geodesics (the two free falling mirrors). Using FNCs attached to one mirror, one has 
\begin{align}\label{fnc1}
 g_{\mu \nu}=\eta_{\mu \nu}+\mathcal{O}\left(x^{2}/\lambda^2\right)\,,\\
u^\mu=\delta_t^\mu+\mathcal{O}\left(x^{2}/\lambda^2\right)\label{fnc2}
\end{align}
where we have used the fact that the mirrors are in free fall ($a^\mu=0$) and the curvature radius of the spacetime (Minkowski plus a gravitational wave signal) is given by the 
signal's wavelength $\lambda$. 

Since the mirrors are at rest in these coordinates, $v^\mu=\delta x^\mu=(0,L^i)$. 
Because the metric is locally flat,
$\delta x^i=L^i$ describes the proper distance 
between the mirrors,
 up to corrections  of fractional order $\mathcal{O}\left(L^{2}/\lambda^2\right)$.
From Eq.~\ref{geodev}, one then obtains
\begin{equation}\frac{\mathrm{d}^{2} L^{i}}{\mathrm{d} t^{2}}=-R_{i t j t}L^{j}\,.\end{equation}
One can now note that the Riemann tensor of a perturbed Minkowski spacetime is a gauge invariant quantity. This follows simply from the fact that the gauge transformation of a tensor is proportional to the Lie derivative of the background tensor along the gauge generator. Since the Riemann tensor vanishes on the Minkowski background, it follows that it is a gauge invariant quantity at linear order.
More explicitly, one can evaluate the Riemann tensor for a perturbed flat spacetime using the results of section~\ref{sec:2}, obtaining
\begin{equation}\label{eq:dec_response}
R_{i t j t}=-\frac{1}{2} \ddot{h}_{i j}^{\mathrm{TT}}+\partial_i \partial_j \psi+\dot{\Sigma}_{(i, j)}-\frac{1}{2} \ddot{\theta} \delta_{i j}.\end{equation}
in terms of the gauge invariant variables introduced in that section.

Using now Eqs.~\ref{laplPsi} and~\ref{pnfields}, it is clear that
all the terms in Eq.~\ref{eq:dec_response} decay as $1/r^3$ (i.e. they correspond to Netwonian and PN tidal forces), and the only term that survives at large distances from a source is the first one. Therefore, one has
\begin{equation}
\label{eq:dec_response2}
\frac{\mathrm{d}^{2} L^{i}}{\mathrm{d} t^{2}}=-R_{i t j t}L^j=\frac{1}{2} \ddot{h}_{i j}^{\mathrm{TT}}L^j+\mathcal{O}\left(\frac{1}{r^3}\right)\,,
\end{equation}
and integrating this equation one obtains  Eq.~\ref{dL}. Note that again, we have implicitly made the assumption that the signal's wavelength is much larger than the 
distance between the two mirrors, because we have neglected terms $\mathcal{O}\left(x^{2}/\lambda^2\right)$
in Eqs.~\ref{fnc1}--\ref{fnc2}.

The phase difference at the photodetector depends of course on the changes in proper length (i.e. light travel time) induced by the gravitational signal on the first  ($\delta L_1$) and second ($\delta L_2$) arm, i.e. in the low frequency (large wavelength) limit
\begin{equation}\label{lowf_resp}
   \frac{{\delta L_1}-{\delta L_2}}{L}=
     h_{\mathrm{TT}}^{ij} D_{ij}=F_{+}  h_++F_{\times} h_{\times}\,,
\end{equation}
where
\begin{equation}
    D^{ij} =\frac{1}{2} (u^i u^j -v^i v^j)
\end{equation}
(with $u^i$ and $v^i$ unit vectors in the arm directions)
is the detector tensor and $F_+, F_{\times}$ are the pattern functions (which encode the detector's response in various directions). The phase difference at the photodetector is then explicitly
\begin{equation}
    \Delta\phi=\frac{4 \pi \nu}{c} \left({\delta L_1}-{\delta L_2}\right)=\frac{4 \pi \nu L}{c} \left(F_{+}  h_++F_{\times} h_{\times}\right)\,,\label{dphi_lowf}
\end{equation}
where $\nu$ is the laser frequency.
(Notice the presence of a factor 2 due to the round trip of the laser.)
\\
\\
{\bf Exercise 5:} {\it Consider an interferometer on the $(x,y)$ plane and a gravitational wave coming from the sky position $\phi,\theta$ (spherical coordinates). Define the wave plus and cross polarizations with respect to two unit vectors $\boldsymbol{e}_x$ and $\boldsymbol{e}_y$ orthogonal to the propagation direction $\boldsymbol{n}$, and such that the triad $(\boldsymbol{e}_x,\boldsymbol{e}_y,\boldsymbol{n})$ is right-handed. Show that the detector's pattern functions are
$$
F_+=\frac{1}{2} \cos 2 \psi  \left(\cos ^2\theta
   +1\right) \cos 2 \phi -\sin 2 \psi  \cos
   \theta  \sin 2 \phi 
$$
and
$$
F_\times = \frac{1}{2} \sin 2 \psi  \left(\cos ^2\theta
   +1\right) \cos 2 \phi +\cos 2 \psi  \cos
   \theta  \sin 2 \phi 
$$
where $\psi<\pi/2$ is the angle between $\boldsymbol{e}_y$ and the intersection between the plane of the detector and the plane orthogonal to the wave propagation direction. $\psi$ is usually referred to as the polarization angle.}

\begin{figure}
    \centering
    \includegraphics[width=0.9\textwidth]{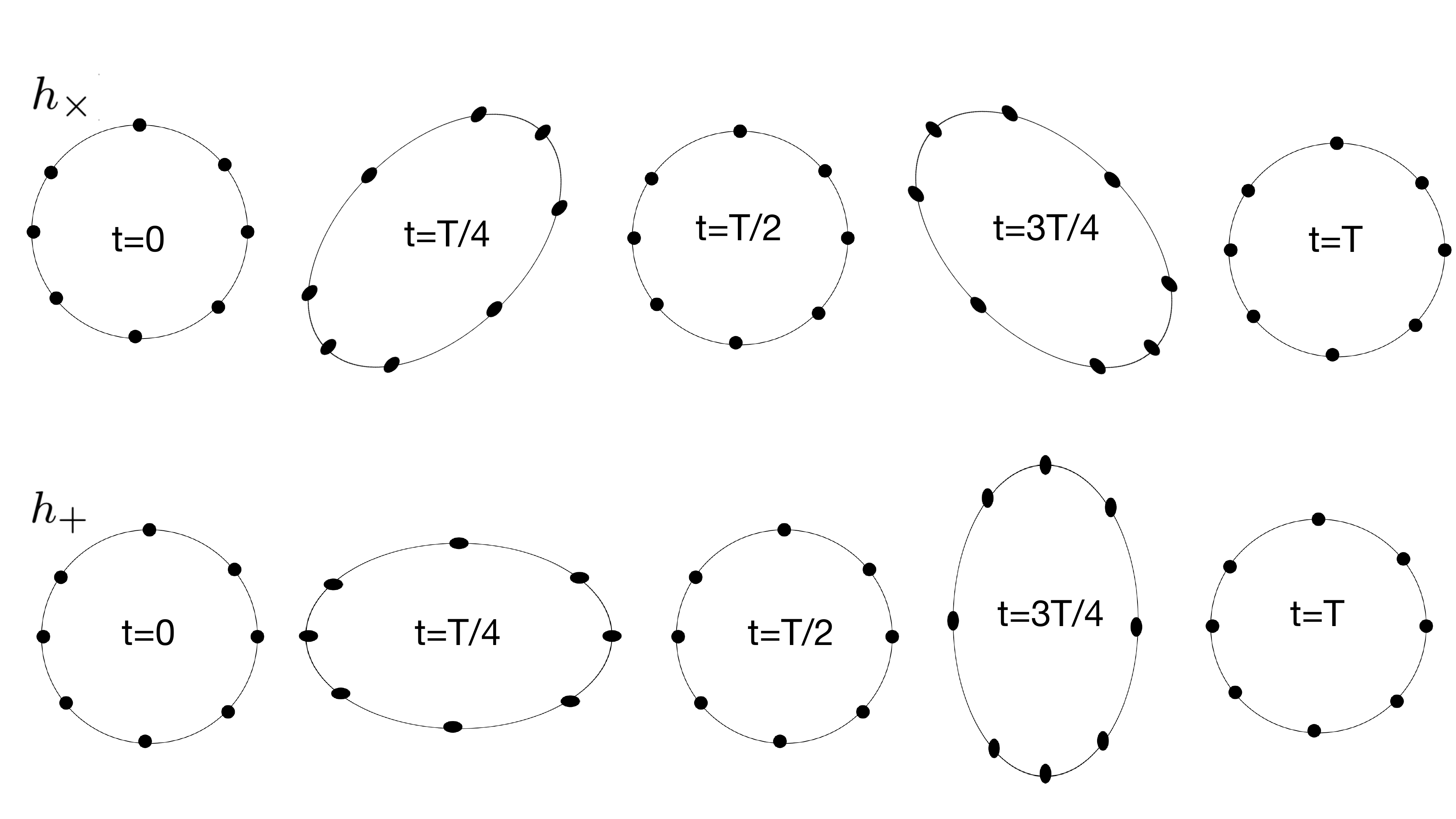}
    \caption{\footnotesize The response of an originally circular ring of particles to a linearly polarized gravitational wave ($h_+$ or $h_\times$) of period $T$ traveling in the orthogonal direction into the page.}
    \label{fig:lin}
\end{figure}

\begin{figure}
    \centering
    \includegraphics[width=0.9\textwidth]{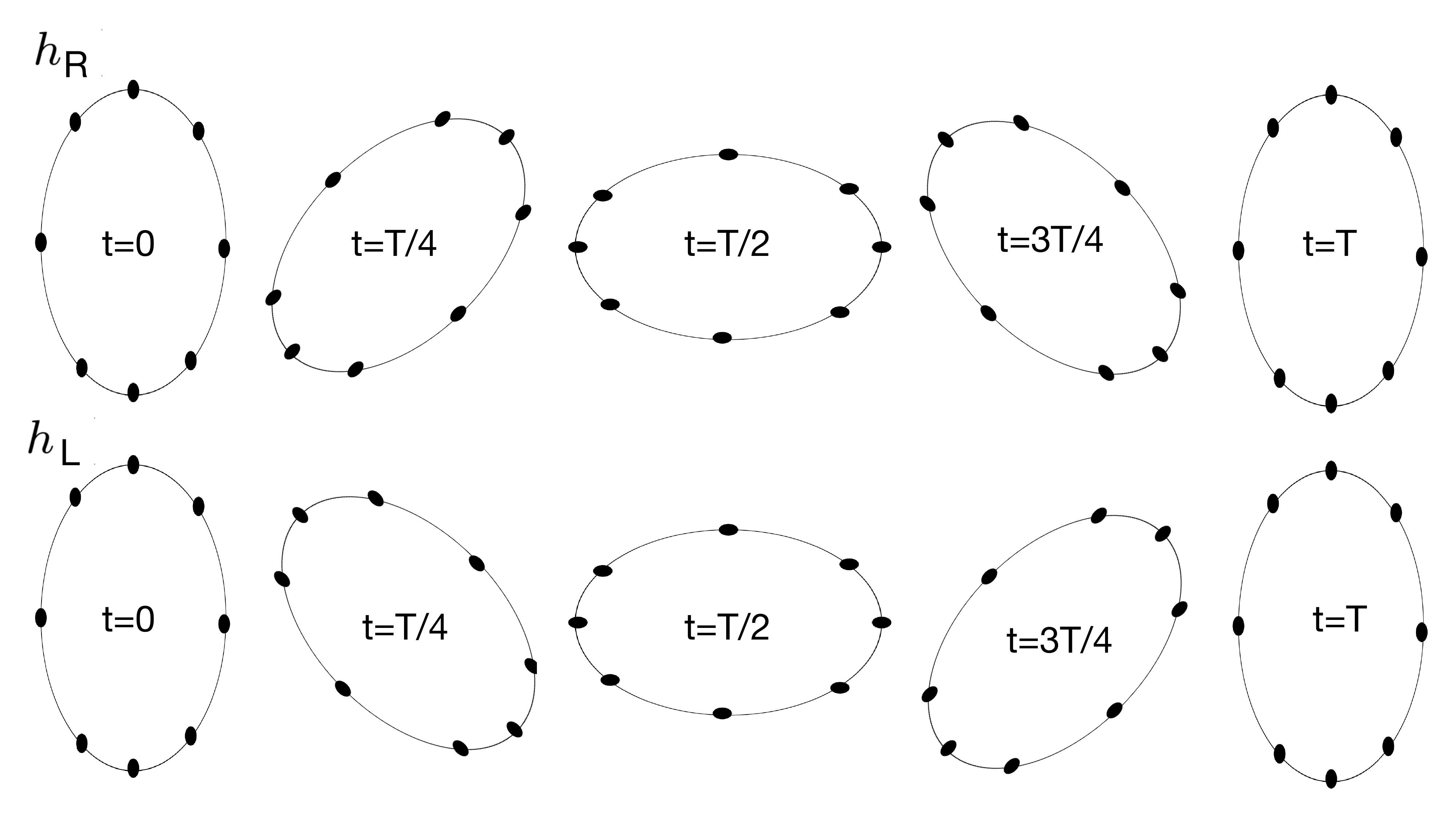}
    \caption{\footnotesize The response of an originally circular ring of particles to a 
    circularly  
    polarized gravitational wave (with right or left helicity) of period $T$ traveling in the orthogonal direction into the page. }
    \label{fig:pol}
\end{figure}

\subsection{A geometric interpretation of the polarizations}

Thanks to Eq.~\ref{lowf_resp}, we can now provide a geometric interpretation of the two polarizations $h_+$ and $h_\times$. If one considers a ring of particles on the $(x,y)$ plane and a signal traveling along the $z$-direction, the $h_+$ and $h_\times$ polarizations deform the ring in a distinct and characteristic way (Fig.~\ref{fig:lin}). Also 
shown, in Fig.~\ref{fig:pol}, is the response of a ring to a right-handed or left-handed circularly polarized wave, i.e. one for which the  $h_+$ and $h_\times$ polarizations have a phase difference of $\pm\pi/2$ and the same amplitude. A generic linear polarization corresponds instead to $h_+$ and $h_\times$ having the same phase (although potentially different amplitudes). 
Note that from Eq.~\ref{hTT}, it follows that for an observer that sees a circular binary face-on ($\iota=0$ or $\iota=\pi$) the signal is circularly polarized; if instead the observer sees the binary edge-on ($\iota=\pi/2$), the polarization is linear.
We also notice that additional polarization patterns are present beyond GR~\cite{eardley}. Those arise because  the scalar and vector degrees of freedom of Eq.~\ref{pnfields} become dynamical (c.f. also section~\ref{dimsec}).

\subsection{The response of a gravitational wave detector: the transfer function}\label{subsec:7_2}

Let us now relax the short wavelength assumption ($\lambda\ll L$) that we made when deriving the detector's response in the previous section. Let us then consider
two free falling mirrors in a 
spacetime with metric
\begin{equation}
    g_{\mu \nu}= \eta_{\mu \nu} + h^{\mathrm{TT}}_{\mu \nu},
\end{equation}
and integrate null geodesics (the lasers) back and forth between them. As derived in section~\ref{subsec:7_1}, the mirrors do not move
(with respect to the coordinates)
if the perturbation is written in the transverse and traceless gauge. 

Let us also make the simplifying assumption that $h_{\mu\nu}^{\mathrm{TT}}$ is given by a plane wave traveling along the $z$-axis: 
\begin{equation}
    h_{\mu\nu}^{\mathrm{TT}} = \begin{pmatrix}
0 & 0 & 0 & 0\\
0 & \cos 2\psi & \sin 2\psi & 0\\
0 & \sin 2\psi & -\cos 2\psi  & 0\\
0 & 0 & 0 & 0\\
\end{pmatrix} h(t-z)\,.
\end{equation}
(Recall  that a generic linear perturbation can always be decomposed in a superposition of plane waves.) It is then clear that the spacetime has three Killing vectors 
\begin{gather}\label{killing}
    k_1=\frac{\partial}{\partial x}\,,\\
     k_2=\frac{\partial}{\partial y}\,,\\
      k_3=\frac{\partial}{\partial t}+\frac{\partial}{\partial z}\,.
\end{gather}
One can then write a tetrad (carried by the mirrors) for the perturbed spacetime
as 
\begin{gather}\label{tetrad0}
    e_{(0)}=\frac{\partial}{\partial t}\,,\\
    e_{(1)}=\left(1-\frac{h}{2}\right) \left(\cos\psi \frac{\partial}{\partial x}+\sin\psi \frac{\partial}{\partial y}\right)+\mathcal{O}(h)^2\,,\\
        e_{(2)}=\left(1+\frac{h}{2}\right) \left(-\sin\psi \frac{\partial}{\partial x}+\cos\psi \frac{\partial}{\partial y}\right)+\mathcal{O}(h)^2\,,\\
    e_{(3)}=\frac{\partial}{\partial z}\,,\label{tetrad}
\end{gather}
and write the null wave-vector
of the laser as
\begin{equation}\label{wavenumber}
\sigma=\nu \left[e_{(0)}+\sin\theta (e_{(1)}\cos\phi+e_{(2)}\sin\phi)+\cos\theta e_{(3)}\right]\,.
\end{equation}
Here, $\nu$ is the laser frequency as measured at the mirror, and  $\theta$ and $\phi$
describe the laser's direction. In particular, $\theta$ is the angle between the arm of the detector and the gravitational wave's propagation direction.

\begin{figure}[!h]
\centering
\includegraphics[width=0.85\textwidth]{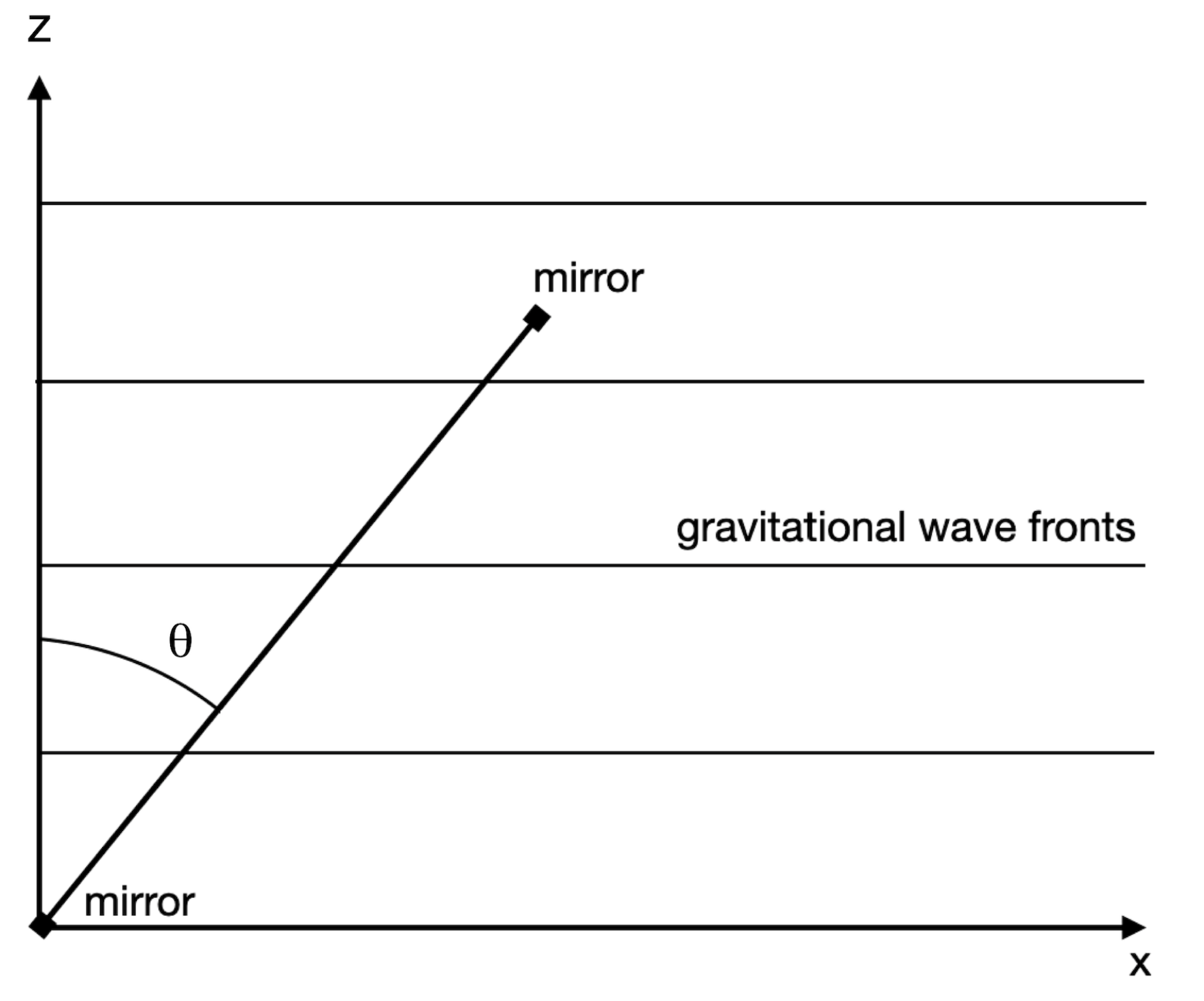}
\caption{\footnotesize Sketch of the geometry of the system leading to Eq.~\ref{dnu}, for $y=0$.}
\label{response_sketch}
\end{figure}

Now, let us recall that the projections of the laser wave-vector on the three Killing vectors are conserved along the laser's null geodesics:
\begin{gather}
    -\sigma \cdot k_3=\nu (1-\cos\theta)=\mbox{const}\,,\\
    \sigma \cdot k_1= \left(1+\frac{h}{2}\right) (\sigma\cdot e_{(1)} )\cos\psi-
    \left(1-\frac{h}{2}\right) (\sigma\cdot e_{(2)}) \sin\psi=\mbox{const}\,,\\
        \sigma \cdot k_2= \left(1+\frac{h}{2}\right) (\sigma\cdot e_{(1)} )\sin\psi+
    \left(1-\frac{h}{2}\right) (\sigma\cdot e_{(2)}) \cos\psi=\mbox{const}\,.  
\end{gather}
This implies also conservation of 
the combination 
$\sqrt{(\sigma\cdot k_1)^2+(\sigma\cdot k_2)^2}$, i.e.
\begin{equation}
   \sqrt{(\sigma\cdot k_1)^2+(\sigma\cdot k_2)^2} =\left[(1+h) (\sigma\cdot e_{(1)} )^2+
   (1-h) (\sigma\cdot e_{(2)} )^2\right]^{1/2}\approx\nu \sin\theta \left(1+\frac{h}{2}\cos2\phi\right)=\mbox{const}\,.
\end{equation}
Note that this conserved quantity can be rewritten in more compact form
as
\begin{gather}
 \nu  \sin\theta \left(1+\frac{h}{2}\cos2\phi\right)=
 \nu  \sin\theta \left(1+\frac{Q}{2}\right)
  =\mbox{const}\,,\\  
  Q= \frac{h^{\mathrm{TT}}_{ij} n^i n^j}{1-\cos^2 \theta}\,,
\end{gather}
where $\boldsymbol{n}=
\sin\theta \cos(\phi+\psi) \partial_x+\sin\theta \sin(\phi+\psi) \partial_y+\cos\theta \partial_z$ is the detector's arm direction (from the first to the second mirror) expressed in Cartesian coordinates [this can be obtained by rewriting
Eq.~\ref{wavenumber} using Eqs.~\ref{tetrad0}--\ref{tetrad} and taking  $h\to0$].

Let us consider now a laser photon starting from first mirror (time ``0''), bouncing on the second mirror (time ``1'') and returning finally to the first mirror (time ``2''). 
From the conservation laws written above
one then obtains
\begin{gather}
\nu_0 (1-\cos\theta_0)=\nu_1 (1-\cos\theta_1)\,,\\
    \nu_0 \sin\theta_0  \left(1+\frac{Q_0}{2}\right)=\nu_1 \sin\theta_1  \left(1+\frac{Q_1}{2}\right)\,.    
\end{gather}
Writing $\cos\theta_1=\cos\theta_0+\delta\cos\theta+ {\cal O}(h)^2$, $\sin\theta_1=\sin\theta_0+\delta\sin\theta +{\cal O}(h)^2$ and $\nu_1=\nu_0+\delta\nu+{\cal O}(h)^2$, with  $\delta\cos\theta$, $\delta\sin\theta$ and $\delta\nu$ of order ${\cal O}(h)$, one can linearize these equations and obtain
\begin{equation}\label{nu1}
\frac{\nu_1-\nu_0}{\nu_0}=\frac{\delta \nu}{\nu_0}=\frac{1+\cos\theta_0}{2} (Q_0-Q_1)   
\end{equation}
for the change in laser frequency between the first and second mirror. Performing the same calculation to compute the change in laser frequency when the photon travels in the opposite direction (from the second mirror to the first), one finds
\begin{equation}\label{nu2}
\frac{\nu_2-\nu_1}{\nu_1}=\frac{1-\cos\theta_0}{2} (Q_1-Q_2)   \,.   
\end{equation}
Note that the different sign of $\cos\theta_0$ in Eqs.~\ref{nu2} and~\ref{nu1} appears because during the return trip the photon's propagation  is along $-\boldsymbol{n}$.

Combining these results, one obtains that
the change $\Delta\nu$ in laser frequency at the first mirror after a round trip is
\begin{equation}\label{dnu}
   \frac{\Delta\nu}{\nu}= \frac{\nu_2-\nu_0}{\nu_0}=
\frac{1}{2}(1+\cos\theta) Q(t)-\cos\theta \,Q\left(t+\frac{\tau (1-\cos\theta)}{2}\right)
   - \frac{1}{2}(1-\cos\theta) Q(t+\tau)
\end{equation}
where we have dropped the index of the angle $\theta$ since the differences among $\cos\theta_0$, $\cos\theta_1$ and $\cos\theta_2$ appear at higher order, and
we have used the fact that $Q_0=Q(t)$, $Q_1=Q(t+{\tau (1-\cos\theta)}/{2})$ and $Q_2=Q(t+\tau)$ with $\tau$ the laser round trip time (c.f. Fig.~\ref{response_sketch}).
From this expression, one can then get the change in the laser phase over a round trip,
\begin{equation}\label{eq:change_phase}
    \Delta\phi=-2\pi\int\Delta\nu \mathrm{d} t\,.
\end{equation}
The quantity measured at the photodetector is then the difference between the phase changes $\Delta \phi$ in the first and second arm of the interferometer.

To compute this phase difference and get some physical insight, let us assume for simplicity  $\theta={\pi}/{2}$, which yields
\begin{equation}\label{eq:period_pulses}
    \Delta\nu=\frac{\nu}{2}  [h(t)-h(t+\tau)]\,,
\end{equation}
with $h=h_{ij}^{\mathrm{TT}} n^i n^j$. 
For a monochromatic wave $h(t)=h_0 \sin{(2\pi f t)}$ one then has
\begin{equation}\label{eq:change_phase_2}
    \Delta \phi=-2\pi\int  \Delta\nu \mathrm{d}t =  \frac{h_0\nu}{f}\sin{(f\pi\tau)} \sin{\Big[2\pi f\Big(t+\frac{\tau}{2}\Big) \Big]}\,,
\end{equation}
which can then be rewritten more simply as
\begin{equation}\label{eq:change_phase_3}
    \Delta \phi= h \left(t+\frac{\tau}{2}\right) \frac{\nu}{f}\sin{(f\pi\tau)}.
\end{equation}

Let us interpret this result in terms of
a proper distance change $\delta L$ between the two mirrors, which produces a change in the laser phase (in a round trip) given by
\begin{equation}\label{dphi}
    \Delta \phi=4\pi\nu \frac{\delta L}{c}\,.
\end{equation}
Let us think about this distance change as due to an ``effective'' strain $h_{\rm eff}$:
\begin{equation*}
    \frac{\delta L}{L}=\frac{1}{2}h_{\rm eff},
\end{equation*}
where the length can be rewritten
in terms of the  round trip time $\tau$, $L={\tau c}/{2}$. Replacing in Eq.~\ref{dphi},
one then obtains
\begin{equation}\label{dphi2}
    \Delta \phi=\pi\nu\tau h_{\rm eff}\,.
\end{equation}
By comparing to Eq.~\ref{eq:change_phase_2},
one  gets
\begin{gather}
    h_{\rm eff}=h\, T(f)\,,\\
    T(f)=\frac{\sin{(f\pi \tau})}{f \pi \tau}\,,\label{Tf}
\end{gather}
where we have introduced the ``transfer function'' $T(f)$. This calculation can be generalized to more generic propagation directions $\theta\neq\pi/2$. 

In summary, we can therefore write the full response of a detector to a Fourier component of frequency $f$ of a gravitational wave signal as
\begin{equation}
    \Delta\phi=\frac{4 \pi \nu L}{c} \left[F_{+}  h_+(f)+F_{\times} h_{\times}(f)\right] T(f)\,.
\end{equation}
As can be seen from Eq.~\ref{Tf}, at low frequencies $T(f)\approx 1$ and this expression reduces to the low frequency response of Eq.~\ref{dphi_lowf}. At high frequencies $f\gg 1/\tau$, the transfer function instead decays as $1/(f\tau)$, modulated by the oscillations of the numerator of Eq.~\ref{Tf}. This explains why
the frequency window of gravitational interferometers scales with their armlength $L=c\tau$.

\section{Gravitational wave data analysis}
As we have seen in the previous section, gravitational wave interferometers measure the interference pattern (i.e.~the phase difference) between laser beams traveling between mirrors. The effect of a gravitational signal, however, is generally so small that it is also crucial to adequately understand the statistical error (``noise'') affecting this measurement, and to devise statistical techniques to characterize the signal and its astrophysical source.

\subsection{Gaussian noise and power spectral density}

\begin{figure}
    \centering
\includegraphics[width=0.7\textwidth,angle=-90]{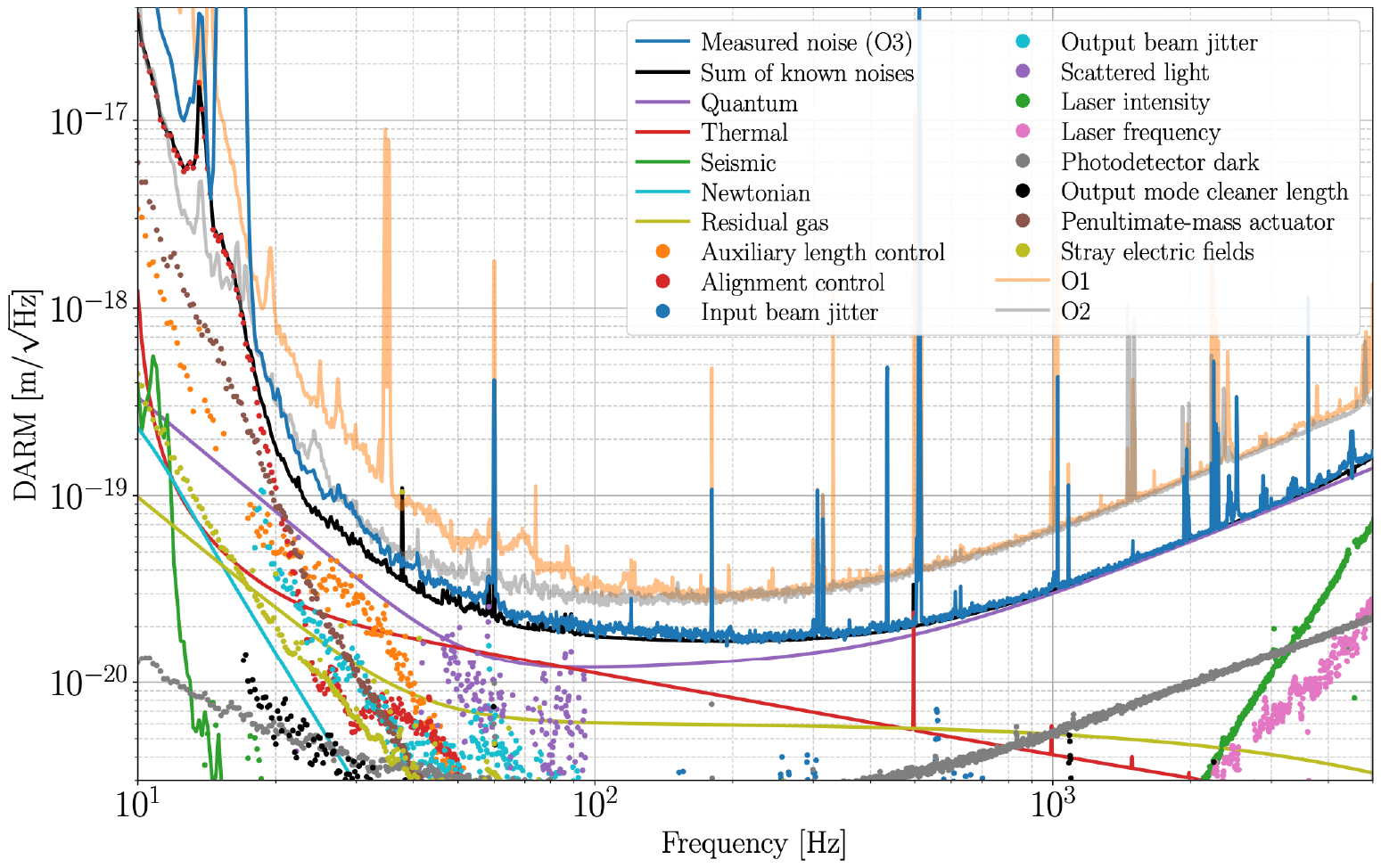}\vskip-0.95cm
    \caption{\footnotesize Various contributions to the Advanced LIGO differential armlength (DARM) caused by the noise. This quantity is proportional to the  
    (square root of the)  power spectral density.  Taken from Ref.~\cite{ligopsd}.}
    \label{fig:ligopsd}
\end{figure}

\begin{figure}
    \centering
\includegraphics[width=0.87\textwidth]{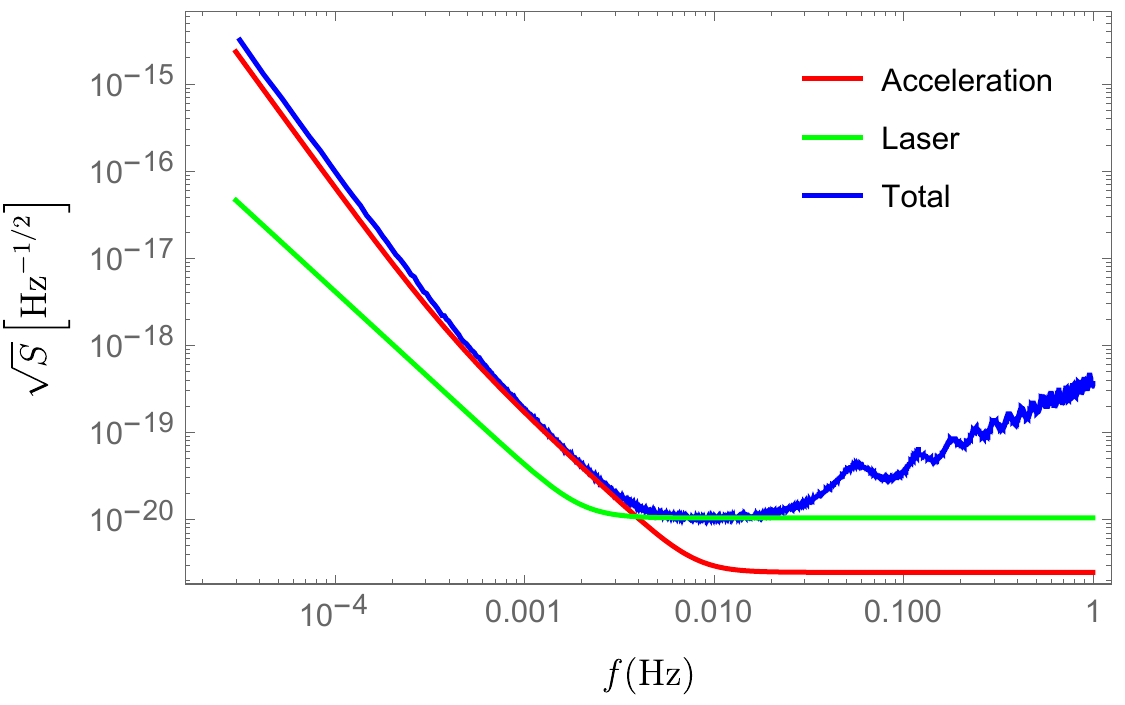}
    \caption{\footnotesize Various contributions to the LISA power spectral density. At high frequencies, one can see the degradation and the oscillations due to the transfer function.}
    \label{fig:lisapsd}
\end{figure}

A detector's output can be written as
$s(t)=h(t)+n(t)$, where  $h(t)$ is the signal and $n(t)$ is the instrumental noise. Defining the Fourier transform of a time series $A(t)$ as
    \begin{equation}
    \tilde{A}(f)=\int_{-\infty}^{\infty} A(t) e^{2\pi i f t} \mathrm{d}t\,,
\end{equation} 
the signal is given by
$\tilde{h}(f)=\left[F_{+}  h_+(f)+F_{\times} h_{\times}(f)\right] T(f)$. As for the noise, a common assumption, which turns out to be a good approximation for gravitational wave interferometers, is that of stationarity and Gaussianity.

Stationarity implies that the statistical properties of the noise are time independent. 
Introducing the ensemble average $\langle\ldots\rangle$, i.e. the average over all possible noise realizations~\footnote{By the ergodic theorem, ensemble averages can be replaced with time averages.}, stationarity implies in particular that $\langle n(t) n(t')\rangle$ should only depend on the time difference $\tau=t'-t$, i.e. $\langle n(t) n(t')\rangle=W(\tau)/2$. (Note also that $\langle n(t)\rangle=0$). Let us see what this implies for the quantity
$\langle \tilde{n}^*(f) \tilde{n}(f')\rangle$. Using the definition of Fourier transform, one has
\begin{align}
   & \langle\tilde{n}^*(f) \tilde{n}(f')\rangle=
    \int\int \mathrm{d} t \mathrm{d} t' \langle n(t) n(t') \rangle
e^{2\pi i (f' t'-f t)}\nonumber\\&=
    \frac12\int\int \mathrm{d} t \mathrm{d} \tau W(\tau) 
e^{2\pi i [(f'-f) t+f'\tau]}=
 \frac12 S(f) \delta(f-f')\,,\label{psd}
\end{align}
where $S(f)$ is the Fourier transform of $W(\tau)$ and is known as the (single-sided) power spectral density of the noise. One can also check easily that 
\begin{equation}
\langle n(t)^2\rangle=\frac{W(0)}{2}=\int_0^\infty S(f) \mathrm{d}f\,.   
\end{equation}

The  assumption of Gaussian noise then amounts to saying that each Fourier component has a Gaussian probability distribution function, with variance given by Eq.~\ref{psd}.
In more detail, because detectors only observe for a finite time $T$ and thus Fourier transforms are implemented as discrete transformations, the frequencies $f_i$ at which measurements are performed are spaced by $\Delta f=1/T$. In terms of these discrete frequencies, Eq.~\ref{psd} becomes
\begin{equation}
    \langle\tilde{n}^*(f_i) \tilde{n}(f_j)\rangle=\frac{1}{2 \Delta f} S(f_i)\delta_{ij}\,,
\end{equation}
and the probability $P(n)$ of having a realization $\tilde{n}(f)$ of the noise is 
\begin{align}
   & P(n)\propto \prod_i \exp\left[-\frac{2 |\tilde{n}(f_i)|^2 \Delta f}{S(f_i)}\right] =     
    \prod_i \exp\left\{-\frac{2 \Delta f[(\text{Re}\,\tilde{n}(f_i))^2+(\text{Im}\,\tilde{n}(f_i))^2]}{S(f_i)}\right\}\nonumber\\&\qquad=
    \exp\left[
    -2\int_0^\infty \mathrm{d}f\frac{ |\tilde{n}(f)|^2 }{S(f)}
        \right]   
    \,.\label{pn}
\end{align}
Here, the index $i$ spans all positive  frequencies in the data [since $n(t)$ is real,
$\tilde{n}^*(f)=\tilde{n}(-f)$ and the signal at negative frequencies follows from that at positive ones].

As can be seen by changing variables in this equation, while the real and imaginary parts of $\tilde{n}(f)$ are Gaussian distributed, the phase of $\tilde{n}(f)$ is uniformly distributed, while
the norm $r(f)=|\tilde{n}(f)|$
follows the Rayleigh distribution $p[r(f_i)]\propto \exp\{-2\Delta f [r(f_i)]^2/S(f_i)\} r(f_i)$.
Introducing the internal product
\begin{equation}\label{eq:int_prod}
    (A \mid B) \equiv 4\, \text{Re} \int^{\infty}_{0} \frac{\mathrm{d}f  \tilde{A}^*(f) \tilde{B}(f)}{S(f)},
\end{equation}
between two real functions $A(t)$ and $B(t)$,  
one can rewrite Eq.~\ref{pn} as
\begin{equation}\label{pnoise}
    P(n)\propto \exp\left[-\frac12 (n|n)\right]\,.
\end{equation}
As we will see,  this internal product will also simplify many expressions below.

The behavior of the power spectral density of the noise depends critically on the detector. For ground based interferometers such as LIGO and Virgo,   at low frequencies the noise originates chiefly from the laser's radiation pressure,  the seismic noise and the thermal noise of the suspensions; at mid-frequencies the main contributions are the thermal noise from the mirror coatings and the laser noise (radiation pressure and shot noise); at high frequencies the shot noise dominates (see Fig.~\ref{fig:ligopsd}). 

For space-borne detectors like LISA (Fig.~\ref{fig:lisapsd}), the limitations to 
the interferometer's
sensitivity come from spurious accelerations of the test masses (due e.g. to cosmic rays, residual gas in the housing, temperature fluctuations)
 at low frequencies; from the laser shot noise at mid-frequencies; and from the antenna transfer function at high frequencies (c.f. the oscillations in Fig.~\ref{fig:lisapsd}, which are due to the numerator of Eq.~\ref{Tf}).

\subsubsection{Detection in the presence of noise}

In order to disentangle the gravitational wave signal from the noise, many techniques have been put forward, among which one of the most popular is match filtering. The latter essentially amounts to cross correlating the detector's noisy output 
\begin{equation}\label{s=h+n}
    s(t)=h(t)+n(t)
\end{equation}
 with a bank of templates $h(t,\boldsymbol{\theta})$. Here, the vector $\boldsymbol{\theta}$ denotes the parameters of the source (i.e. for quasi-circular binaries, the masses, the spins, the distance, the initial phase, the merger time, the inclination, the sky position and the polarization angle). It is quite intuitive that the cross correlation 
\begin{equation}
    \int \mathrm{d}t\, s(t) h(t,\boldsymbol{\theta}) 
\end{equation}
will ``on average'' be maximized if the template matches the signal. Taking in fact the average of this equation, one obtains
\begin{equation}
    \int \mathrm{d}t\, \langle s(t) h(t,\boldsymbol{\theta})\rangle =
    \int \mathrm{d}t\,  h(t) h(t,\boldsymbol{\theta})\,, 
\end{equation}
and the second integral is small if  the signal and template do not match (because in that case the integrand is highly oscillatory).

Let us try to formalize this statement (see e.g. Ref.~\cite{maggiore}) by defining a generic filter
\begin{equation}
    \hat{A}=\int \mathrm{d}t\, A(t) K(t)
\end{equation}
where $A(t)$ is the time series being filters, and $K(t)$ a filter function. Let us try to define a signal-to-noise ratio $S/N$ based on this filter. It is natural to take the signal $S$ as the filter of the detector's output $s(t)$, averaged over many realizations of the noise:
\begin{equation}
    S=\langle \hat{s} \rangle=\int \mathrm{d}t \,\langle s(t)\rangle K(t)=\int \mathrm{d}t \,h(t) K(t)=\int \mathrm{d}f \, \tilde{h}(f) \tilde{K}^*(f)\,.
\end{equation}
As for the denominator $N$, since $\langle \hat{n} \rangle=0$, we can define instead
\begin{align}\label{eq:rmsN}
    N^2&\,=\langle \hat{n}^2 \rangle=\int \mathrm{d}t \mathrm{d}t' \,K(t) K(t') \langle n(t) n(t')\rangle =
    \frac12 \int \mathrm{d}t \mathrm{d}t'\, K(t) K(t') W(t'-t)\nonumber\\=&\quad
    \frac12\int \mathrm{d}t \mathrm{d}t'  \mathrm{d}f\,K(t) K(t') S(f) e^{-2 \pi i f (t'-t)}=\frac12\int\mathrm{d}f\,|\tilde{K}(f)|^2  S(f) 
    \,.
\end{align}
The ratio $S/N$ is then
\begin{equation}
    \frac{S}{N}=\frac{\int_{-\infty}^{+\infty} \mathrm{d}f \, \tilde{h}(f) \tilde{K}^*(f)}{\left[\int_{-\infty}^{+\infty}  \mathrm{d}f (1/2) S(f) |K(f)|^2\right]^{1/2}}\,.
\end{equation}
By introducing the internal product of Eq.~\ref{eq:int_prod}, and using the fact that for a real function $A$ one has $\tilde{A}(-f)=\tilde{A}^*(f)$, one can rewrite
\begin{equation}
    \frac{S}{N}=\frac{(h|u)}{(u|u)^{1/2}}\,,\label{snr}
\end{equation}
where we have defined 
\begin{equation}
    \tilde{u}(f)=\frac12 S(f) \tilde{K}(f)\,.
\end{equation}
Clearly, Eq.~\ref{snr} is maximized if $u$ and $h$
are parallel, i.e. the filter yielding the optimal signal-to-noise ratio is
\begin{equation}
    \tilde{K}(f)\propto \frac{\tilde{h}(f)}{S(f)}\,.
\end{equation}
Therefore, if one is searching for a signal, the optimal filter is the template perfectly matching the signal, weighted by the noise power spectral density. 
This is known as Wiener's optimal filter theorem. 
Replacing back into  Eq.~\ref{snr},  one obtains that the optimal 
signal-to-noise ratio is simply given by
\begin{equation}
    \left(\frac{S}{N}\right)^2={(h|h)}=4 \int_0^\infty\mathrm{d}f\, \frac{|\tilde{h}(f)|^2}{S(f)}\,.\label{snr_opt}
\end{equation}

\subsubsection{Statistical significance and false alarm probability}
\label{sec:FAR}

In the previous section, we derived the optimal matched filter and the optimal signal-to-noise ratio (Eq.~\ref{snr_opt}) under the assumption that the signal $h(t)$ was perfectly known and that we only wanted to quantify, in the presence of Gaussian noise, its best linear estimator. In a real search, however, the presence of the signal itself is what we are trying to establish: the question we actually need to answer is then to what extent a filtered output above a certain threshold can be ascribed to a genuine GW signal, rather than to a  fluctuation of the noise.

To address this question, let us follow again~\cite{maggiore} and regard the matched-filter output
\begin{equation}
\hat{s}\equiv \int \mathrm{d}t \,[h(t)+n(t)]\,K(t)
\end{equation}
as a {\it random variable}, dependent on the specific realization of the noise $n(t)$, rather than as an ensemble-averaged quantity as in the previous section. The signal-to-noise ratio associated with a given realization is then given by the dimensionless quantity
\begin{equation}\label{eq:rho_def}
\rho= \frac{\hat{s}}{N}\,,
\end{equation}
with $N$ the rms of the filter output in the absence of a signal (i.e. Eq.~\ref{eq:rmsN}). Note that Eq.~\ref{eq:rho_def} defines $\rho$ as the {\it realized} signal-to-noise ratio for a specific noisy stretch of data, as opposed to the ensemble-averaged signal-to-noise ratio $S/N=\langle \hat{s}\rangle/N$ that we used in the matched-filtering derivation. One has, by construction, $S/N=\langle\rho\rangle$.

When there is no GW signal in the data ($h=0$), $\rho$ is a linear functional of the Gaussian noise $n(t)$, and is therefore itself a Gaussian random variable with zero mean and unit variance (by construction). Its probability distribution is thus the standard unit Gaussian
\begin{equation}
\label{eq:p_rho_nosignal}
p(\rho\,|\,h=0)\,\mathrm{d}\rho = \frac{1}{\sqrt{2\pi}}\,e^{-\rho^2/2}\,\mathrm{d}\rho\,.
\end{equation}
A large positive value of $\rho$ corresponds to an output $\hat{s}$ that is strongly correlated with the template $h$, and is therefore what one should regard as a candidate signal. A large {\it negative} value, on the other hand, would mean that the data are anti-correlated with the template, i.e. a poor match rather than a signal-like excursion, and it should not be counted as a candidate. The relevant detection criterion is therefore $\rho>\rho_t$, and the probability that Gaussian noise alone produces such a  value is the {\it false alarm probability}
\begin{equation}
\label{eq:pFA}
p_{\rm FA}(\rho_t) = \int_{\rho_t}^\infty \mathrm{d}\rho\,p(\rho\,|\,h=0)  =\int_{\rho_t}^\infty \frac{\mathrm{d}\rho}{\sqrt{2\pi}}\,e^{-\rho^2/2}= \frac{1}{2}\,\mathrm{erfc}\!\left(\frac{\rho_t}{\sqrt{2}}\right)\,,
\end{equation}
where $\mathrm{erfc}(z)$ is the complementary error function.

The detection threshold $\rho_t$ is chosen by requiring the total number of false alarms expected over the full search to remain small (typically of order  unity per observing run). This translates into a constraint on the per-trial false alarm probability~\ref{eq:pFA}, which however must be weighted by the number of independent trials  that the search performs on the data. Let us see how this plays out for the specific case of the search for  an inspiraling compact binary in a single ground-based  detector.

Since the signal depends on the astrophysical parameters of the source (essentially the masses and spins of the two components), the search is implemented by filtering the data through a discrete ``bank'' of templates $h(t,\boldsymbol\theta)$ that  covers the relevant region of parameter space with sufficient resolution. Typical template banks contain of order $N_{\rm templ}\sim 10^5$ independent waveforms. For each template, the search must also scan over the (unknown) arrival time of the putative signal, with a time resolution of a few ms set by the inverse bandwidth of the detector near its peak kHz sensitivity. Over a one-year observation ($T_{\rm obs}\sim 3\times 10^7$ s), this corresponds to $N_{\rm time}\sim T_{\rm obs}/(3\,{\rm ms})\sim 10^{10}$ independent starting times per template, and therefore to a total of $N_{\rm trials}\sim N_{\rm templ}\,N_{\rm time}\sim 10^{15}$ independent filter evaluations over the course of a year. Requiring that this enormous number of trials yield at most a handful of accidental false alarms translates into a per-trial false-alarm probability $p_{\rm FA}\lesssim 1/N_{\rm trials}\sim 10^{-15}$, which from Eq.~\ref{eq:pFA} fixes the single-detector detection threshold to $\rho_t\approx 8$, with only a weak dependence on $N_{\rm trials}$. Using a lower threshold would flood the candidate list with accidental noise fluctuations, while using a higher one would throw away a significant fraction of the real signals.

A single-detector threshold $\rho_t\approx 8$ is already a rather severe  requirement on the astrophysical signal: via Eq.~\ref{snr_opt}, it translates into a maximum  horizon distance beyond which the source  cannot be told apart from noise. A powerful way to relax the requirement is to demand that the candidate event appear {\it in coincidence} in (at least) two detectors whose noise processes are independent. If the two noises are  uncorrelated, the probability of finding a  simultaneous spurious excursion above $\rho_t$ in both instruments is the product of the individual false-alarm probabilities, i.e. the  square of Eq.~\ref{eq:pFA}. The  same overall accidental rate $\sim 10^{-15}$ per trial can then be reached with a per-detector false-alarm probability of order $\sqrt{10^{-15}}\sim 10^{-7.5}$, which from Eq.~\ref{eq:pFA} yields the considerably milder per-detector threshold $\rho_t\approx 5.5$. This quantitative  gain is the main reason why modern GW searches are run as coincidences between at least two interferometers, whenever more than one is operating, and is one concrete  illustration of the  advantage of a detector network over a single  instrument.

The above (frequentist) discussion of the statistical significance of a candidate signal can also be recast in a Bayesian framework, where the question becomes how much the data  prefer the noise-only hypothesis  over a signal hypothesis (or the other way around). Let us consider the two alternatives
\begin{itemize}
\item $H_1$: the data $s(t)$  contain only noise, $s(t) = n(t)$;
\item $H_2$: the data  contain a GW signal $h(t;\boldsymbol{\theta})$ parametrized by some unknown set of parameters $\boldsymbol{\theta}$, on top of the noise, $s(t) = h(t;\boldsymbol{\theta}) + n(t)$.
\end{itemize}
Since the noise is Gaussian (cf. Eq.~\ref{pnoise}), its probability distribution takes  the form
\begin{equation}
p(n)\propto \exp\left[-\frac{1}{2}(n|n)\right]\,,
\end{equation}
where $(\cdot\,|\,\cdot)$ is the noise internal product of Eq.~\ref{eq:int_prod}. Since $n=s$ under $H_1$ and $n=s-h(\boldsymbol{\theta})$ under $H_2$, the likelihood of the data under the two hypotheses  reads
\begin{equation}
p(s\,|\,H_1) \propto \exp\left[-\frac{1}{2}(s|s)\right]\,,\qquad p(s\,|\,H_2,\boldsymbol{\theta}) \propto \exp\left[-\frac{1}{2}(s-h|s-h)\right]\,.
\end{equation}
Under $H_2$ the signal parameters are not  known {\it a priori}, and must be integrated over with a suitable prior $\pi(\boldsymbol{\theta})$, yielding the  marginalized likelihood
\begin{equation}
p(s\,|\,H_2) = \int \mathrm{d}\boldsymbol{\theta}\,\pi(\boldsymbol{\theta})\,p(s\,|\,H_2,\boldsymbol{\theta})\,.
\end{equation}
The relative Bayesian evidence of the two hypotheses is then given by the Bayes factor
\begin{equation}
\label{eq:bayes_factor}
\Lambda \equiv \frac{p(s\,|\,H_1)}{p(s\,|\,H_2)} = \left\{\int \mathrm{d}\boldsymbol{\theta}\,\pi(\boldsymbol{\theta})\,\exp\left[(s|h(\boldsymbol{\theta})) - \frac{1}{2}(h(\boldsymbol{\theta})|h(\boldsymbol{\theta}))\right]\right\}^{-1}\,,
\end{equation}
where we have used  $(s-h|s-h) = (s|s) - 2(s|h) + (h|h)$ and  the cancellation of the $(s|s)$ factor between numerator and denominator. With these conventions, $\Lambda>1$ means that the data prefer the noise-only hypothesis, while $\Lambda<1$ means that the data prefer the signal hypothesis: the {\it smaller} $\Lambda$, the stronger the evidence of  a GW signal in the data.

It is instructive to evaluate Eq.~\ref{eq:bayes_factor} in two complementary regimes. In both cases we will need the statistical properties of the ``noise'' contribution $(n|h)$. The latter is a linear functional of the Gaussian, zero-mean noise $n$, and is thus itself a zero-mean Gaussian random variable. A short calculation using the definition of the internal product~\ref{eq:int_prod} and the noise 2-point correlation function~\ref{psd} shows that its variance equals $(h|h)$, i.e. 
\begin{equation}
\label{eq:rms_nh}
\langle(n|h)^2\rangle^{1/2} = (h|h)^{1/2} = \rho\,.
\end{equation}

\medskip

\noindent{\it Strong-signal regime.} Consider first  a realization of the data that {\it does} contain a GW signal, $s=h+n$, and assume for simplicity that the template exactly matches the signal. Then one has \begin{equation}
(s|h) = (h|h) +(n|h) = \rho^2 + \mathcal{O}(\rho)\,,
\end{equation}
and the exponent in Eq.~\ref{eq:bayes_factor} is
\begin{equation}
(s|h) - \frac{1}{2}(h|h) = \rho^2 - \frac{\rho^2}{2} +\mathcal{O}(\rho) = \frac{\rho^2}{2} +\mathcal{O}(\rho)\,,
\end{equation}
where the $\mathcal{O}(\rho)$ noise fluctuation is subleading with respect to the $\mathcal{O}(\rho^2)$ signal term for $\rho\gg 1$. The integrand in Eq.~\ref{eq:bayes_factor} is then sharply peaked around this template, and can be evaluated by a saddle-point (Laplace) approximation (which contributes corrections from the prior volume and from the Fisher determinant at the peak). The result is
\begin{equation}
\label{eq:LambdaStrong}
\Lambda \approx \exp\!\left(-\frac{\rho^2}{2}\right)\qquad (\text{signal  present})\,,
\end{equation}
i.e. when a GW signal  is present in the data, the Bayes factor is  exponentially small in the signal-to-noise ratio, and the signal hypothesis is exponentially preferred over the noise-only one.

\medskip

\noindent{\it Weak-signal / noise regime.} Consider now the opposite case, in which the data contain {\it no} GW signal, $s=n$. In this case
the exponent in Eq.~\ref{eq:bayes_factor} is
\begin{equation}
(s|h) - \frac{1}{2}(h|h) = 
(n|h)- \frac{1}{2}(h|h)\approx- \frac{\rho^2}{2}
\end{equation}
because the term $\langle(n|h)\rangle=0$. Therefore, 
\begin{equation}
\label{eq:LambdaWeak}
\Lambda \approx \exp\!\left(+\frac{\rho^2}{2}\right)\qquad (\text{no signal})\,,
\end{equation}
i.e. a stretch of pure noise data, when tested against a signal template of SNR $\rho$, returns a Bayes factor that is exponentially large in the signal-to-noise ratio and {\it favors} the noise-only hypothesis.

\medskip

Comparing with the false-alarm probability~\ref{eq:pFA}, we see that $|\log\Lambda|$ and $-\log p_{\rm FA}$ both scale as $\rho^2/2$ at large SNR: a larger signal-to-noise ratio simultaneously implies a smaller probability that the event is due to a noise fluctuation {\it and} an exponentially larger  Bayesian preference for
the signal hypothesis. The frequentist and Bayesian pictures are, in this sense,  quantitatively consistent.

\subsection{The signal-to-noise ratio for inspiraling binaries}
Equation~\ref{snr_opt} allows one to compute the optimal signal-to-noise ratio, if the frequency domain signal is known. In the special case of inspiraling binaries, the time domain signal is given, at the lowest PN order, by Eq.~\ref{hTT}. The evolution of the gravitational wave frequency $f_{\rm gw}$ is instead described by Eq.~\ref{fdot}, which yields
\begin{equation}\label{fgw}
    f_{\rm gw}=\frac{1}{\pi} M_c^{-5/8}\left(\frac{5}{256 (t_c-t)}\right)^{3/8}\,,
\end{equation}
where we have identified  the integration constant $t_c$
with the 
 the coalescence time. 
 This identification is justified because 
it yields a frequency diverging at $t=t_c$.  
 This divergence is not physical, but simply signals that the quadrupole formula, used to derive Eq.~\ref{fdot}, breaks down. It is  natural to assume that this breakdown happens  at the merger, since the assumption of a binary system, implicit in the quadrupole formula, ceases to apply there.
 
As we have seen in section~\ref{sec:det}, the detector is sensitive to a linear combination
of $h_+$ and $h_\times$, modulated by the transfer function. Neglecting for the moment  
the transfer function -- because its effect is 
small for LIGO/Virgo, while for LISA it is usually
included in the power spectral density of the noise (c.f. Fig.~\ref{fig:lisapsd}) --
we can write the detector's response, using
Eq.~\ref{hTT}, as
\begin{equation}\label{h(t)}
    h(t)=F_+ h_++ F_\times h_\times=\frac{4 M_c^{5/3} [ \pi f_{\rm gw}(t)]^{2/3}}{D} \left[F_+ \frac{1+\cos^2\iota}{2} \cos\Phi(t) +\cos\iota F_\times \sin\Phi(t)\right]\, 
\end{equation}
where $\Phi(t)=\int_{t_0}^t \mathrm{d} t' 2 \pi f_{\rm gw}(t')$.

Note that although we derived Eq.~\ref{h(t)}
in flat space, it is possible to show (see e.g. Ref.~\cite{maggiore}) that it is valid also in a Robertson-Walker spacetime, provided that the frequency $f_{\rm gw}$ is the one measured at the detector [where it is redshifted by a factor $1/(1+z)$ relative to the frequency at the source], the masses are  redshifted, i.e. $M_c\to M_c (1+z)$,
and $D$ is interpreted as the luminosity distance $D_L$.

To perform the Fourier transform analytically, one can use the stationary phase approximation. 
Considering first the $h_+$ polarization, 
let us rewrite Eq.~\ref{h(t)} as
\begin{equation}
   h(t) =\frac{A(t)}{2} \left[e^{i \Phi(t)}+e^{-i \Phi(t)}\right]
\end{equation}
by introducing an amplitude $A(t)$, the Fourier transform is 
\begin{equation}\label{ht3}
\tilde{h}(f)=\int \mathrm{d}t\, \frac{A(t)}{2} e^{2\pi i  f t} \left[e^{i \Phi(t)}+e^{-i \Phi(t)}\right] \,.
\end{equation}
The main contribution to this integral comes from the regions the phase does not change rapidly (the contribution of the other regions is negligible because the fast oscillations average to zero).
Note that we are only interested in computing $\tilde{h}(f)$ for $f>0$ since $\tilde{h}(-f)=\tilde{h}^*(f)$ [because $h(t)$ is real]. Then, because $\dot{\Phi}=f_{\rm gw}>0$, the first term in square brackets in Eq.~\ref{ht3}
has no stationary points, and we can thus neglect it. As for the second term, let us denote by $t_*$ the time at which the phase is stationary, i.e. $\dot{\Phi}(t_*)=2 \pi f$. 
Taylor expanding near this stationary point, one obtains
\begin{align}
&\tilde{h}(f)\approx \frac12 A(t_*)e^{i[2 \pi  f t_*-\Phi(t_*)]}\int \mathrm{d}t\,  e^{-i\ddot{\Phi}(t_*) (t-t_*)^2/2}=\frac12A(t_*)e^{i[2 \pi  f t_*-\Phi(t_*)]}\sqrt{\frac{2}{\ddot{\Phi}(t_*)}}\int_{-\infty}^{\infty} \mathrm{d}x\, e^{-i \,x^2}\nonumber\\
&\qquad=\frac12 A(t_*)e^{i[2 \pi  f t_*-\Phi(t_*)-\pi/4]}\sqrt{\frac{2\pi}{\ddot{\Phi}(t_*)}}\,.\label{ht4}
\end{align}
One can then express $t_*$ as a function of $f$ by solving $f=\dot\Phi(t_*)/(2\pi)=f_{\rm gw}(t_*)$. Using Eq.~\ref{fgw} one then has
\begin{equation}
    t_c-t_*=\frac{5}{256}M_c^{-5/3} (\pi f)^{-8/3}\,,\label{tstar}
\end{equation}
which can be used into Eq.~\ref{ht4}
to obtain the amplitude and phase as functions of frequency. Doing the same calculation for $h_\times$ yields the same result, but with an additional factor $i$ (i.e. a $\pi/2$ phase difference). The final result then reads
\begin{gather}\label{hf}
    \tilde{h}(f)=\frac{1}{D_L}\left(\frac{5}{24}\right)^{1/2} \pi^{-2/3} M_c^{5/6} f^{-7/6} Q e^{i \psi}\,,\\
    \psi=2 \pi f t_*-\Phi(t_*)-\pi/4
\end{gather}
with $t_*$ given by Eq.~\ref{tstar} and
\begin{equation}
    Q=F_+ \frac{1+\cos^2\iota}{2}  +i \cos\iota F_\times \,.\label{Q}
\end{equation}
Note that we have expressed this final result in terms of $D_L$, and thus $f$ and $M_c$ must be interpreted as the detector-frame frequency and the redshifted chirp mass, respectively.

One can then compute the signal-to-noise ratio using Eq.~\ref{snr_opt} as
\begin{equation}
    \left(\frac{S}{N}\right)^2=\frac{5}{6} \frac{1}{\pi^{4 / 3}} \frac{1}{D_L^2}\left(\frac{G M_c}{c^3}\right)^{5 / 3}|Q|^2 \int_0^{f_{\rm max}} d f \frac{f^{-7 / 3}}{S(f)}\,,
\end{equation}
where for clarity we have reinstated $G$ and $c$. The factor $Q$ depends on the sky position and orientation of the source (c.f. Eq.~\ref{Q}). For   sources with optimal sky position and inclination ($\iota=\theta=0$) one has $Q=1$, while an average over sky position and inclination (assuming isotropic distributions) yields $\langle Q^2\rangle=4/25$. Signal-to-noise ratios allowing detection are typically $S/N\gtrsim 8$.
\\
\\
{\bf Exercise 6: }{\it Using the stationary phase approximation, show that the signal-to-noise ratio of a quasi-monochromatic signal 
\begin{equation}
h(t)=F_+ h_+(t)+F_\times h_\times(t)\equiv \sqrt{2} A \cos\left[2 \pi \left(f_0+\frac{\dot{f_0}t}{2}\right) t+\phi_0\right]
\end{equation}
is given by
\begin{equation}
        \left(\frac{S}{N}\right)^2=\frac{2 A^2 T}{S(f_0)}\,,
\end{equation}
where $T$ is the observation time. [Hint: the maximum integration frequency depends on the observation time.]}

\subsection{Parameter estimation}

Let us now suppose that a gravitational wave signal has been detected in the data $s$,
and let us examine whether the parameters $\boldsymbol{\theta}$ of the source can be inferred.
In principle, the probability distribution function for the parameters conditional on the data, i.e. the ``posterior distribution'' $P(\boldsymbol{\theta}|s)$, is given by Bayes theorem in terms of the
likelihood $P(s|\boldsymbol{\theta})$ and the ``prior distribution'' $P(\boldsymbol{\theta})$:
\begin{equation}\label{bayes}
    P(\boldsymbol{\theta}|s)\propto P(s|\boldsymbol{\theta}) P(\boldsymbol{\theta})\,.
\end{equation}
Here, the prior distribution encodes the knowledge on the parameters before data are taken (e.g. it can be taken to be uniform if one wants to remain agnostic). As for the likelihood, i.e. the probability to obtain the data $s$ given the waveform parameters $\boldsymbol{\theta}$, it can be computed from Eq.~\ref{pnoise} by replacing $n=s-h$ as per Eq.~\ref{s=h+n}:
\begin{equation}
P(s|\boldsymbol{\theta})\propto \exp\left[-\frac12 (s-h_{\boldsymbol{\theta}}|s-h_{\boldsymbol{\theta}})\right]\propto \exp\left[-\frac12 (h_{\boldsymbol{\theta}}|h_{\boldsymbol{\theta}})+(h_{\boldsymbol{\theta}}|s)\right]\,,
\end{equation}
where $h_{\boldsymbol{\theta}}$ is the template.

Given the large dimensionality of the parameter space (up to 15 parameters for quasi-circular binaries), the posteriors predicted by Eq.~\ref{bayes}
cannot be plotted or sampled by brute force. A more sophisticated approach based on Markov-Chain-Monte-Carlo or nested sampling techniques is needed, which is outside the scope of these simple notes. It is possible, however, to expand the posteriors around their maximum in a Taylor series, obtaining the Fisher matrix of the system.
In more detail, let consider the case of uninformative (uniform) priors, $P(\boldsymbol{\theta})=$\, const, and assume that the posterior distribution peaks at $\boldsymbol{\theta}=\boldsymbol{\bar{\theta}}$. Imposing $\partial P(s|\boldsymbol{\theta})/\partial \theta^i=0$ at the peak, one then gets
\begin{equation}
    \left(\frac{\partial h_{\boldsymbol{\theta}}}{\partial \theta^i}(\boldsymbol{\bar{\theta}})\Bigg\vert s\right)-\left(\frac{\partial h_{\boldsymbol{\theta}}}{\partial \theta^i}(\boldsymbol{\bar{\theta}})\Bigg\vert h_{\boldsymbol{\theta}}(\boldsymbol{\bar{\theta}})\right)=0\,.
\end{equation}
Expanding now the posterior distribution at quadratic order near the peak and assuming a high signal-to-noise ratio [which allows for approximating $s\approx h_{\boldsymbol{\theta}}({\boldsymbol{\bar{\theta}}}$)], one obtains
\begin{equation}
    P(\boldsymbol{\theta}|s)\propto \exp\left[-\frac12 \Gamma_{ij} (\theta^i-\bar{\theta}^i)(\theta^j-\bar{\theta}^j)\right]\,,
\end{equation}
where we have introduced the Fisher matrix
\begin{equation}
    \Gamma_{ij}= \left(\frac{\partial h_{\boldsymbol{\theta}}}{\partial \theta^i}(\boldsymbol{\bar{\theta}})\Bigg\vert \frac{\partial h_{\boldsymbol{\theta}}}{\partial \theta^j}(\boldsymbol{\bar{\theta}})\right)\,.
\end{equation}
In the high signal-to-noise ratio limit,
 the posterior distribution
near the peak is therefore 
described by a multivariate Gaussian with
 covariance matrix 
\begin{equation}
    C_{ij}= (\Gamma^{-1})_{ij}\,.
\end{equation}

While not useful to analyze real data, the Fisher matrix formalism can be used to obtain predictions about how well the parameters are measurable by a given detector. This information is encoded in the covariance matrix. Note that eliminating the row and column corresponding to a parameter in the covariance matrix amounts to marginalizing over that parameter. Instead, eliminating the row and column corresponding to a parameter in the Fisher matrix corresponds to assuming that parameter is known exactly.
\\
\\
{\bf Exercise 7: }{\it Consider a quasicircular system of two black holes with masses $m_{1,2}$ of $30$ and $35 M_\odot$ at 400 {\rm Mpc} luminosity distance. Compute the signal-to-noise ratio for a detector with power spectral density
$$S=  10^{-49} [x^{-4.14} - 5 x^{-2} + 111 (1 - x^2 + x^4/2)/(1 + x^2/2)] {\rm Hz}^{-1},$$ with  $x = f/(215 \,{\rm Hz})$, tapering the inspiral waveform
of Eq.~\ref{hf} at the ISCO of a particle with mass $\mu=m_1 m_2/(m_1+m_2)$ around a Schwarzschild black hole of mass $M=m_1+m_2$. Assume optimal inclination and sky position, and no spins.
}
\\\\
{\bf Exercise 8: }{\it For the quasi-monochromatic signal of  Exercise 6, assume $f_0=80$ {\rm Hz}, $\dot{f}_0$ from Eq.~\ref{fdot} and compute the amplitude $A$ that gives $S/N=20$ for an observation time of one year and the detector of Exercise 7. For such a source, compute the covariance matrix of the parameters ($A$, $f_0$ and $\dot{f}_0$).
}

\section{Stochastic gravitational-wave backgrounds}
\label{sec:sgwb}

So far, we have focused on the gravitational-wave signals produced by individual,  well-modeled sources, such as the inspiral, merger and ringdown of compact binaries.
In practice, however, gravitational-wave detectors also record a second, conceptually distinct class of signals: {\it stochastic gravitational-wave backgrounds} (SGWBs), i.e. the incoherent superposition of waves originating from many individually unresolved sources~\cite{maggiore,Christensen:2018iqi}. Such backgrounds may have an astrophysical origin, e.g. the combined emission from the cosmological population of supermassive black hole binaries, from galactic white dwarf binaries, or from unresolved stellar-mass compact binary coalescences; alternatively, they may have a cosmological origin, e.g. primordial vacuum fluctuations amplified during inflation, first-order phase transitions in the early Universe, or  cosmic strings and other topological defects.
While a cosmological SGWB has not yet been observed, a background at nHz frequencies, most likely produced by the cosmological population of inspiraling supermassive black hole binaries, was recently reported by several pulsar timing array (PTA) experiments~\cite{nanograv15,epta15,ppta15,cpta15}.

Unlike deterministic signals, a stochastic background is a random process and can only be characterized statistically, through the correlators of the metric perturbation. In this section, we first discuss how an (isotropic) SGWB is parametrized, following~\cite{maggiore}; we then write down the likelihood for its detection with a pair of cross-correlated interferometers~\cite{CornishRomano}; and we  finally derive the celebrated Hellings--Downs correlation~\cite{HellingsDowns} -- the characteristic angular fingerprint of an isotropic SGWB on a pulsar timing array -- building on the frequency-shift formulae that we had already derived in section~\ref{subsec:7_2}. We conclude  by sketching the Bayesian  likelihood actually used to analyze PTA data~\cite{vanHaasteren09,vanHaasterenVallisneri,nanograv15,SmarraThesis}.

\subsection{Characterization of stochastic backgrounds}
\label{sec:sgwb_char}

Let us start by generalizing the plane-wave decomposition of a vacuum metric perturbation (cf.  section~\ref{flat_prop}) to  a superposition of waves of arbitrary frequency and propagation direction. In the transverse-traceless gauge, we can write
\begin{equation}
\label{eq:planewaveSGWB}
h_{ij}(t,\vec{x}) = \sum_{A=+,\times} \int_{-\infty}^{+\infty}\!\mathrm{d}f\int \mathrm{d}^2\hat{n}\, \tilde{h}_A(f,\hat{n})\, e^A_{ij}(\hat{n})\, e^{-2\pi i f(t-\hat{n}\cdot\vec{x})}\,,
\end{equation}
where $\hat{n}$ is the direction of propagation, while the two transverse-traceless polarization tensors $e^A_{ij}(\hat n)$ can be built from any right-handed orthonormal triad $(\hat{u},\hat{v},\hat{n})$ as
\begin{equation}
\label{eq:epol_def}
e^+_{ij}(\hat{n}) = \hat{u}_i\hat{u}_j -\hat{v}_i\hat{v}_j\,,\qquad e^\times_{ij}(\hat{n}) = \hat{u}_i\hat{v}_j + \hat{v}_i\hat{u}_j\,,
\end{equation}
which generalizes to arbitrary $\hat{n}$ the special case ($\hat{n}=\hat{z}$) worked out in section~\ref{flat_prop}. The  normalization in Eq.~\ref{eq:epol_def} implies the sum rule $\sum_{A} e^A_{ij}(\hat{n})\, e^{A\, ij}(\hat{n})=4$, which we will use often in the following. The reality of $h_{ij}$ furthermore requires  $\tilde{h}_A(-f,\hat{n})=\tilde{h}_A^*(f,\hat{n})$. For a stochastic background, the Fourier amplitudes $\tilde{h}_A(f,\hat{n})$ are not deterministic, but are instead random variables drawn from an underlying ensemble, whose statistical properties are characterized  by the ensemble averages $\langle \ldots \rangle$ over  all possible realizations.~\footnote{In practice, only a single realization of the Universe is  accessible. By the ergodic theorem, the ensemble average can then be replaced by an average over sufficiently long time windows.}

Following~\cite{maggiore}, we will  make the following four working assumptions on the statistical properties of the background, which are standard in the SGWB literature:
\begin{itemize}
\item {\it Stationarity}: the statistics of the Fourier amplitudes is invariant under time translations, so that any 2-point or higher-point correlator depends only on differences of the time arguments. In Fourier space, this requirement forces the 2-point correlator $\langle\tilde{h}^*_A(f,\hat{n})\tilde{h}_{A'}(f',\hat{n}')\rangle$ to be proportional to $\delta(f-f')$. The assumption is  well-justified: the time-scale on which any realistic stochastic  background can evolve -- set by the expansion of the Universe for a cosmological background, or by the characteristic time-scale of the underlying source population for an astrophysical one -- is in both cases far longer than any realistic observing run.

\item {\it Gaussianity}: the real and imaginary parts of the Fourier amplitudes $\tilde{h}_A(f,\hat{n})$ are Gaussian random variables; equivalently, the full statistics of the background is  determined by its 2-point correlator alone, with all higher-order correlators following from the latter. This  is a consequence of the central limit theorem whenever the background is generated by the incoherent superposition of a large number of individually weak, independent contributions, as  generically happens for  cosmological backgrounds or for  dense astrophysical populations. It may  instead fail  if only a few ``loud'' sources contribute significantly to each frequency bin, in which case  non-Gaussian features remain.

\item {\it Isotropy}: the background's energy distributes uniformly over the celestial sphere. Up to an overall normalization, this implies that the 2-point correlator depends on the propagation directions $\hat{n}$ and $\hat{n}'$ only through their relative angle, and vanishes for distinct directions, $\langle\tilde{h}^*_A(f,\hat{n})\tilde{h}_{A'}(f',\hat{n}')\rangle\propto \delta^2(\hat{n},\hat{n}')\equiv \delta(\cos\theta-\cos\theta')\delta(\phi-\phi')$, with a proportionality constant that is independent of $\hat{n}$. This is a natural working  assumption for  cosmological backgrounds, by analogy with the near-isotropy of the CMB; it clearly breaks down  for astrophysical backgrounds with significant spatial clustering (e.g. the galactic population of unresolved white-dwarf binaries, which traces the galactic plane).

\item {\it No polarization preference}: the background does not select any privileged linear combination of the two GW polarizations. The 2-point correlator is therefore  diagonal in the polarization labels, $\langle\tilde{h}^*_A\tilde{h}_{A'}\rangle\propto \delta_{AA'}$. This is again a natural starting point both in a cosmological context, and for astrophysical backgrounds obtained by combining many sources with random orientations relative to the observer.
\end{itemize}

Under these assumptions, the two-point correlator is fully characterized by a single function of frequency, the (single-sided) spectral density $S_h(f)$, through
\begin{equation}
\label{eq:Sh_def}
\langle \tilde{h}^*_A(f,\hat{n})\, \tilde{h}_{A'}(f',\hat{n}')\rangle = \delta(f-f')\, \frac{\delta^2(\hat{n},\hat{n}')}{4\pi}\, \delta_{AA'}\, \frac{1}{2}\,S_h(f)\,.
\end{equation}
Inserting Eq.~\ref{eq:planewaveSGWB} into  $\langle h_{ij} h^{ij}\rangle$ and using Eq.~\ref{eq:Sh_def}, one finds
\begin{equation}
\label{eq:hijhij_Sh}
\langle h_{ij}(t,\vec{x}) h^{ij}(t,\vec{x})\rangle = 4\int_0^\infty \mathrm{d}f\, S_h(f)\,.
\end{equation}

The 2-point correlator~\ref{eq:Sh_def} provides a complete statistical characterization of the background, but it is still a somewhat abstract quantity. At the level of physical intuition, it is more convenient to translate $S_h(f)$ into the energy density that the background carries, which can be computed  from the stress-energy tensor~\ref{SETgw2} of a linear  TT perturbation. Its $tt$-component reads
\begin{equation}
\rho_{\rm gw} = \frac{1}{32\pi} \langle \dot{h}^{\rm TT}_{ij}\dot{h}_{\rm TT}^{ij}\rangle\,,
\end{equation}
and proceeding as above  one can express  the right-hand side in terms of $S_h(f)$. A short calculation yields
\begin{equation}
\label{eq:rhogw_integral}
\rho_{\rm gw} = \frac{\pi}{2}\int_0^\infty \mathrm{d}f\, f^2\, S_h(f)\,.
\end{equation}
It is then natural to normalize the energy density to the critical density of the Universe,
\begin{equation}
\rho_c = \frac{3 H_0^2}{8\pi}\,,
\end{equation}
with $H_0$ the present-day Hubble expansion rate, and to define the dimensionless spectral energy density
\begin{equation}
\label{eq:Omgw_def}
\Omega_{\rm gw}(f) \equiv \frac{1}{\rho_c} \frac{\mathrm{d}\rho_{\rm gw}}{\mathrm{d}\log f}\,,
\end{equation}
which represents the fraction of the critical density contributed by  gravitational waves per logarithmic frequency interval.  Combining Eqs.~\ref{eq:rhogw_integral} and~\ref{eq:Omgw_def}, one obtains the useful relation
\begin{equation}
\label{eq:Omgw_Sh}
\Omega_{\rm gw}(f) = \frac{4\pi^2}{3 H_0^2}\, f^3\, S_h(f)\,.
\end{equation}
In the following, $\Omega_{\rm gw}(f)$ (or, equivalently, $S_h(f)$) will be our target for detection and characterization. Different cosmological and astrophysical mechanisms  predict different  shapes of $\Omega_{\rm gw}(f)$: a quasi-scale-invariant spectrum ($\Omega_{\rm gw}\sim$ const, up to  small deviations produced by the slow-roll parameters) is expected, e.g., for an inflationary background; a population of quasi-circular, Newtonian inspiraling supermassive black hole binaries gives instead $\Omega_{\rm gw}\propto f^{2/3}$~\cite{Phinney01}; first-order cosmological phase transitions typically produce broken  power laws peaking at a frequency set by the temperature of the transition; and so on.
Let us also note that, because gravitational waves interact extremely weakly with matter after their production, all cosmological SGWBs travel essentially unimpeded  to us, and carry  information about the very epochs at which they were generated. By contrast, the  early Universe is opaque to electromagnetic radiation up to the  epoch of CMB decoupling. The detection of a cosmological SGWB would therefore open a qualitatively new window on the very early Universe, well beyond what electromagnetic probes can reach.
\\
\\
{\bf Exercise 9:} {\it Starting from Eq.~\ref{eq:planewaveSGWB} and using the 2-point function~\ref{eq:Sh_def}, derive Eqs.~\ref{eq:hijhij_Sh} and~\ref{eq:rhogw_integral} explicitly. [Hint: you will need $\sum_A e^A_{ij}(\hat{n}) e^{A\,ij}(\hat{n}) = 4$.]}

\subsection{Cross-correlation likelihood for two detectors}
\label{sec:sgwb_2det}

Let us now consider the problem of detecting a SGWB in the data of a pair of gravitational-wave detectors, such as LIGO Hanford and LIGO Livingston. The key physical idea  is that, while the instrumental noise in the two detectors is to excellent approximation uncorrelated (its sources -- seismic, thermal, quantum, etc. -- being local to each  instrument), both detectors see the {\it same} gravitational-wave sky, so a SGWB shows up as a {\it correlated} component in their outputs. Cross-correlating the two data streams  thus averages the instrumental noise  away while building up coherently on the SGWB, and is the fundamental tool on which  SGWB searches with a network of  detectors  are based.

Each detector measures the (low-frequency) response
\begin{equation}
h_k(t) = D_k^{ij}\, h^{\rm TT}_{ij}(t,\vec{x}_k)\,,
\end{equation}
where $D_k^{ij}$ and $\vec{x}_k$ are the detector tensor and the location of the $k$-th detector, respectively (cf. Eq.~\ref{lowf_resp}). Using the plane-wave expansion~\ref{eq:planewaveSGWB}, we can write
\begin{equation}
\label{eq:hk_sgwb}
h_k(t) = \sum_{A}\int\mathrm{d}f\int \mathrm{d}^2\hat{n}\, F^A_k(\hat{n})\,\tilde{h}_A(f,\hat{n})\, e^{-2\pi i f(t - \hat{n}\cdot\vec{x}_k)}\,,
\end{equation}
where $F^A_k(\hat{n})= D_k^{ij}e^A_{ij}(\hat{n})$ are the pattern functions  defined in section~\ref{subsec:7_1}. Because $\tilde{h}_A(f,\hat{n})$ is a Gaussian random variable with 2-point function~\ref{eq:Sh_def}, so is the detector response $h_k(t)$. A short calculation gives its Fourier-space 2-point function as
\begin{equation}
\label{eq:hkhl_2pt}
\langle \tilde{h}^*_k(f)\tilde{h}_l(f')\rangle = \delta(f-f')\, \frac{1}{2}\, \gamma_{kl}(f)\, S_h(f)\,,
\end{equation}
with the {\it overlap reduction function}~\footnote{Some references (e.g.~\cite{maggiore}) include an additional average  $\int \mathrm{d}\psi/(2\pi)$ over the polarization angle $\psi$ that rotates the basis $(\hat{u},\hat{v})$ in the plane orthogonal to $\hat{n}$. In our derivation this is not necessary: under such a rotation, the pattern functions $F^A_k=D_k^{ij}e^A_{ij}$ transform as the components of an $SO(2)$ vector in $(+,\times)$ space (rotating by the double angle $2\psi$, because of the spin-2 nature of the GW polarization), and $\sum_A F^A_k(\hat{n}) F^A_l(\hat{n})$ is their Euclidean scalar product, hence invariant. The $\psi$ average therefore gives trivially 1.}
\begin{equation}
\label{eq:orf}
\gamma_{kl}(f) = \int\frac{\mathrm{d}^2\hat{n}}{4\pi}\sum_A F^A_k(\hat{n}) F^A_l(\hat{n})\, e^{2\pi i f \hat{n}\cdot(\vec{x}_l - \vec{x}_k)}\,.
\end{equation}
The function $\gamma_{kl}(f)$ measures to what extent two detector responses remain coherent when driven by the same isotropic stochastic signal. 
In the idealized situation of two co-located and co-aligned interferometers, $\gamma_{kl}(f)$ reduces to an angular efficiency factor that only depends on the detector geometry (and equals $2/5$ at low frequencies for a 90-degree-arm interferometer, see e.g.~\cite{maggiore}). As soon as the two instruments are separated by a distance comparable to (or larger than) the signal wavelength, $f\,|\vec{x}_l-\vec{x}_k|\gtrsim 1$, the exponential in Eq.~\ref{eq:orf} oscillates many times as $\hat{n}$ sweeps the sphere, and the integral averages out: the two detectors then see essentially unrelated stochastic GW signals. The overlap is also reduced if the detector arms are misaligned with one another, as different orientations weight the two polarizations differently; see e.g.~\cite{AllenRomano,maggiore} for explicit computations of $\gamma_{kl}(f)$ for various real detector pairs.

The output of the $k$-th detector is the sum of the stochastic GW response and the instrumental noise,
\begin{equation}\label{eq:sk_sgwb}
s_k(t) = h_k(t) + n_k(t)\,,
\end{equation}
with $n_k(t)$ assumed to be  Gaussian, stationary, and uncorrelated across detectors, and characterized by a single-sided PSD $S_{n,k}(f)$ (cf. Eq.~\ref{psd}). Our task is  to build the likelihood  of the data $\{s_k\}$ conditional on a given  SGWB spectrum $S_h(f)$. To this purpose, we will follow the hierarchical Bayesian derivation of~\cite{CornishRomano}, which treats the specific realization $\{h_k\}$ of the stochastic signal as a nuisance parameter to be integrated out.

Let us start from the  likelihood of the observed data {\it conditional} on the specific realization of the stochastic signal.  For a given realization $\{\tilde{h}_k(f)\}$, the residual $\tilde{r}_k(f)\equiv\tilde{s}_k(f)-\tilde{h}_k(f)$ must be a realization of the instrumental noise alone.   The  multi-detector generalization of Eq.~\ref{pnoise}, using the fact that the noise in different detectors is uncorrelated, then reads
\begin{equation}
\label{eq:pnoise_SGWB}
p(\{s_k\}\,|\,\{h_k\}, \{S_{n,k}\}) \propto \prod_i \exp\left[-2\Delta f\,\tilde{r}^*_k(f_i)\,[\mathcal{C}'(f_i)]^{-1}_{kl}\,\tilde{r}_l(f_i)\right]\,,
\end{equation}
where the product runs over positive frequencies $f_i = i/T$ (with $T$ the observation time and $\Delta f = 1/T$), and the  noise covariance  is diagonal in the detector indices, $\mathcal{C}'_{kl}(f)= \delta_{kl}\,S_{n,k}(f)$.

Next, we note that the Gaussian prior on the realization of the stochastic signal -- with the hyperparameter $S_h(f)$ held fixed --  is  dictated by the statistics of Eq.~\ref{eq:Sh_def} and takes the same form in the detector indices,
\begin{equation}
\label{eq:signal_prior_SGWB}
\pi(\{h_k\}\,|\,S_h) \propto \prod_i \exp\left[-2\Delta f\,\tilde{h}^*_k(f_i)\,[\mathcal{K}(f_i)]^{-1}_{kl}\,\tilde{h}_l(f_i)\right]\,,
\end{equation}
with the (generically non-diagonal) signal covariance  $\mathcal{K}_{kl}(f) = \gamma_{kl}(f)\,S_h(f)$, as follows from Eq.~\ref{eq:hkhl_2pt}.

The marginalized likelihood -- i.e. the likelihood of the data given the hyperparameters $S_h(f),\{S_{n,k}(f)\}$, independently of the particular realization of the stochastic signal -- is obtained by integrating over the real and imaginary parts of each Fourier mode of the signal,
\begin{equation}
\label{eq:marg_int}
p(\{s_k\}\,|\,S_h, \{S_{n,k}\}) = \int \prod_{k,i} \mathrm{d}[\mathrm{Re}\,\tilde{h}_k(f_i)]\,\mathrm{d}[\mathrm{Im}\,\tilde{h}_k(f_i)]\, p(\{s_k\}\,|\,\{h_k\}, \{S_{n,k}\})\,\pi(\{h_k\}\,|\,S_h)\,.
\end{equation}
The integrand is Gaussian in the real and imaginary parts of  $\tilde{h}_k(f_i)$, and the integral can be performed frequency by frequency by completing the square in the exponential. After a standard (but somewhat tedious) manipulation~\cite{CornishRomano}, one obtains the compact result
\begin{equation}
\label{eq:likelihood_2det}
p(\{s_k\}\,|\,S_h, \{S_{n,k}\}) \propto \prod_i \exp\left[-2\Delta f\,\tilde{s}^*_k(f_i)\,[\mathcal{C}(f_i)]^{-1}_{kl}\,\tilde{s}_l(f_i)\right]\,,
\end{equation}
with the combined covariance
\begin{equation}
\label{eq:C_SGWB_2det}
\mathcal{C}_{kl}(f) = \delta_{kl}\,S_{n,k}(f) + \gamma_{kl}(f)\,S_h(f)\,.
\end{equation}
Equation~\ref{eq:C_SGWB_2det} is the central result of this subsection: {\it the effect of a SGWB on a pair of cross-correlated detectors is to add a coherent off-diagonal component to the data covariance, proportional to the common spectral density $S_h(f)$ and modulated by the overlap reduction function $\gamma_{kl}(f)$}. The full likelihood~\ref{eq:likelihood_2det}--\ref{eq:C_SGWB_2det} can then be maximized (to obtain a point estimate of $S_h(f)$) or  sampled via MCMC techniques (to obtain posterior distributions over the parameters of some model, such as a power law  $\Omega_{\rm gw}(f)=\Omega_*(f/f_*)^\alpha$).
\\
\\
{\bf Exercise 10:} {\it Starting from Eqs.~\ref{eq:pnoise_SGWB}--\ref{eq:signal_prior_SGWB},   perform the Gaussian integral~\ref{eq:marg_int} explicitly  by completing the square in $\tilde{h}_k(f)$, and recover Eqs.~\ref{eq:likelihood_2det}--\ref{eq:C_SGWB_2det}. }

\subsection{Pulsar timing arrays and the Hellings--Downs correlation}
\label{sec:HD}

At frequencies much below those accessible to ground-based and space-based interferometers, say between $\sim 10^{-9}$ and $\sim 10^{-7}$ Hz, gravitational waves are detected by exploiting the natural clocks provided by millisecond pulsars. Millisecond pulsars are  extremely stable rotators: over decade-long observing campaigns, the pulse arrival times of the best millisecond pulsars can be measured  with  residuals of order $\sim 100$ ns, and a gravitational wave crossing the line of sight to the pulsar perturbs these apparent arrival times in a characteristic way. A {\it pulsar timing array} (PTA) monitors many  tens of millisecond pulsars for a long baseline (typically 10--20 years) and exploits the spatially correlated pattern of their timing residuals to detect gravitational waves. The operating PTAs (NANOGrav, EPTA+InPTA, PPTA and CPTA) routinely combine their data in the International Pulsar Timing Array~\cite{nanograv15,epta15,ppta15,cpta15,ipta}.

The fundamental observable in a PTA is the (scalar) frequency shift of the radio pulses emitted by a pulsar as measured by an Earth-based observer. As we have already computed this shift in section~\ref{subsec:7_2} in the context of the transfer function of interferometers, we can readily re-use our results. In particular, the one-way frequency shift of a photon traveling between two points in a plane-wave gravitational-wave background is given by Eqs.~\ref{nu1}--\ref{nu2}. Since radio waves propagate along null geodesics just as laser photons do, we can apply these equations directly to the trajectory of a radio pulse traveling from a pulsar to the Earth.

More precisely, let us identify the pulse propagation direction introduced in section~\ref{subsec:7_2} with $-\hat{p}_a$, where $\hat{p}_a$ is the unit vector pointing from the observer (at the solar-system barycenter, which we take  as the origin) to pulsar $a$. Letting $\theta_a$ denote the angle between $-\hat{p}_a$ and the direction of propagation $\hat{k}$ of the gravitational wave (so that $\cos\theta_a = -\hat{k}\cdot\hat{p}_a$), Eq.~\ref{nu1}, which describes the change in the pulse frequency between emission at the pulsar and reception on Earth, can then be rewritten as~\cite{maggiore_vol2}
\begin{equation}
\label{eq:PTA_redshift}
z_a(t) \equiv \frac{\nu_{\rm emit} - \nu_{\rm obs}}{\nu_{\rm obs}} = \frac{1}{2}\,\frac{\hat{p}^i_a \hat{p}^j_a}{1+\hat{k}\cdot\hat{p}_a}\,\Big[h_{ij}(t,\vec{x}=0) - h_{ij}(t-\tau_a, \vec{x}_a)\Big]\,,
\end{equation}
where $\vec{x}_a=\tau_a\,\hat{p}_a$ (in $c=1$ units) is the location of the pulsar and $\tau_a\sim {\rm kpc}/c$ is the light travel time from the pulsar to the observer.  Equation~\ref{eq:PTA_redshift} is the fundamental equation of pulsar timing detection of gravitational waves. It shows that the instantaneous redshift of the pulses carries {\it two} ``echoes'' of the gravitational wave: an {\it Earth term}, $h_{ij}(t,\vec{x}=0)$, evaluated at the observer's location at the time of observation; and a {\it pulsar term}, $h_{ij}(t-\tau_a,\vec{x}_a)$, evaluated at the location of the pulsar at the retarded time $t-\tau_a$ at which the pulse was emitted.

The quantity actually reconstructed from the radio-telescope data is not the instantaneous redshift, but the time-integrated {\it timing residual}~\cite{maggiore_vol2}
\begin{equation}
\label{eq:timing_residual}
R_a(t) = \int_0^t \mathrm{d}t'\, z_a(t')\,,
\end{equation}
which represents the departure of the observed pulse arrival times from the prediction of a deterministic timing model (that fits the pulsar's spin-down, position, proper motion, binary orbit if any, and propagation effects). Typical timing residuals in a PTA are  of order $\sim 100$ ns, as anticipated above, and their  cross-correlations between pairs of pulsars are the smoking-gun signature of a gravitational wave background, as we  now  show.

Let us  insert the plane-wave expansion~\ref{eq:planewaveSGWB} into Eq.~\ref{eq:PTA_redshift}. After a short calculation, $z_a(t)$ can be written as~\cite{maggiore_vol2}
\begin{equation}
\label{eq:z_a_expansion}
z_a(t) = \sum_{A}\int\mathrm{d}f\int\mathrm{d}^2\hat{n}\,F^A_a(\hat{n})\,\tilde{h}_A(f,\hat{n})\, e^{-2\pi i f t}\,\Big[1 - e^{2\pi i f\tau_a(1+\hat{n}\cdot\hat{p}_a)}\Big]\,,
\end{equation}
where the pulsar ``pattern function'' reads
\begin{equation}
\label{eq:F_pulsar}
F^A_a(\hat{n}) = \frac{\hat{p}^i_a \hat{p}^j_a \,e^A_{ij}(\hat{n})}{2(1+\hat{n}\cdot\hat{p}_a)}\,,
\end{equation}
(recall that, in the notation of Eq.~\ref{eq:planewaveSGWB}, $\hat{n}$ is the GW propagation direction). The bracketed factor in Eq.~\ref{eq:z_a_expansion} encodes the ``Earth minus pulsar'' structure of Eq.~\ref{eq:PTA_redshift}.

Let us now consider two pulsars $a$ and $b$ and compute the cross-correlation $\langle z_a(t) z_b(t)\rangle$ under the SGWB assumption of the previous subsection. Using the  2-point function~\ref{eq:Sh_def} in Eq.~\ref{eq:z_a_expansion}, we obtain~\cite{maggiore_vol2}
\begin{equation}
\label{eq:zz_integral}
\langle z_a(t) z_b(t)\rangle = \frac{1}{2}\int_{-\infty}^{+\infty}\mathrm{d}f\,S_h(f)\int\frac{\mathrm{d}^2\hat{n}}{4\pi}\,K_{ab}(f,\hat{n})\,\sum_A F^A_a(\hat{n}) F^A_b(\hat{n})\,,
\end{equation}
with
\begin{equation}
K_{ab}(f,\hat{n}) = \Big[1-e^{-2\pi i f\tau_a(1+\hat{n}\cdot\hat{p}_a)}\Big]\Big[1-e^{2\pi i f\tau_b(1+\hat{n}\cdot\hat{p}_b)}\Big]\,.
\end{equation}
Each of the two square brackets in $K_{ab}$ contains an ``Earth'' term (the constant $1$) and a ``pulsar'' term  (the exponential), so the product in $K_{ab}$ generates four  contributions.  As discussed in~\cite{maggiore_vol2,Anholm09}, three of these four  contributions turn out to be  negligible for a realistic PTA. Indeed, PTAs observe GWs at frequencies $f\sim$ few$\times 10^{-9}$--$10^{-8}$ Hz, while the typical millisecond pulsars of a PTA are at distances $\tau_a\sim 0.4$--$3\,\text{kpc}/c$, so that the dimensionless product  $f\tau_a$ ranges from $\sim 10^2$ to $\sim 10^3$ and is therefore always large. The three remaining terms therefore oscillate rapidly as $f$ varies over the PTA band, and average out in the frequency integral of Eq.~\ref{eq:zz_integral}. Keeping only the ``Earth--Earth'' term, we can therefore replace, to very good accuracy,
\begin{equation}
K_{ab}(f,\hat{n}) \longrightarrow 1\,.
\end{equation}
In this approximation, the frequency and angular integrals in Eq.~\ref{eq:zz_integral} factorize, yielding
\begin{equation}
\label{eq:zz_factorized}
\langle z_a(t) z_b(t)\rangle = C(\theta_{ab})\int_0^\infty \mathrm{d}f\,S_h(f)\,,
\end{equation}
where $\theta_{ab}$ is the angular separation between the two pulsars as seen from Earth, and
\begin{equation}
\label{eq:HD_integral}
C(\theta_{ab}) \equiv \int\frac{\mathrm{d}^2\hat{n}}{4\pi}\sum_A F^A_a(\hat{n}) F^A_b(\hat{n})\,.
\end{equation}
By construction, $C(\theta_{ab})$ is the PTA analog of the overlap reduction function~\ref{eq:orf}, with the caveat that the travel-time factor has already been discarded in the limit $K_{ab}\to 1$, so that it depends only on the angular separation  between the two pulsars.

To close the calculation, we now have to evaluate the angular integral in Eq.~\ref{eq:HD_integral}. Parametrizing $\hat{n}$ in spherical coordinates, $\hat{n}(\theta,\phi)=(\sin\theta\cos\phi,\sin\theta\sin\phi,\cos\theta)$, a  natural choice for the right-handed orthonormal triad entering the polarization tensors~\ref{eq:epol_def} is
\begin{equation}
\hat{u} = (\sin\phi,-\cos\phi,0)\,,\qquad \hat{v} = (\cos\theta\cos\phi,\cos\theta\sin\phi,-\sin\theta)\,.
\end{equation}
With this choice, the integral in Eq.~\ref{eq:HD_integral} can be performed in closed form, although the algebra is rather long; a particularly clean way to organize it is to exploit  the residual rotation freedom in Eq.~\ref{eq:HD_integral} so as to put one of the two pulsars along the $z$ axis, with the other at polar angle $\theta_{ab}$. The explicit computation is carried out in detail in Ref.~\cite{Anholm09}, and the final result is the celebrated {\it Hellings--Downs curve}~\cite{HellingsDowns}
\begin{equation}
\label{eq:HD}
C(\theta_{ab}) = x_{ab}\,\log x_{ab} - \frac{x_{ab}}{6} + \frac{1}{3}\,,\qquad x_{ab}\equiv \frac{1-\cos\theta_{ab}}{2}\,.
\end{equation}

\begin{figure}
\centering
\includegraphics[width=0.7\textwidth]{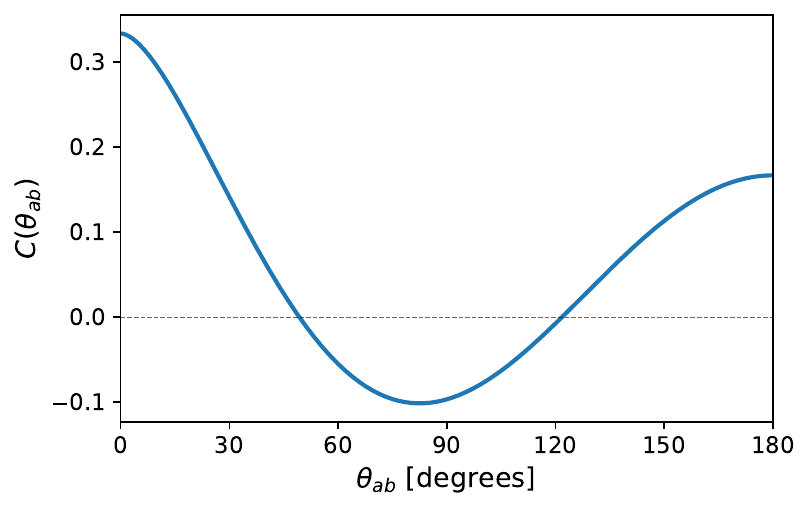}
\caption{\footnotesize The Hellings--Downs correlation curve $C(\theta_{ab})$ of Eq.~\ref{eq:HD}.}
\label{fig:HD}
\end{figure}

The Hellings--Downs curve is plotted in Fig.~\ref{fig:HD}. It starts from a positive auto-correlation value at zero separation, reaches a negative minimum near $\theta_{ab}\approx 90^\circ$ (where the two pulsars become anti-correlated), and rises back to a positive value at the antipodal direction. This distinctive ``quadrupolar'' angular pattern, with two zero crossings and a negative minimum, is a direct consequence of the spin-2 nature of the graviton, and it is what allows one to separate a genuine SGWB from competing sources of spatially-correlated timing residuals, such as clock errors (which produce a constant monopolar correlation), solar-system ephemeris errors (dipolar correlation), or per-pulsar intrinsic spin noise (diagonal, $\propto\delta_{ab}$)~\cite{nanograv15}. Observing the Hellings--Downs pattern across the pulsars of a PTA is therefore the smoking-gun criterion for declaring the detection of a nHz SGWB, as indeed happened with the 2023 data releases of NANOGrav, EPTA+InPTA, PPTA and CPTA~\cite{nanograv15,epta15,ppta15,cpta15}.

Let us finally translate Eq.~\ref{eq:zz_factorized} into the corresponding statement for the timing residuals $R_a(t)$. Substituting Eq.~\ref{eq:z_a_expansion} into the definition~\ref{eq:timing_residual}, computing the Gaussian average using~\ref{eq:Sh_def}, and taking $K_{ab}\to 1$ as above, one finds~\cite{maggiore_vol2}
\begin{equation}
\label{eq:Rab}
\langle R_a(t) R_b(t)\rangle = 2 C(\theta_{ab})\int_0^\infty \mathrm{d}f\,\frac{S_h(f)}{(2\pi f)^2}\,\big[1 -\cos(2\pi f t)\big]\,.
\end{equation}
The factor $1/(2\pi f)^2$  stems from the time integral in the definition of the timing residual, while the $[1-\cos(2\pi f t)]$ piece reflects the choice of $t=0$ implicit in Eq.~\ref{eq:timing_residual}. In practice, the frequencies that enter Eq.~\ref{eq:Rab} are confined to a finite range, bounded from below by the inverse observing baseline and from above by the cadence of the measurements; over that range the bracketed factor is of order unity. It is then useful to rewrite Eq.~\ref{eq:Rab} in terms of the {\it timing-residual power spectrum}
\begin{equation}
\label{eq:Pg_def}
P_g(f) \equiv \frac{2}{3}\,\frac{S_h(f)}{(2\pi f)^2}\,,
\end{equation}
which is the  quantity routinely  quoted in the PTA literature in place of $S_h(f)$.
For a power-law background $\Omega_{\rm gw}(f)\propto f^{2\alpha}$, Eqs.~\ref{eq:Omgw_Sh} and~\ref{eq:Pg_def} imply a timing-residual spectrum $P_g(f)\propto f^{2\alpha-5}$. Defining the spectral index $\gamma$ through the  PTA convention $P_g(f)\propto f^{-\gamma}$ (with an explicit minus sign, so that $\gamma>0$ corresponds to a ``red'' spectrum), this becomes $\gamma = 5 - 2\alpha$. For the population of quasi-circular, Newtonian-inspiral supermassive black hole binaries one has $\Omega_{\rm gw}\propto f^{2/3}$~\cite{Phinney01}, i.e. $\alpha = 1/3$, and therefore the benchmark spectral index
\begin{equation}
\gamma_{\rm SMBHB} = \frac{13}{3}\,,
\end{equation}
corresponding to a timing-residual spectrum $P_g(f)\propto f^{-13/3}$. This  is the  fiducial slope against which PTA results are routinely compared~\cite{nanograv15}.

\subsection{The pulsar timing array likelihood}
\label{sec:PTA_like}

In real PTA analyses, one does not work directly with the ensemble-average relations~\ref{eq:zz_factorized}--\ref{eq:Rab}, but rather constructs  a Bayesian likelihood for the observed  timing residuals and samples it (via MCMC or nested-sampling techniques) to infer the hyperparameters that  control the SGWB and the per-pulsar noise processes.  Let us  briefly  sketch the derivation of this likelihood, following~\cite{vanHaasteren09,vanHaasterenVallisneri,nanograv15,SmarraThesis}.

Consider first  a single pulsar, and drop the  pulsar label to lighten the notation. The observed TOAs are first fit against a  deterministic timing model accounting for the rotational spin-down of the pulsar, its astrometric parameters (sky location, proper motion, parallax) and -- if the pulsar is in a binary -- also its orbital elements. Let $\vec{\beta}_0$  denote the best-fit parameters. If the timing model were exact, the residuals $\vec{\delta t}\equiv \vec{t}_{\rm TOA}- \vec{t}_{\rm tim}(\vec{\beta}_0)$ would vanish.~\footnote{The $n$-th component of $\vec{\delta t}$ is  the time integral~\ref{eq:timing_residual} evaluated at the arrival time of the $n$-th pulse; $\vec{\delta t}$ is thus the discretization of the continuous-time timing residual $R_a(t)$ of the previous subsection.} In practice, however, $\vec{\beta}_0$ will differ from the true parameters  by a small offset $\vec{\xi}$, and additional physical effects not captured by the timing model (GW signals, intrinsic noise, etc.)\ will also be present.  Collecting the latter into a vector $\vec{t}_{\rm other}$ and linearizing in $\vec{\xi}$, we can write
\begin{equation}
\label{eq:dt_decomp}
\vec{\delta t} = M\,\vec{\xi} + \vec{t}_{\rm other}\,,
\end{equation}
where the  {\it design matrix} $M$ has entries $M_{nj}=\partial t_{{\rm tim},n}/\partial\beta_j\big|_{\vec{\beta}_0}$.

Following common  PTA practice, the low-frequency (``red'') component of $\vec{t}_{\rm other}$ is modeled as $F\,\vec{w}$, where $F$ is a matrix of sines and cosines at the frequencies $f_j=j/T$ ($j=1,\ldots,N_f$), where $T$ is the observing span,
\begin{equation}
F_{n,\,2j-1} =\cos(2\pi f_j t_n)\,,\qquad F_{n,\,2j} = \sin(2\pi f_j t_n)\,,
\end{equation}
and $\vec{w}$ is a vector of $2N_f$ Fourier coefficients (one cosine and one sine amplitude per harmonic), drawn from a zero-mean Gaussian distribution with diagonal covariance,
\begin{equation}
\label{eq:phi_diag}
\langle w_k w_l\rangle = \varphi_k\,\delta_{kl}\,,\qquad \varphi_{2j-1}=\varphi_{2j}=S(f_j)\,\Delta f\,,
\end{equation}
where $S(f)$ is the single-sided PSD of the red-noise process and $\Delta f\equiv 1/T$. The Fourier-basis expansion is a very efficient representation of the red-noise covariance: for typical PTA datasets, $N_f\sim 30$ is more than sufficient to describe the spectrum in the frequency range of interest~\cite{LentatiFourier}. For a power-law spectrum,
\begin{equation}
\label{eq:PL_spectrum}
S(f) = \frac{A^2}{12\pi^2}\left(\frac{f}{f_{\rm ref}}\right)^{-\gamma}\,f_{\rm ref}^{-3}\,,
\end{equation}
the amplitude $A$ and slope $\gamma$ are the two hyperparameters of the red-noise process; this same power-law parametrization will later be used both for the per-pulsar intrinsic spin noise and for the common SGWB signal.

After subtracting the timing-model and red-noise contributions, the remaining (``white'') residual $\vec{r}\equiv\vec{\delta t}-M\vec{\xi}-F\vec{w}$ is a zero-mean Gaussian process with covariance $N\equiv\langle\vec{r}\,\vec{r}^T\rangle$, parametrized by a small number of hyperparameters known as EFAC, EQUAD and ECORR in the PTA jargon~\cite{nanograv15,SmarraThesis}, characterizing respectively a multiplicative scaling of the nominal TOA uncertainty, a white-noise variance added in quadrature, and a jitter-like noise correlated across TOAs taken in the same observing epoch.

Stacking  the timing-model and red-noise contributions in a single matrix $\mathcal{T}\equiv [M\;\,F]$, and  the corresponding coefficient vectors in $\vec{b}\equiv(\vec{\xi},\,\vec{w})^T$, Eq.~\ref{eq:dt_decomp} becomes simply
\begin{equation}
\vec{\delta t} = \mathcal{T}\vec{b} + \vec{r}\,.
\end{equation}
Collectively denoting all hyperparameters (white-noise, red-noise amplitudes and slopes) by $\vec{\eta}$, the likelihood of $\vec{\delta t}$ conditional on $\vec{b}$ and $\vec{\eta}$ is the multivariate Gaussian
\begin{equation}
\label{eq:LH_b}
\mathcal{L}(\vec{\delta t}\,|\,\vec{b},\vec{\eta}) \propto \frac{1}{\sqrt{\det N}}\exp\left[-\frac{1}{2}(\vec{\delta t}-\mathcal{T}\vec{b})^T N^{-1}(\vec{\delta t}-\mathcal{T}\vec{b})\right]\,.
\end{equation}
The prior on $\vec{b}$ is Gaussian in the red-noise coefficients (with covariance given by Eq.~\ref{eq:phi_diag}). For the timing-model parameters $\vec{\xi}$, it is customary to impose an improper uniform prior ($\pi(\vec{\xi})=\text{const}$), since the inference on these parameters is heavily likelihood-dominated~\cite{vanHaasterenVallisneri,SmarraThesis}. The combined prior can then be written as
\begin{equation}
\label{eq:prior_b}
\pi(\vec{b}\,|\,\vec{\eta}) \propto \exp\left[-\frac{1}{2}\vec{b}^T B^{-1}\vec{b}\right]\,,\qquad B = \begin{pmatrix} \infty\,\mathbf{1}_{m\times m} & 0 \\ 0 & \varphi\end{pmatrix}\,,
\end{equation}
where $m$ is the number of timing-model parameters, $\varphi$ is the $2N_f\times 2N_f$ diagonal matrix with entries $\varphi_k$ given by Eq.~\ref{eq:phi_diag}, and the $\infty$ in the timing-model block is to be understood in the sense that $B^{-1}$ has a  vanishing block there (i.e.\ the prior exerts no constraint on $\vec{\xi}$).

Since only the hyperparameters $\vec{\eta}$ are of physical interest, we marginalize  the posterior $P(\vec{b},\vec{\eta}\,|\,\vec{\delta t}) \propto \mathcal{L}\,\pi\,p(\vec{\eta})$ over the nuisance vector $\vec{b}$. The resulting Gaussian integral~\cite{vanHaasteren09,vanHaasterenVallisneri} yields
\begin{equation}
\label{eq:marg_like_single}
P(\vec{\eta}\,|\,\vec{\delta t}) \propto \frac{p(\vec{\eta})}{\sqrt{\det C}}\exp\left[-\frac{1}{2}\vec{\delta t}^T\, C^{-1}\,\vec{\delta t}\right]\,,
\end{equation}
with the combined (white + red) covariance
\begin{equation}
\label{eq:C_pta}
C = N + \mathcal{T}\,B\,\mathcal{T}^T\,.
\end{equation}
The structure is analogous to the two-detector result~\ref{eq:C_SGWB_2det}: the white-noise covariance $N$ plays the role of the per-detector instrumental noise $\delta_{kl}S_{n,k}$, while $\mathcal{T} B \mathcal{T}^T$ encodes all the low-frequency stochastic processes (timing-model uncertainties, intrinsic spin noise, and GW-induced red noise), in the same way as $\gamma_{kl}(f)\,S_h(f)$ encodes the SGWB contribution in the two-interferometer case.

Since $\dim\vec{b}\ll N_{\rm TOA}$, the inverse of $C$ can be computed efficiently via the Woodbury identity~\cite{vanHaasteren09}
\begin{equation}
\label{eq:Woodbury}
C^{-1} = N^{-1} - N^{-1}\,\mathcal{T}\,\Sigma^{-1}\,\mathcal{T}^T\,N^{-1}\,,\qquad \Sigma \equiv B^{-1} + \mathcal{T}^T N^{-1}\mathcal{T}\,,
\end{equation}
which only requires the factorization of the much smaller matrix $\Sigma$ ($\dim\vec{b}\times\dim\vec{b}$), while $N^{-1}$ is trivially computed since $N$ is (quasi-)diagonal. This is the form implemented in publicly available PTA inference pipelines such as \texttt{ENTERPRISE}~\cite{enterprise}.

The generalization to the full array of $N_p$ pulsars is straightforward~\cite{nanograv15,SmarraThesis}. The key new ingredient is the cross-pulsar covariance of the Fourier coefficients~\cite{nanograv15},
\begin{equation}
\label{eq:B_PTA}
\Phi_{ab}(f_k) = \delta_{ab}\, S^{\rm int}_a(f_k)\,\Delta f + \Gamma_{ab}\,S^{\rm GWB}(f_k)\,\Delta f\,,
\end{equation}
where $S^{\rm int}_a(f)$ is the PSD of the intrinsic spin noise of pulsar $a$ and $S^{\rm GWB}(f)$ is the common SGWB timing-residual PSD, both modeled as power laws~\ref{eq:PL_spectrum}. The spatial correlation is encoded in the normalized Hellings--Downs function~\cite{nanograv15}
\begin{equation}
\label{eq:Gamma_PTA}
\Gamma_{ab} = \frac{3}{2}\,C(\theta_{ab}) + \frac{1}{2}\,\delta_{ab}\,,
\end{equation}
with $C(\theta_{ab})$ the Hellings--Downs curve of Eq.~\ref{eq:HD}. The $\delta_{ab}/2$ term accounts for the pulsar-term contribution to the auto-correlation: for $a=b$ the pulsar--pulsar term in the factor $K_{ab}$ of Eq.~\ref{eq:zz_integral} does not average out, and contributes the same angular integral as the Earth--Earth term, doubling the auto-correlation power relative to the Earth-term-only value $C(0)=1/3$~\cite{maggiore_vol2}. The prefactor $3/2$ is then a conventional rescaling  chosen so that $\Gamma_{aa}=(3/2)(1/3)+1/2=1$~\cite{nanograv15}.

The entries $\Phi_{ab}(f_k)$ generalize the single-pulsar variances $\varphi_k$ in the red-noise block of $B$ (Eq.~\ref{eq:prior_b}): for each frequency $f_k$, the diagonal entries ($a=b$) receive both the intrinsic spin noise $S^{\rm int}_a$ and the SGWB contribution, while the off-diagonal entries ($a\neq b$) carry only the SGWB part, weighted by the Hellings--Downs function $\Gamma_{ab}$. The multi-pulsar marginalized likelihood retains the same form as Eqs.~\ref{eq:marg_like_single}--\ref{eq:C_pta},
\begin{equation}
\label{eq:marg_like_full}
P(\vec{\eta}\,|\,\vec{\delta t}) \propto \frac{p(\vec{\eta})}{\sqrt{\det C}}\exp\left[-\frac{1}{2}\vec{\delta t}^{\,T}\, C^{-1}\,\vec{\delta t}\right]\,,\qquad C = N + \mathcal{T}\,B\,\mathcal{T}^T\,,
\end{equation}
where the vectors and matrices are now obtained by stacking the single-pulsar quantities over the $N_p$ pulsars of the array~\cite{vanHaasterenVallisneri}. Concretely, $\vec{\delta t}$ is a vector of length $N_{\rm TOA}=\sum_a N_a$, with $N_a$ the number of TOAs of pulsar $a$; $N$ ($N_{\rm TOA}\times N_{\rm TOA}$) and $\mathcal{T}\equiv[M\;\,F]$ ($N_{\rm TOA}\times d$, with $d=\sum_a(m_a+2N_f)$, $m_a$ being the number of timing-model parameters of pulsar $a$) are block-diagonal in the pulsar index, each pulsar contributing independently; and $B$ ($d\times d$) retains the block structure of Eq.~\ref{eq:prior_b} (improper prior on the timing-model parameters, Gaussian prior on the Fourier coefficients), but its red-noise block now acquires the cross-pulsar entries $B_{(a,k)(b,l)} = \delta_{kl}\,\Phi_{ab}(f_k)$ of Eq.~\ref{eq:B_PTA}, which couple different pulsars via the Hellings--Downs function $\Gamma_{ab}$. Since $d\ll N_{\rm TOA}$, the Woodbury identity~\ref{eq:Woodbury} again reduces the inversion of the large matrix $C$ to that of the much smaller $\Sigma$ ($d\times d$), with the (quasi-)diagonal $N$ inverted trivially. Equation~\ref{eq:marg_like_full} thus encodes the full signature of an isotropic SGWB on the array: the only new ingredient relative to the single-pulsar case is the off-diagonal structure of $B$, which introduces the Hellings--Downs spatial correlations.

By sampling the posterior in the hyperparameters $\vec{\eta}$ -- which include the per-pulsar intrinsic-noise amplitudes and slopes $(A_a,\gamma_a)$, the common SGWB amplitude and slope $(A_{\rm GWB},\gamma_{\rm GWB})$, and the white-noise parameters -- and computing Bayes factors between a ``noise-only'' model ($A_{\rm GWB}=0$, i.e.\ no common red-noise process), a ``common spectrum, uncorrelated'' model ($\Gamma_{ab}\to\delta_{ab}$, i.e.\ common spectral shape but no spatial correlations), and the full Hellings--Downs model (with $\Gamma_{ab}$ given by Eq.~\ref{eq:Gamma_PTA}), one can assess the statistical significance  of a SGWB detection and measure its spectral properties. It is this Bayesian model-comparison framework that was used to report the first evidence for a nHz SGWB  from the 2023 PTA data releases~\cite{nanograv15,epta15,ppta15,cpta15}.

\section*{Acknowledgements}
I would like to thank a number of friends and colleagues for insightful conversations over the years on topics related to the physics of gravitational waves, and in particular Stas Babak,  Emanuele Berti, Luc Blanchet, Alessandra Buonanno, Vitor Cardoso, Guillaume Faye, Scott Hughes, Luis Lehner, Eric Poisson and Luciano Rezzolla. Special thanks should also be extended
to Flavio Riccardi and Marika Giulietti, who transcribed my course at SISSA  in May 2020 into a set of notes, the structure of some parts of which  resembles these.
This work was supported by the European Union's H2020 ERC Consolidator Grant ``GRavity from Astrophysical to Microscopic Scales'' (Grant No.  GRAMS-815673), and  by the EU Horizon 2020 Research and Innovation Programme under the Marie Sklodowska-Curie Grant Agreement No. 101007855.


\end{document}